\documentclass[10pt,aps,prd,superscriptaddress,preprintnumbers,longbibliography,
amsfonts,amssymb,amsmath,showpacs,twocolumn]{revtex4-1}
\usepackage[utf8]{inputenc}
\usepackage[english]{babel}
\usepackage[unicode=true,bookmarksopen=true]{hyperref}

\usepackage{etoolbox}
\apptocmd{\sloppy}{\hbadness 10000\relax}{}{}

\usepackage{mathtools}

\let\originalleft\left
\let\originalright\right
\renewcommand{\left}{\mathopen{}\mathclose\bgroup\originalleft}
\renewcommand{\right}{\aftergroup\egroup\originalright}

\makeatletter
\newenvironment{equations}[1][]{\subequations\ifx\relax#1\relax\else\label{#1}
\fi\align\ignorespaces}{\endalign\ignorespacesafterend\endsubequations}
\def\@spliteq#1{\begin{equation}\begin{split}#1\end{split}\end{equation}}
\def\splitequation{\collect@body\@spliteq}

\makeatother

\usepackage{lmodern}
\usepackage[T1]{fontenc}
\DeclareMathAlphabet{\mathbfi}{OML}{cmm}{b}{it}
\DeclareFontShape{OMX}{cmex}{m}{n}{
  <-7.5> cmex7
  <7.5-8.5> cmex8
  <8.5-9.5> cmex9
  <9.5-> cmex10
}{}
\SetSymbolFont{largesymbols}{normal}{OMX}{cmex}{m}{n}
\SetSymbolFont{largesymbols}{bold}  {OMX}{cmex}{m}{n}

\renewcommand{\vec}[1]{{\ifnum9<1#1\mathbf{#1}
\else\ifcat\noexpand#1\relax\boldsymbol{#1}\else\mathbfi{#1}\fi\fi}}

\newcommand{\eqend}[1]{\,#1}
\newcommand{\mathe}{\mathrm{e}}
\newcommand{\mathi}{\mathrm{i}}
\newcommand{\total}{\mathop{}\!\mathrm{d}}
\newcommand{\abs}[1]{{\left\lvert{#1}\right\rvert}}
\newcommand{\symp}[2]{\left\langle{#1},{#2}\right\rangle}
\newcommand{\hankel}[2]{\mathop{}\mathrm{H}^{(#1)}_{#2}}
\newcommand{\bessel}[1]{\mathop{}\mathrm{J}_{#1}}
\newcommand{\hypergeom}[2]{{}_{#1}\mathrm{F}_{#2}}
\newcommand{\sgn}{\operatorname{sgn}}
\newcommand{\laplace}{\mathop{}\!\bigtriangleup}
\newcommand{\bigo}[1]{\mathcal{O}\left({#1}\right)}
\newcommand{\ket}[1]{\left\vert{#1}\right\rangle}
\newcommand{\bra}[1]{\left\langle{#1}\right\vert}

\allowdisplaybreaks

\date{\today}

\begin{document}

\title{Mode-sum construction of the covariant graviton two-point function in the 
Poincar{\'e} patch of de~Sitter space}

\author{Markus B. Fr\"ob}
\email{mfroeb@itp.uni-leipzig.de}
\affiliation{Institut f\"ur Theoretische Physik, Universit\"at Leipzig, 
Br\"uderstra\ss e 16, 04103 Leipzig, Germany}
\author{Atsushi Higuchi}
\email{atsushi.higuchi@york.ac.uk}
\affiliation{Department of Mathematics, University of York, Heslington, York 
YO10 5DD, United Kingdom}
\author{William C. C. Lima}
\email{william.correadelima@york.ac.uk}
\affiliation{Department of Mathematics, University of York, Heslington, York 
YO10 5DD, United Kingdom}


\begin{abstract}
We construct the graviton two-point function for a two-parameter family of 
linear covariant gauges in $n$-dimensional de~Sitter space. The construction is 
performed via the mode-sum method in the Bunch-Davies vacuum in the 
Poincar{\'e} 
patch, and a Fierz-Pauli mass term is introduced to regularize the infrared (IR) 
divergences. The resulting two-point function is de~Sitter invariant and free 
of IR divergences in the massless limit (for a certain range of parameters), 
although analytic continuation with respect to the mass for the pure-gauge sector 
of the two-point function is necessary for this result. This general result 
agrees with the propagator obtained by analytic continuation from the sphere 
[Phys.
~Rev.~D \textbf{34}, 3670 (1986); Class.~Quant.~Grav. \textbf{18}, 4317 (2001)]. 
However, if one starts with strictly zero mass theory, the IR divergences are 
absent only for a specific value of one of the two parameters, with the other 
parameter left generic. These findings agree with recent calculations in the 
Landau (exact) gauge [J.~Math.~Phys. \textbf{53}, 122502 (2012)], where IR 
divergences do appear in the spin-two (tensor) part of the two-point function. 
However, 
we find the strength (including the sign) of the IR divergence to be different 
from the one found in this reference.
\end{abstract}

\maketitle


\section{Introduction}
\label{sec:introduction}

Quantum fields in de~Sitter space have received increasing attention in recent 
years due to the accumulating experimental evidence for an inflationary epoch 
during the early moments of our Universe~\cite{smoot_ajl_1992,hinshaw_ajss_2007,
guth_prd_1981,linde_plb_1982,albrecht_steinhardt_prl_1982}. Moreover, the 
observations showing that the expansion of our Universe is accelerating suggest 
that it might attain a de~Sitter stage in the 
future~\cite{ries_aj_1998,perlmutter_aj_1999}. On more theoretical grounds, the 
interest in de~Sitter space stems from the fact that it is the maximally 
symmetric solution of Einstein equation with positive cosmological constant. Its 
isometries, very much like in the Minkowskian case, are expect to be reflected 
in the structure of the theory. This is indeed the case for massive scalar 
fields~\cite{chernikov_tagirov_ap_1968,schomblond_spindel_ahp_1976,
bunch_davies_prsa_1978,allen_prd_1985}.

The analysis of gravitons propagating on the de~Sitter geometry in this 
theoretical context is particularly relevant since their observable implications 
can serve as a probe to the inflationary era as well as a window to low-energy 
quantum gravity. The question of the existence of a state for free gravitons in 
de~Sitter space that shares the background symmetries, however, is still a 
matter of contention in the literature. This thirty-year long controversy stems 
mainly from the fact that the natural graviton modes in the spatially flat 
coordinate patch of the de~Sitter space (the Poincar{\'e} patch), which is the 
part relevant for inflationary cosmology, resemble those of a massless, 
minimally coupled (MMC) scalar field~\cite{ford_parker_prd_1977a}. As is 
well known, the natural vacuum state (the Bunch-Davies state) is IR-divergent 
for the MMC scalar, and no de~Sitter--invariant state 
exists~\cite{ford_parker_prd_1977a, allen_prd_1985}. (Note, however, that there 
is a unitary representation of the de~Sitter group corresponding to the free MMC 
scalar field~\cite{naimark_amstr_1957, thomas_am_1941,dixmier_bsmf_1961,
kuriyan_mukunda_sudarshan_cmp_1968,ottoson_cmp_1968,higuchi_jmp_1987} and that 
one can realize this representation by removing the mode responsible for the IR 
divergence~\cite{kirsten_garriga_prd_1993}.) The same is true for other 
cosmological (Friedman-Lema{\^\i}tre-Robertson-Walker, FLRW) spacetimes, of 
which the Poincar{\'e} patch of de~Sitter space is a special case. The 
similarities between gravitons and scalar field modes led Ford and Parker to 
conclude that in all these spacetimes the graviton two-point function is 
IR-divergent, although the divergences they report do not appear in the physical 
quantities they studied~\cite{ford_parker_prd_1977b}.

A noteworthy distinction between the scalar field and linearized gravity is that 
the latter possesses gauge symmetries arising from the diffeomorphism invariance 
of general relativity. Therefore, it is important to settle whether these IR 
divergences are restricted to the gauge sector of linearized gravity, or they 
appear also in the physical sector, the exact definition of which is the source 
of disagreement in the literature. This question was addressed in Refs.~\cite{
higuchi_npb_1987,allen_npb_1987}, where it was shown that in the 
traceless-transverse-synchronous gauge the IR-divergent part of the graviton 
two-point function can be expressed in a nonlocal pure-gauge form. The 
discussion was taken further by the authors of 
Ref.~\cite{higuchi_marolf_morrison_cqg_2011}, who observed that a local gauge 
transformation on the graviton modes is sufficient to eliminate the IR 
divergences plaguing the graviton two-point function in that gauge. Moreover, 
examples of other gauges and coordinate systems in which the graviton two-point 
function is IR finite have been worked out~\cite{allen_prd_1986, 
allen_turyn_npb_1987, hawking_hertog_turok_prd_2000, higuchi_kouris_cqg_2001a, 
higuchi_kouris_cqg_2001b, higuchi_weeks_cqg_2003, 
bernar_crispino_higuchi_prd_2014}.

Another interesting test for the gauge nature of these IR divergences is 
provided by the linearized Weyl tensor, which is a local and gauge-invariant 
observable in the linearized theory. It was shown that the two-point function of 
the linearized Weyl tensor is IR finite even if computed using a 
de~Sitter--noninvariant graviton two-point function with an IR 
cutoff~\cite{mora_tsamis_woodard_prd_2012,froeb_roura_verdaguer_jcap_2014}, and 
that the result agrees with the one of Ref.~\cite{kouris_cqg_2001}, which was 
calculated from the covariant two-point 
function~\cite{allen_turyn_npb_1987,higuchi_kouris_cqg_2001a,
higuchi_kouris_cqg_2001b}. Moreover, it was shown that this two-point function 
is IR finite also in slow-roll inflationary FLRW spacetimes, where the IR 
divergences of the graviton two-point functions are usually worse than in exact 
de~Sitter space, as long as the slow-roll parameter is not too 
large~\cite{froeb_jcap_2014}. Thus, the finiteness of the two-point function of 
the linearized Weyl tensor is not an accident due to the maximal symmetry of 
de~Sitter space.

The covariant graviton two-point function was also shown to be physically 
equivalent to the transverse-traceless-synchronous one constructed on global 
de~Sitter space~\cite{faizal_higuchi_prd_2012} (in the sense that they produce 
the same two-point function of any local gauge-invariant tensor linear in the 
graviton field), and the latter is known to be 
IR finite~\cite{higuchi_weeks_cqg_2003}. Therefore, it does not come as a 
surprise that within the gauge-invariant formulation of linearized gravity of 
Ref.~\cite{fewster_hunt_rmp_2013}, the graviton and the Weyl-tensor two-point 
functions were shown to contain the same gauge-invariant information in the 
Poincar{\'e} patch~\cite{higuchi_arxiv_2012}. Furthermore, the existence of a 
de~Sitter--invariant Hadamard state for the graviton was recently verified in 
Ref.~\cite{morrison_arxiv_2013}, although the definition of such a state in this 
paper is technically different from the one originally proposed in 
Ref.~\cite{fewster_hunt_rmp_2013}.

Nevertheless, over the years many authors have employed IR-regulated graviton 
propagators which explicitly break de~Sitter invariance in computing loop 
corrections. From these studies it has been claimed, e.g., that the 
cosmological constant decreases in time~\cite{tsamis_woodard_plb_1993, 
tsamis_woodard_ap_1995, tsamis_woodard_cqg_2009} (see 
Refs.~\cite{garriga_tanaka_prd_2008, tsamis_woodard_prd_2008} for a criticism of 
these results and the rebuttal), that geometrical fluctuations can grow in the 
IR in inflationary spacetimes~\cite{giddings_sloth_jcap_2011a, 
giddings_sloth_prd_2011b, giddings_sloth_prd_2012}, and that some coupling 
constants change in time~\cite{kitamoto_kitazawa_prd_2013a, 
kitamoto_kitazawa_prd_2013b}. Overall, these results suggest that the IR sector 
of the interacting theory could harbor important physical effects which break 
de~Sitter invariance. They are rather intriguing, nonetheless, given the 
pure-gauge form of the de~Sitter--breaking terms in the free propagator, on 
which those effects rely. Thus, loop calculations starting from 
de~Sitter--invariant graviton propagators would be important to decide whether 
or not those effects are gauge artifacts.

However, as mentioned above, there are still some objections in the literature 
concerning the derivation of de~Sitter--invariant graviton 
propagators~\cite{woodard_ijmp_2014}. Indeed, a graviton two-point function that 
breaks de~Sitter invariance was obtained recently in the so-called exact 
de~Donder gauge~\cite{miao_tsamis_woodard_jmp_2011, khaya_miao_woodard_jmp_2012,
mora_tsamis_woodard_jmp_2012} by means of a formalism of covariant projection 
operators acting on scalar quantities. Then, in a follow up paper, Morrison 
took advantage of some freedom in the solutions of the differential 
equations satisfied by those scalar quantities in that formalism and 
constructed a de~Sitter--invariant graviton two-point function, which is 
equivalent to the de~Sitter--noninvariant one when smeared with 
transverse-traceless tensors of compact support~\cite{morrison_arxiv_2013}, in 
order to extract the gauge-invariant content~\cite{fewster_hunt_rmp_2013}. This 
work of Morrison's was criticized in Ref.~\cite{miao_mora_tsamis_woodard_2014}, 
with the main criticism being that the freedom argued to occur in 
Ref.~\cite{morrison_arxiv_2013} is not actually present if one were to derive 
the two-point function from a mode sum.

In this paper, we revisit the question of the existence of a de~Sitter--invariant 
state for gravitons, performing canonical quantization of the graviton field in 
the Poincar{\'e} patch of the de~Sitter space and then constructing the 
corresponding two-point function via mode sums. We consider a two-parameter 
family of covariant linear gauges with a Fierz-Pauli mass 
term~\cite{fierz_pauli_prsl_1939} which serves as an IR regulator. We then study 
the IR behavior of the two-point function thus obtained and analyze the 
convergence of the mode sums when we take the mass to zero after performing the 
momentum integrals [and after analytic continuation with respect to the mass in 
the pure-gauge vector (spin-1) sector]. Note that while in flat space the van 
Dam-Veltman-Zakharov discontinuity prevents this limit from being 
smooth~\cite{vandam_veltman_npb_1970, zakharov_jetplett_1970}, there is no such 
problem in de~Sitter space~\cite{higuchi_npb_1987, higuchi_npb_1989}, at least 
in the linear regime. We also consider the result obtained by taking the zero 
mass limit before the momentum integration. We recover the de~Sitter--invariant 
and IR-finite results of Ref.~\cite{morrison_arxiv_2013} if we take the massless 
limit after the momentum integration. Interestingly, if we use the procedure of 
taking the massless limit before performing the momentum integrals, the result 
differs from both the IR-divergent and IR-finite results in the literature. We 
do agree with the IR-divergent results~\cite{miao_tsamis_woodard_jmp_2011, 
khaya_miao_woodard_jmp_2012, mora_tsamis_woodard_jmp_2012} in the sense that for 
their choice of gauge parameters IR divergences appear in the tensor (spin-2) 
sector. However we disagree even on the sign of the IR divergence. Our complete 
results reveal that there exists a de~Sitter--invariant, IR-finite graviton 
two-point function in the massless limit for a generic choice of gauge 
parameters if this limit is taken after the momentum integrals (and after a 
certain analytic continuation in the pure-gauge sector) and for a particular 
value of one of the gauge parameters if it is taken before the momentum 
integrals.

The rest of the paper is organized as follows: we start in 
Sec.~\ref{subsec:symplectic_product_method} with a brief discussion on the 
canonical quantization of a general free field theory via the symplectic product 
method. In Sec.~\ref{subsec:lagrangian}, we present the Lagrangian density for 
the graviton, with the mass and gauge-fixing terms in addition to the 
gauge-invariant part, and decompose the field in its scalar, vector, and 
transverse-tensor sectors. We canonically quantize each sector with the aid of 
symplectic product method in Secs.~\ref{subsec:scalar_sector}, 
\ref{subsec:vector_sector}, and \ref{subsec:tensor_sector} and calculate the 
corresponding two-point functions through their mode-sum definitions. The 
massless limit of the two-point function thus obtained is then studied in 
Sec.~\ref{subsec:massless_limit}, and its behavior for large 
separations is obtained in Sec.~\ref{subsec:large_z}. The convergence of the 
momentum integrals in the mode sums in the IR (with the massless limit taken 
before or after the momentum integration) is investigated in detail in 
Sec.~\ref{sec:ir_issues}. We summarize and discuss our results and make some 
remarks on 
them in Sec.~\ref{sec:discussion}. Throughout this paper we use units such that 
$\hbar = c = 1$, and use the mostly plus convention for the metric.


\section{Canonical quantization}
\label{sec:canonical_quantization}

\subsection{Overview of the symplectic product method}
\label{subsec:symplectic_product_method}

Let us consider a free field theory described by a Lagrangian density 
$\mathcal{L}$ which is defined on a spacetime with background metric $g_{ab}$ 
and is a function of the symmetric tensor field $h_{ab}$ and its covariant 
derivative $\nabla_c h_{ab}$. The canonical conjugate momentum current $p^{abc}$ 
is defined from $\mathcal{L}$ by
\begin{equation}
\label{momentum_current}
p^{abc} \equiv \frac{1}{\sqrt{-g}} \frac{\partial \mathcal{L}}{\partial( 
\nabla_a h_{bc})} \eqend{,}
\end{equation}
where $g$ is the determinant of the background metric. The Euler-Lagrange 
equation for $h_{ab}$ can then be written as
\begin{equation}
\label{euler_lagrange_eq}
\nabla_a p^{abc} - \frac{1}{\sqrt{-g}} \frac{\partial \mathcal{L}}{\partial 
h_{bc}} = 0 \eqend{.}
\end{equation}
For any two solutions $h^{(1)}_{ab}$ and $h^{(2)}_{ab}$ of 
Eq.~\eqref{euler_lagrange_eq}, it follows that the current
\begin{equation}
J^{(1,2)a} \equiv - \mathi \left( h^{(1)*}_{bc} p^{(2)abc} - h^{(2)}_{bc} 
p^{(1)abc*} \right) \label{symp-prod}
\end{equation}
(with a star denoting complex conjugation) is conserved, i.e., $\nabla_a 
J^{(1,2)a} = 0$. Hence, the symplectic product
\begin{equation}
\label{symplectic_prod}
\symp{h^{(1)}_{ab}}{h^{(2)}_{cd}} \equiv \int_\Sigma J^{(1,2)a} n_a \total 
\Sigma \eqend{,}
\end{equation}
with $\Sigma$ a Cauchy surface, $n_a$ the future-directed unit normal to it, and 
$\total \Sigma$ the normalized surface element, is time-independent and thus 
independent of the choice of $\Sigma$.

If the symplectic product \eqref{symplectic_prod} is nondegenerate, i.e., if 
there is no solution with zero symplectic product with all solutions, we 
quantize the classical field $h_{ab}$ by imposing the usual canonical 
commutation relations on the field operator $\hat{h}_{ab}$ and its canonical 
conjugate momentum $n_a \hat{p}^{abc}$:
\begin{equation}
\label{field_momentum_commutator}
\left.\left[ \hat{h}_{ab}(\vec{x}), n_e \hat{p}^{ecd}(\vec{x}') 
\right]\right\rvert_\Sigma = - \mathi \delta_{(a}^c \delta_{b)}^d \delta\left( 
\vec{x}, \vec{x}' \right) \eqend{,}
\end{equation}
with all other equal-time commutators vanishing. The $\delta$ distribution 
appearing here is the one associated with the spatial sections, defined by
\begin{equation}
\int \delta\left( \vec{x}, \vec{x}' \right) f(\vec{x}') N(\vec{x}') \total 
\Sigma' = f(\vec{x}) \eqend{,}
\end{equation}
where $N = n_a (\partial/\partial t)^a$ is the lapse function, with $t$ 
parametrizing the Cauchy surfaces and increasing towards the future. In order to 
construct a representation for the field operator, one then chooses a complete 
set of 
modes $\left\{ h_{ab}^{(k)}\right\}_{k\in\mathcal{I}}$ and $\left\{ 
h_{ab}^{(k)*}\right\}_{k\in\mathcal{I}}$ of the Euler-Lagrange 
equation~\eqref{euler_lagrange_eq}, with $\mathcal{I}$ an appropriate set of 
quantum numbers, such that $h_{
ab}^{(k)}$ and $h_{ab}^{(k')*}$ have vanishing symplectic product. Then one 
expands the quantum field $\hat{h}_{ab}$ in terms of this complete set of modes 
as
\begin{equation}
\label{field_op_expansion}
\hat{h}_{ab} = \sum_{k\in\mathcal{I}} \left( \hat{A}_k h_{ab}^{(k)} + 
\hat{A}_k^\dagger h_{ab}^{(k)*} \right) \eqend{.}
\end{equation}
From the symplectic product~\eqref{symplectic_prod} and the commutation 
relations~\eqref{field_momentum_commutator}, it is straightforward to show that
\begin{equation}
\label{M_matrix}
\left[ \left\langle h^{(k)}_{ab}, \hat{h}_{cd} \right\rangle, \left\langle 
\hat{h}_{ab}, h^{(l)}_{cd} \right\rangle \right] =
\symp{h^{(k)}_{ab}}{h^{(l)}_{cd}} \equiv M_{kl} \eqend{,}
\end{equation}
where the (Hermitian) matrix $M_{kl}$ is invertible, thanks to the completeness 
of our set of modes. On the other hand, by employing the 
expansion~\eqref{field_op_expansion} one obtains
\begin{equation}
\left[ \symp{h^{(k)}_{ab}}{\hat{h}_{cd}}, \symp{\hat{h}_{ab}}{h^{(l)}_{cd}} 
\right] = \sum_{k', l' \in \mathcal{I}} M_{kk'} \left[
\hat{A}_{k'}, \hat{A}_{l'}^\dagger \right] M_{l'l} \eqend{.}
\end{equation}
Therefore, one concludes that the operators $\hat{A}_k$ and $\hat{A}_k^\dagger$ 
must satisfy the following commutation relations:
\begin{equation}
\label{a_a_dagger_commutator}
\left[ \hat{A}_k, \hat{A}_l^\dagger \right] = \left( M^{-1} \right)_{kl} 
\eqend{.}
\end{equation}
By a similar calculation with $h^{(l)}_{cd}$ replaced by $h^{(l)*}_{cd}$ we find
\begin{equation}
\left[ \hat{A}_k, \hat{A}_l\right] = 0 \eqend{.}
\end{equation}
Thus, in the state $\ket{0}$ annihilated by all the $\hat{A}_k$, the Wightman 
two-point function,
\begin{equation}
\Delta_{abc'd'}(x,x') \equiv \bra{0} \hat{h}_{ab}(x) \hat{h}_{c'd'}(x') \ket{0} 
\eqend{,}
\end{equation}
is given by the mode sum
\begin{equation}
\label{wightman_mode_sum}
\Delta_{abc'd'}(x,x') = \sum_{k,l \in \mathcal{I}} \left( M^{-1} \right)_{kl} 
h^{(k)}_{ab}(x) h^{(l)}_{c'd'}(x') \eqend{,}
\end{equation}
where the primed indices refer to quantities defined on the point $x'$ here and 
below. For a more detailed exposition of this method, which is completely 
equivalent to the usual canonical quantization scheme but is technically easier 
to 
use if the field equations for $\hat{h}_{ab}$ are complicated, see, e.g., 
Ref.~\cite{higuchi_npb_1989}.

\subsection{The Lagrangian}
\label{subsec:lagrangian}

We start from the Einstein-Hilbert Lagrangian density for gravity plus a 
positive cosmological constant $\Lambda$ in an $n$-dimensional spacetime,
\begin{equation}
\label{L_grav}
\mathcal{L}_\text{grav} \equiv \frac{1}{\kappa^2} \left( \bar{R} - 2 \Lambda 
\right) \sqrt{-\bar{g}} \eqend{,}
\end{equation}
where $\kappa^2 \equiv 16\pi G_\text{N}$ with Newton's constant $G_\text{N}$, 
$\bar{g}_{ab}$ denotes the full metric (background plus perturbations), and 
$\bar{R}$ and $\bar{g}$ are the corresponding scalar curvature and metric 
determinant,
 respectively. We take the background metric $g_{ab}$ to be the de~Sitter 
metric, which in the Poincar{\'e} patch reads
\begin{equation}
\label{de_sitter_metric}
g_{ab} = a(\eta)^2 \eta_{ab} \eqend{,}
\end{equation}
with $\eta_{ab}$ the flat (Minkowski) metric, $a(\eta) \equiv (-H\eta)^{-1}$ the 
conformal factor, $\eta\in(-\infty,0)$ the conformal time, and $H$ the Hubble 
constant.

We write the full metric as $\bar{g}_{ab} = g_{ab} + \kappa h_{ab}$ and expand 
the Lagrangian~\eqref{L_grav} up to second order in the perturbation $h_{ab}$. 
Choosing the cosmological constant to be $2 \Lambda = (n-1)(n-2) H^2$, one 
finds 
that the part of the Lagrangian density linear in the metric perturbations 
vanishes up to a boundary term. The second-order part can then be cast, up to a 
boundary term, into the form
\begin{splitequation}
\label{L_inv}
\mathcal{L}_\text{inv} &\equiv - \frac{1}{4} \bigg[ \nabla_c h_{ab} \nabla^c 
h^{ab} - \nabla_c h \nabla^c h \\
&\ + 2 \nabla_a h^{ab} \nabla_b h - 2 \nabla^a h_{ac} \nabla_b h^{bc} \\
&\ + 2 H^2 \left( h_{ab} h^{ab} + \frac{n-3}{2} h^2 \right) \bigg] \sqrt{-g} 
\eqend{,}
\end{splitequation}
where we have used that the Riemann tensor of de~Sitter space is $R_{abcd} = H^2 
\left( g_{ac} g_{bd} - g_{ad} g_{bc} \right)$. As usual, $h \equiv h^a{}_a$ 
denotes the trace of $h_{ab}$, and indices are raised and lowered with the 
background metric $g_{ab}$. It can readily be verified that 
$\mathcal{L}_\text{inv}$ is invariant (up to a boundary term) under the gauge 
transformation
\begin{equation}
\label{gauge_transformation}
h_{ab} \rightarrow h_{ab} - \nabla_a \xi_b - \nabla_b \xi_a
\end{equation}
for any vector field $\xi^a$.

As is well known, because of this gauge symmetry a direct quantization of 
$\mathcal{L}_\text{inv}$ is not possible. The reason is that the symplectic 
product~\eqref{symplectic_prod} between a pure gauge mode and any other solution 
of the linearized Einstein equation is identically zero~\footnote{See 
Ref.~\cite{friedman_cmp_1978} for a formal proof of this statement.}, which 
implies that $M$ is a degenerate matrix and that the inverse $M^{-1}$ in 
Eq.~\eqref{a_a_dagger_commutator} does not exist. This difficulty can be 
circumvented by adding a gauge-fixing term in the Lagrangian density in such a 
way that $M$ becomes an invertible matrix. In our case, we add the following 
most general linear covariant gauge-fixing term to the Lagrangian density:
\begin{equation}
\label{L_gf}
\mathcal{L}_\text{gf} \equiv - \frac{1}{2\alpha} G_b G^b \sqrt{-g} \eqend{,}
\end{equation}
with
\begin{equation}
\label{gb_def}
G_b \equiv \nabla^a h_{ab} - \frac{1+\beta}{\beta} \nabla_b h,
\end{equation}
where $\alpha$ and $\beta$ are real parameters. The analogue of the Feynman 
gauge ($\xi=1$) in electromagnetism, for which the gauge-fixing term is $- 
1/(2\xi) (\nabla_a A^a)^2$, is achieved for $\alpha = 1$ and $\beta = -2$, while 
a one-parameter family analogous to the Landau gauge can be obtained in the 
limit $\alpha \to 0$. (These gauges are called the exact gauge, and, for the 
special case $\beta = -2$, the exact de Donder gauge in 
Refs.~\cite{miao_tsamis_woodard_jmp_2011, khaya_miao_woodard_jmp_2012}.) 
However, these gauges may not always be the most useful ones. An example of a 
less well known but useful gauge from electromagnetism is the Fried-Yennie 
gauge, a covariant gauge with gauge parameter $\xi = 
(n-1)/(n-3)$~\cite{fried_yennie_pr_1958}. This gauge makes individual Feynman 
diagrams IR finite (as opposed to only their sum), and is extremely useful in 
bound-state calculations~\cite{eides_grotch_shelyuto_pr_2001}. Note that our 
$\beta$ is the same as the one used in Refs.~\cite{higuchi_kouris_cqg_2001a, 
higuchi_kouris_cqg_2001b}, while it is related to the parameter $b$ used in 
Ref.~\cite{mora_tsamis_woodard_jmp_2012} (for which we have to take in addition 
$\alpha = 0$) by
\begin{equation}
\beta = \frac{2}{b-2} \eqend{.}
\end{equation}
We also introduce in the Lagrangian density the Fierz-Pauli mass 
term~\cite{fierz_pauli_prsl_1939},
\begin{equation}
\label{L_mass}
\mathcal{L}_\text{mass} \equiv - \frac{m^2}{4} \left( h_{ab} h^{ab} - h^2 
\right) \sqrt{-g} \eqend{,}
\end{equation}
which will serve to regulate the IR behavior of the theory. We will take the $m 
\to 0$ limit in the end.

The total Lagrangian density
\begin{equation}
\label{L_total}
\mathcal{L} \equiv \mathcal{L}_\text{inv} + \mathcal{L}_\text{gf} + 
\mathcal{L}_\text{mass}
\end{equation}
leads to the field equation
\begin{equation}
\label{field_eq}
L^{(\text{inv})cd}_{ab} h_{cd} + L^{(\text{gf})cd}_{ab} h_{cd} - m^2 \left( 
h_{ab} - g_{ab} h \right) = 0 \eqend{,}
\end{equation}
where we have defined the differential operators $L^{(\text{inv})cd}_{ab}$ and 
$L^{(\text{gf})cd}_{ab}$ by
\begin{widetext}
\begin{equation}
\label{inv_op}
L^{(\text{inv})cd}_{ab} h_{cd} \equiv 2 \int \frac{\delta 
\mathcal{L}_\text{inv}}{\delta h^{ab}} \total^n x = \nabla^2 h_{ab} + g_{ab}
\left( \nabla_c \nabla_d h^{cd} - \nabla^2 h \right) + \nabla_a \nabla_b h - 2 
\nabla_{(a} \nabla^c h_{b)c} - H^2 \big[ 2 h_{ab} + (n-3)
g_{ab} h \big] \eqend{,}
\end{equation}
with the abbreviation $\nabla^2 = \nabla_a \nabla^a$, and
\begin{equation}
\label{gf_op}
L^{(\text{gf})cd}_{ab} h_{cd} \equiv 2 \int \frac{\delta 
\mathcal{L}_\text{gf}}{\delta h^{ab}} \total^n x = \frac{2}{\alpha} \left(
\nabla_{(a} G_{b)} - \frac{1+\beta}{\beta} g_{ab} \nabla_c G^c \right) \eqend{,}
\end{equation}
\end{widetext}
with $G_b$ defined by Eq.~\eqref{gb_def}. For the total Lagrangian density, 
Eq.~\eqref{L_total}, the canonical conjugate momentum current defined by
Eq.~\eqref{momentum_current} can be written in the form
\begin{equation}
\label{canonical_momentum}
p^{abc} = p^{abc}_\text{inv} + p^{abc}_\text{gf} \eqend{,}
\end{equation}
with
\begin{splitequation}
\label{inv_momentum}
p^{abc}_\text{inv} \equiv - \frac{1}{2} \bigg[ &\nabla^a h^{bc} + g^{a(b} \left( 
\nabla^{c)} h - 2 \nabla_d h^{c)d} \right) \\
&\quad+ g^{bc} \left( \nabla_d h^{ad} - \nabla^a h \right) \bigg]
\end{splitequation}
coming from the gauge-invariant Lagrangian density $\mathcal{L}_\text{inv}$, 
Eq.~\eqref{L_inv}, and with
\begin{equation}
\label{gf_momentum}
p^{abc}_\text{gf} \equiv - \frac{1}{\alpha} \left( g^{a(b} G^{c)} - 
\frac{1+\beta}{\beta} g^{bc} G^a \right)
\end{equation}
following from the gauge-fixing term $\mathcal{L}_\text{gf}$, Eq.~\eqref{L_gf}.

In order to obtain the two-point function, we have to find a complete set of 
modes $h^{(k)}_{ab}$ that satisfy Eq.~\eqref{field_eq}. This task is facilitated 
by decomposing the metric perturbation into (covariant) scalar, vector, and 
tensor sectors as
\begin{equation}
\label{graviton_decomposition}
h_{ab} = h_{ab}^\text{(S)} + h_{ab}^\text{(V)} + h_{ab}^\text{(T)} \eqend{.}
\end{equation}
The scalar sector is defined in terms of two scalars, $B$ and $\Psi$, as
\begin{equation}
\label{scalar_sector}
h_{ab}^\text{(S)} \equiv \nabla_a \nabla_b B + g_{ab} \Psi \eqend{,}
\end{equation}
the vector sector is given by
\begin{equation}
\label{vector_sector}
h_{ab}^\text{(V)} \equiv \nabla_a v_b + \nabla_b v_a \eqend{,}
\end{equation}
with the vector field $v_a$ satisfying the divergence-free condition, $\nabla_a 
v^a = 0$, and the tensor sector is transverse and traceless, i.e., $h^\text{(T)} 
= \nabla^a h_{ab}^\text{(T)} = 0$. It is here that the symplectic product 
method shows its advantages, as it is not necessary to determine the canonical 
conjugate momentum for each of these sectors taking into account these 
constraints, which would be a tedious task. Instead, one can simply restrict 
the 
canonical conjugate momentum current~\eqref{canonical_momentum} to each of the 
sectors. Since the scalar, vector, and tensor sectors are orthogonal to each 
other with respect to the symplectic product~\eqref{symplectic_prod}, the 
matrix 
$M$ defined by Eq.~\eqref{M_matrix} becomes block-diagonal and we can study each 
sector separately. [This is of course equivalent to the fact that the three 
sectors decouple in the Lagrangian~\eqref{L_total}].


\section{The two-point function} \label{sec:two-point-function}

\subsection{Scalar sector}
\label{subsec:scalar_sector}

We begin the construction of the field modes in the scalar sector. For that 
sector, the field equation~\eqref{field_eq} can be cast in the form
\begin{equation}
\label{scalar_sector_field_eq}
\nabla_a \nabla_b \Phi_1 + g_{ab} \Phi_2 = 0 \eqend{,}
\end{equation}
with $\Phi_1$ and $\Phi_2$ defined by
\begin{splitequation}
\Phi_1 &\equiv - \frac{2}{\alpha \beta} \left[ \nabla^2 - (n-1) \beta H^2 + 
\frac{\alpha \beta}{2} m^2 \right] B \\
&\quad+ \frac{(n-2) \alpha \beta - 2 n - 2 (n-1) \beta}{\alpha \beta} \Psi
\end{splitequation}
and
\begin{splitequation}
\Phi_2 &\equiv - (n-2) \nabla^2 \Psi + \frac{2 (1+\beta) [ n + (n-1)\beta 
]}{\alpha \beta^2} \nabla^2 \Psi \\
&\quad+ \frac{2 (1+\beta)}{\alpha \beta^2} \left[ \nabla^2 - (n-1) \beta H^2 + 
\frac{\alpha\beta}{2} m^2 \right] \nabla^2 B \\
&\quad+ (n-1) [ m^2 - (n-2) H^2 ] \Psi - \frac{1}{\beta} m^2 B \eqend{.}
\end{splitequation}
Obviously, Eq.~\eqref{scalar_sector_field_eq} is satisfied if $\Phi_1 = \Phi_2 = 
0$, which leads to the equations
\begin{equation}
\label{B_mass_eq}
\left[ \nabla^2 - (n-1) \beta H^2 + \frac{\alpha \beta}{2} m^2 \right] B = - 
\left( n + \frac{\lambda \beta}{2} \right) \Psi
\end{equation}
and
\begin{splitequation}
\label{Psi_mass_eq}
&\left[ \nabla^2 - (n-1) \beta H^2 + \frac{\alpha \beta}{2} m^2 \right] \Psi \\
&\quad= - \frac{\beta m^2}{n-2} \left[ \frac{\alpha m^2}{2} - (n-1) H^2 \right] 
B - \frac{n + \lambda \beta}{n-2} m^2 \Psi \eqend{,}
\end{splitequation}
where we have defined $\lambda \equiv 2 (n-1) - (n-2) \alpha$. Due to the 
presence of the mass, Eqs.~\eqref{B_mass_eq} and~\eqref{Psi_mass_eq} are coupled 
and cannot be solved in a straightforward manner. Nevertheless, since we are 
interested in the massless limit and since the scalar sector is well behaved in 
the IR if $\beta > 0$ (as noted before, see 
Refs.~\cite{higuchi_kouris_cqg_2001a,higuchi_kouris_cqg_2001b,
mora_tsamis_woodard_jmp_2012}), we make this 
assumption for $\beta$ and let $m = 0$ straight away. This leads to the 
equations
\begin{equation}
\label{B_eq}
\left[ \nabla^2 - (n-1) \beta H^2 \right] B = - \left( n + \frac{\lambda 
\beta}{2} \right) \Psi
\end{equation}
and
\begin{equation}
\label{Psi_eq}
\left[ \nabla^2 - (n-1) \beta H^2 \right] \Psi = 0 \eqend{,}
\end{equation}
which are coupled Klein-Gordon equations with squared mass $M^2_\beta \equiv 
(n-1) \beta H^2$. One can prove that, if $h_{ab}$ satisfies the field 
equation~\eqref{field_eq}, then the trace $h$ and the scalar $\nabla^a \nabla^b 
h_{ab}$ can 
be reproduced by $h_{ab}^{(S)}$ of the form~\eqref{scalar_sector} with $B$ and 
$\Psi$ satisfying Eqs.~\eqref{B_eq} and \eqref{Psi_eq}. This implies that one 
can write $h_{ab} = h_{ab}^{(V+T)} + h_{ab}^{(S)}$, where $h_{ab}^{(V+T)}$ is 
traceless and satisfies $\nabla^a \nabla^b h_{ab}^{(V+T)} = 0$.

The scalar sector of our two-point function will be expressed in terms of that 
of the Klein-Gordon scalar field of squared mass $M^2_\beta$. It is known that 
the scalar-field two-point function is well behaved in the IR if and only if 
the 
mass is strictly positive in the Poincar{\'e} patch. For this reason, we require 
$\beta > 0$ as we stated before (and exclude, e.g., the analogue of the Feynman 
gauge).

In the Poincar{\'e} patch, the Klein-Gordon equation for a scalar field $\phi$ 
with mass $M$ reads
\begin{equation}
\label{KG_eq}
\left( \partial_\eta^2 - \frac{n-2}{\eta} \partial_\eta - \laplace + 
\frac{M^2}{H^2 \eta^2} \right) \phi = 0 \eqend{,}
\end{equation}
where $\laplace \equiv \eta^{\mu\nu} \partial_\mu \partial_\nu$ is the 
flat-space Laplace operator, and admits mode solutions of the form
\begin{equation}
\label{scalar_mode}
\phi^{M^2}_\vec{p}(\eta, \vec{x}) \equiv f_{M^2}(\eta, \vec{p}) \mathe^{\mathi 
\vec{p} \vec{x}} \eqend{.}
\end{equation}
Here, $f_{M^2}(\eta, \vec{p})$ is a function that is used extensively in this 
work and takes the following form:
\begin{equation}
\label{scalar_mode_temporal_part}
f_{M^2}(\eta, \vec{p}) \equiv \sqrt{\frac{\pi}{4 H}} (-H\eta)^\frac{n-1}{2} 
\mathe^{\mathi \mu \frac{\pi}{2}} \hankel1\mu\left(-\abs{\vec{p}}\eta \right)
\end{equation}
with $\hankel1\mu$ the Hankel function of the first kind and the parameter
\begin{equation}
\label{mu}
\mu \equiv \sqrt{\frac{(n-1)^2}{4} - \frac{M^2}{H^2}} \eqend{.}
\end{equation}

Two linearly independent sets of solutions to Eqs.~\eqref{B_eq} 
and~\eqref{Psi_eq}, which together with their complex conjugates form a complete 
set, can then be constructed from the mode solutions Eq.~\eqref{scalar_mode}. In 
terms of 
these modes, these two linearly independent sets of solutions are
\begin{equations}[set_1]
B^{(1)}_\vec{p}(\eta, \vec{x}) &= \phi^{M^2_\beta}_\vec{p}(\eta, \vec{x}) 
\eqend{,} \\
\Psi^{(1)}_\vec{p}(\eta, \vec{x}) &= 0 \eqend{,}
\end{equations}
and
\begin{equations}[set_2]
B^{(2)}_\vec{p}(\eta, \vec{x}) &= - \left( n + \frac{\lambda \beta}{2} \right) 
\left. \frac{\partial}{\partial M^2} \phi^{M^2}_\vec{p}(\eta, \vec{x}) 
\right\rvert_{M^2=M^2_\beta} \eqend{,} \\
\Psi^{(2)}_\vec{p}(\eta, \vec{x}) &= \phi^{M^2_\beta}_\vec{p}(\eta, \vec{x}) 
\eqend{.}
\end{equations}
Recalling the definition of $h_{ab}^\text{(S)}$ in terms of $B$ and $\Psi$, 
Eq.~\eqref{scalar_sector}, we thus obtain the modes of the scalar sector:
\begin{equation}
\label{mode_1}
h_{\vec{p}ab}^{(\text{S},1)}(\eta, \vec{x}) \equiv \nabla_a \nabla_b 
\phi^{M^2_\beta}_\vec{p}(\eta, \vec{x})
\end{equation}
and
\begin{splitequation}
\label{mode_2}
&h_{\vec{p}ab}^{(\text{S},2)}(\eta, \vec{x}) \equiv g_{ab} 
\phi^{M^2_\beta}_\vec{p}(\eta, \vec{x}) \\
&\quad- \left( n + \frac{\lambda \beta}{2} \right) \nabla_a \nabla_b \left. 
\frac{\partial}{\partial M^2} \phi^{M^2}_\vec{p}(\eta, \vec{x}) 
\right\rvert_{M^2 = M^2_\beta} \eqend{.}
\end{splitequation}
\vspace{0.4cm}
The expansion of $h_{ab}^\text{(S)}$ in terms of these modes reads
\begin{equation}
\label{field_expansion_scalar_sector}
\hat{h}_{ab}^{(\text{S})} = \sum_{k=1}^2 \int \left[ 
\hat{a}^{(\text{S})}_k(\vec{p}) h_{\vec{p}ab}^{(\text{S},k)} + \textrm{H.c.} 
\right] \frac{\total^{n-1} p}{(2\pi)^{n-1}} \eqend{,}
\end{equation}
where H.c.\ stands for the Hermitian conjugate of the preceding term. The 
commutators for the operators $\hat{a}^{(\text{S})}_k(\vec{p})$ and 
$\hat{a}^{(\text{S})\dagger}_l(\vec{p})$ are obtained from the commutation 
relations~\eqref{a_a_dagger_commutator} with the matrix $M$ determined through 
the symplectic product of the modes~\eqref{M_matrix}.

We now have to determine this matrix. As we have explained before, the invariant 
symplectic product, obtained by letting $p^{abc} = p^{abc}_{\text{inv}}$ in 
Eq.~\eqref{symplectic_prod}, vanishes identically between solutions of 
$L^{(\text{
inv})cd}_{ab} h_{cd} = 0$ and gauge modes of the form $h_{ab} = 2 \nabla_{(a} 
\xi_{b)}$ for arbitrary vectors $\xi_b$. Since the modes $\nabla_a \nabla_b B$ 
are of gauge form, and solve $L^{(\text{inv})cd}_{ab} \nabla_c \nabla_d B = 0$ 
identically, they do not make a contribution to the invariant symplectic 
product. Hence we have
\begin{widetext}
\begin{splitequation}
\label{symp_prod_hs_hs}
\symp{h_{ab}^{(\text{S})(1)}}{h_{cd}^{(\text{S})(2)}} = - \mathi \int_\Sigma 
\bigg[ &\nabla_b \nabla_c B^{(1)*} p^{(2)abc}_\text{inv,$\Psi$} - \nabla_b 
\nabla_c B^{(2)} p^{(1)abc*}_\text{inv,$\Psi$} + \nabla_b \nabla_c B^{(1)*} 
p^{(2)abc}_\text{gf} - \nabla_b \nabla_c B^{(2)} p^{(1)abc*}_\text{gf} \\
&\quad+ g_{bc} \Psi^{(1)*} p^{(2)abc} - g_{bc} \Psi^{(2)} p^{(1)abc*} \bigg] n_a 
\total \Sigma \eqend{,}
\end{splitequation}
\end{widetext}
where $p^{(k)abc}_\text{inv,$\Psi$}$ is the ``$\Psi$ part'' of the invariant 
canonical momentum current, obtained by substituting $h_{ab} = g_{ab} 
\Psi^{(k)}$ into Eq.~\eqref{inv_momentum}. Using the field 
equations~\eqref{B_eq} and~\eqref{Psi_eq}, we obtain
\begin{equations}
&p^{(k)abc}_\text{inv,$\Psi$} + p^{(k)abc}_\text{gf} = - \frac{n-2}{2\beta} 
g^{bc} \nabla^a \Psi^{(k)} \eqend{,} \\
&g_{bc} p^{(k)abc} = - \frac{n-2}{2} \nabla^a \left[ (n-1) H^2 B^{(k)} + 
\frac{n}{\beta} \Psi^{(k)} \right] \eqend{,}
\end{equations}
and the symplectic product~\eqref{symp_prod_hs_hs} between two scalar modes 
reduces to
\begin{splitequation}
\symp{h_{ab}^{(\text{S})(1)}}{h_{cd}^{(\text{S})(2)}} &= \frac{(n-1)(n-2)}{2} 
H^2 \symp{B^{(1)}}{\Psi^{(2)}}_\text{KG} \\
&\quad+ \frac{(n-1)(n-2)}{2} H^2 \symp{\Psi^{(1)}}{B^{(2)}}_\text{KG} \\
&\quad- \frac{(n-2) \lambda}{4} \symp{\Psi^{(1)}}{\Psi^{(2)}}_\text{KG} 
\raisetag{1.1\baselineskip} \eqend{,}
\end{splitequation}
with the Klein-Gordon symplectic product
\begin{splitequation}
\label{KG_symplectic_prod}
&\symp{\phi^{(1)}}{\phi^{(2)}}_\text{KG} \\
&\quad \equiv \mathi \int_\Sigma \left( \phi^{(1)*} \nabla^a \phi^{(2)} - 
\phi^{(2)} \nabla^a \phi^{(1)*}\right) n_a \total \Sigma \eqend{.}
\end{splitequation}
By taking the hypersurface $\Sigma$ to be $\eta = \text{const}$, which entails 
$\total \Sigma = (-H \eta)^{-(n-1)} \total^{n-1} x$ and $n^a = (-H\eta) 
\delta_0^a$, one can readily check that the scalar field 
modes~\eqref{scalar_mode} are 
normalized with respect to this symplectic product
\begin{equation}
\label{scalar_modes_norm}
\symp{\phi_\vec{p}^{M^2}}{\phi_\vec{q}^{M^2}}_\text{KG} = (2\pi)^{n-1} 
\delta^{n-1}(\vec{p}-\vec{q}) \eqend{.}
\end{equation}
Then, by noticing that
\begin{splitequation}
&\frac{\partial}{\partial M^2} \left[ 
\symp{\phi^{M^2}_\vec{p}}{\phi^{M^2_\beta}_\vec{q}} + 
\symp{\phi^{M^2_\beta}_\vec{p}}{\phi^{M^2}_\vec{q}} \right]_{M^2 = M^2_\beta} \\
&= \frac{\partial}{\partial M^2} \left[ 
\symp{\phi^{M^2}_\vec{p}}{\phi^{M^2}_\vec{q}} \right]_{M^2 = M^2_\beta} = 0 
\eqend{,}
\raisetag{1.1\baselineskip}
\end{splitequation}
one concludes that the symplectic product of the mode solutions~\eqref{mode_1} 
and~\eqref{mode_2} is
\begin{equation}
\label{modes_symplectic_prod}
\symp{h_{\vec{p}ab}^{(\text{S},k)}}{h_{\vec{q}cd}^{(\text{S},l)}} = N_{kl} 
(2\pi)^{n-1} \delta^{n-1}(\vec{p}-\vec{q}) \eqend{,}
\end{equation}
where the matrix $N_{kl}$ is given by
\begin{equation}
\label{N_matrix}
N_{kl} \equiv \frac{n-2}{4} \begin{pmatrix} 0 & 2 (n-1) H^2 \\ 2 (n-1) H^2 & - 
\lambda \end{pmatrix} \eqend{.}
\end{equation}
According to formula~\eqref{a_a_dagger_commutator}, the canonical commutation 
relations for the scalar-sector creation and annihilation operators are given by
\begin{equation}
\label{commutator_a_a_dagger_scalar_sector}
\left[ \hat{a}^{(\text{S})}_k(\vec{p}), \hat{a}^{(\text{S})\dagger}_l(\vec{q}) 
\right] = \left( N^{-1} \right)_{kl} (2\pi)^{n-1} \delta^{n-1}(\vec{p}-\vec{q}) 
\eqend{,}
\end{equation}
where the inverse matrix $(N^{-1})_{kl}$ reads
\begin{equation}
\label{inverse_N}
(N^{-1})_{kl} = \frac{1}{(n-1)^2 (n-2) H^2} \begin{pmatrix} \lambda H^{-2} & 2 
(n-1) \\ 2 (n-1) & 0 \end{pmatrix}
\end{equation}
and all other commutators vanish. The vacuum state annihilated by the 
$\hat{a}^{(\text{S})}_k$ is the Bunch-Davies vacuum, which reduces to the usual 
Minkowski vacuum as $H \to 0$. According to the general 
formula~\eqref{wightman_mode_sum},
 the scalar sector of the graviton two-point function is then given by
\begin{widetext} 
\begin{equation}
\label{scalar_sector_two_point_function_mode_sum}
\Delta^{(\text{S})}_{abc'd'}(x,x') = \sum_{k,l=1}^2 (N^{-1})_{kl} \int 
h_{\vec{p}ab}^{(\text{S},k)}(x) h_{\vec{p}c'd'}^{(\text{S},l)*}(x') 
\frac{\total^{n-1} p}{(2\pi)^{n-1}} \eqend{.}
\end{equation}

The mode sum~\eqref{scalar_sector_two_point_function_mode_sum} converges for 
$\beta > 0$, and we can write the two-point function of the scalar sector as
\begin{splitequation}
\label{scalar_sector_two_point_function}
\Delta_{abc'd'}^{(\text{S})}(x,x') &= \frac{1}{(n-1)^2 (n-2) H^4} \nabla_a 
\nabla_b \nabla_{c'} \nabla_{d'} \left[ \lambda \Delta_{M^2}(x,x') + (n-1) H^2 
\left( 2 n + \lambda \beta \right) \frac{\partial}{\partial M^2} 
\Delta_{M^2}(x,x') \right]_{M^2 = M^2_\beta} \\
&\quad+ \frac{2}{(n-1) (n-2) H^2} \left( g_{ab} \nabla_{c'} \nabla_{d'} + 
g_{c'd'} \nabla_a \nabla_b \right) \Delta_{M^2_\beta}(x,x') \eqend{,}
\end{splitequation}
\end{widetext}
with the scalar two-point function of mass $M$
\begin{equation}
\label{scalar_field_two_point_function}
\Delta_{M^2}(x,x') = \int \phi_\vec{p}^{M^2}(\eta,\vec{x}) 
\phi_\vec{p}^{M^2*}(\eta',\vec{x}') \frac{\total^{n-1} p}{(2\pi)^{n-1}} 
\eqend{.}
\end{equation}
Inserting the concrete form of the scalar mode functions~\eqref{scalar_mode}, we 
obtain
\begin{equation}
\label{delta_m2_as_i_mu}
\Delta_{M^2}(x,x') = \frac{H^{n-2}}{(4\pi)^\frac{n}{2}} I_\mu(x,x') \eqend{,}
\end{equation}
where $\mu$ is defined by Eq.~\eqref{mu} and $I_\mu$ is found as
\begin{splitequation}
\label{I_integral_form}
I_\mu(x,x') &\equiv 2^{n-2} \pi^\frac{n+2}{2} (\eta \eta')^\frac{n-1}{2} \\
&\times \int \hankel1\mu\left( - \abs{\vec{p}} \eta \right) \hankel2\mu\left( - 
\abs{\vec{p}} \eta' \right) \mathe^{\mathi \vec{p} (\vec{x}-\vec{x}')} 
\frac{\total^{n-1} p}{(2\pi)^{n-1}}
\end{splitequation}
using that 
\begin{equation}
\mathe^{\mathi \mu \frac{\pi}{2}} \hankel1\mu(x) \left[ 
\mathe^{\mathi \mu \frac{\pi}{2}} \hankel1\mu(x') \right]^* = \hankel1\mu(x) 
\hankel2\mu(x')
\end{equation}
for all real $x$ and $x'$, and $\mu$ either real or purely 
imaginary. The integral $I_\mu$ is well 
known~\cite{schomblond_spindel_ahp_1976}, and for $n$ spacetime dimensions it 
was evaluated in closed form after inserting a factor of $\mathe^{-\epsilon 
\abs{\vec{p}}}$ to ensure convergence for large $\abs{\vec{p}}$ in 
Ref.~\cite{froeb_higuchi_jmp_2014}. The result only depends on the
de~Sitter invariant
\begin{equation}
\label{Z}
Z(x,x') \equiv 1 - \frac{\vec{r}^2 - (\eta-\eta')^2}{2 \eta \eta'} \eqend{,}
\end{equation}
where $\vec{r} \equiv \vec{x} - \vec{x}'$. The result reads
\begin{splitequation}
\label{I}
I_\mu(x,x') &= \frac{\Gamma\left( a_+ \right) \Gamma\left( a_- 
\right)}{\Gamma\left( \frac{n}{2} \right)} \\
&\quad \times \hypergeom{2}{1}\left[ a_+, a_-; \frac{n}{2}; \frac{1+Z}{2} - 
\mathi \epsilon \sgn(\eta-\eta')
\right] \eqend{,}
\end{splitequation}
where $\hypergeom{2}{1}$ is the Gau{\ss} hypergeometric function and $a_\pm 
\equiv (n-1)/2 \pm \mu$. Note that the convergence factor supplied the necessary 
prescription for treating the singularity of the Gau{\ss} hypergeometric 
function 
as $Z \to 1$ so that $I_\mu(x,x')$ becomes a well-defined bidistribution. We 
will omit this explicit prescription below to simplify the notation unless it 
needs to be emphasized, and by abuse of notation write also $I_\mu(x,x') = 
I_\mu(Z)
$.

We obtain the scalar-field two-point function necessary for our purposes here by 
letting $M^2 = M^2_\beta$. Thus it is convenient to define a new parameter 
$\mu_\text{S}$ to be equal to $\mu$ in Eq.~\eqref{mu} after the substitution 
$M^2 = 
M^2_\beta = (n-1) \beta H^2$,
\begin{equation}
\label{mu_S}
\mu_\text{S} \equiv \sqrt{\frac{(n-1)^2}{4} - (n-1) \beta} \eqend{.}
\end{equation}
Furthermore, to express the two-point 
function~\eqref{scalar_sector_two_point_function} more explicitly, we define the 
set of functions
\begin{splitequation}
\label{I_k}
I^{(k)}_\mu(Z) &\equiv \frac{\Gamma\left( a_+ + k \right) \Gamma\left( a_- + k 
\right)}{2^k \Gamma\left(\frac{n}{2}+k \right)} \\
&\quad\times \hypergeom{2}{1}\left( a_+ + k, a_- + k; \frac{n}{2}+k; 
\frac{1+Z}{2} \right) \eqend{.}
\end{splitequation}
For positive integer $k$, this is the $k$-th derivative of $I_\mu(Z)$ with 
respect to $Z$, but this function is, in fact, well defined for all complex $k$ 
and $\mu$ provided that $\frac{n-1}{2} \pm \mu + k \not\in \{ 0, -1, -2, \ldots 
\}$, a 
fact that we will use later on. From hypergeometric identities~\cite{dlmf}, one 
can readily show that it satisfies the relation
\begin{splitequation}
\label{hypergeometric_diff_eq_I}
&(1-Z^2) I^{(k+2)}_\mu(Z) - (n+2k) Z I^{(k+1)}_\mu(Z) \\
&\quad+ \left[ \mu^2 - \frac{(n-1)^2}{4} - k (n+k-1) \right] I^{(k)}_\mu(Z) = 0 
\eqend{,}
\end{splitequation}
which for $k = 0$ is nothing but the Klein-Gordon equation. We also need an 
expression for the derivative of $I_\mu^{(k)}(Z)$ with respect to the parameter 
$\mu$. Thus, we define the function
\begin{equation}
\label{I_tilde_k}
\tilde{I}^{(k)}_\mu(Z) \equiv - \frac{1}{2\mu} \frac{\partial}{\partial \mu} 
I^{(k)}_\mu(Z) \eqend{,}
\end{equation}
the factor $-1/(2\mu)$ has been chosen so that
\begin{equation}
\tilde{I}^{(k)}_\mu(Z) = H^2 \frac{\partial}{\partial M^2} I^{(k)}_\mu(Z)
\end{equation}
if the parameter $\mu$ is given in terms of $M^2$ by Eq.~\eqref{mu}. For these 
functions, we also obtain a recursion relation by differentiating the 
relation~\eqref{hypergeometric_diff_eq_I} with respect to $\mu$, which leads to
\begin{splitequation}
\label{hypergeometric_diff_eq_tilde_I}
&(1-Z^2) \tilde{I}^{(k+2)}_\mu(Z) - (n+2k) Z \tilde{I}^{(k+1)}_\mu(Z) \\
&\quad+ \left[ \mu^2 - \frac{(n-1)^2}{4} - k (n+k-1) \right] 
\tilde{I}^{(k)}_\mu(Z) = I^{(k)}_\mu(Z) \eqend{.}
\end{splitequation}

For later use, it is convenient to have an explicit expression of the two-point 
function~\eqref{scalar_sector_two_point_function} in terms of 
de~Sitter--invariant bitensors and scalar functions. A complete set of such 
bitensors symmetric in 
the index pairs $ab$ and $c'd'$ is given by
\begin{equations}[T_set]
T^{(1)}_{abc'd'} &\equiv g_{ab} g_{c'd'} \label{T_1} \eqend{,} \\
T^{(2)}_{abc'd'} &\equiv H^{-2} \left( g_{ab} Z_{;c'} Z_{;d'} + g_{c'd'} Z_{;a} 
Z_{;b} \right) \label{T_2} \eqend{,} \\
T^{(3)}_{abc'd'} &\equiv H^{-4} Z_{;a} Z_{;b} Z_{;c'} Z_{;d'} \label{T_3} 
\eqend{,} \\
T^{(4)}_{abc'd'} &\equiv 4 H^{-4} Z_{;(a} Z_{;b)(c'} Z_{;d')} \label{T_4} 
\eqend{,} \\
T^{(5)}_{abc'd'} &\equiv 2 H^{-4} Z_{;a(c'} Z_{;d')b} \label{T_5} \eqend{,}
\end{equations}
and since we calculate the Wightman two-point function, the derivatives in 
Eq.~\eqref{scalar_sector_two_point_function} can be taken as if $I_\mu$ were 
just a function of $Z$, which would not be the case were we dealing with the 
Feynman 
propagator. The prescription for the singularity of the Gau{\ss} hypergeometric 
function given in Eq.~\eqref{I} corresponds to the Wightman two-point function, 
and does not generate additional local terms upon differentiation as can also 
be checked explicitly.

Using the relations
\begin{equations}[Z_relations]
Z_{;ab} &= - H^2 Z g_{ab} \eqend{,} \label{ab} \\
Z^{;a} Z_{;a} &= H^2 (1-Z^2) \eqend{,} \label{nabla_Z_contraction_1} \\
Z^{;a} Z_{;ab'} &= - H^2 Z Z_{;b'} \eqend{,} \label{nabla_Z_contraction_2} \\
Z^{;ab'} Z_{;ac'} &= H^4 \delta^{b'}_{c'} - H^2 Z^{;b'} Z_{;c'} 
\label{nabla_Z_contraction_3}
\end{equations}
for the covariant derivatives of $Z$ [which can be obtained, e.g., by direct 
calculations starting from the definition of $Z$, Eq.~\eqref{Z}, in the 
Poincar{\'e} patch], it follows that
\begin{equation}
\label{scalar_sector_2pf_bitensors}
\Delta_{abc'd'}^{(\text{S})}(x,x') = \frac{H^{n-2}}{(4\pi)^\frac{n}{2}} 
\sum_{k=1}^5 T^{(k)}_{abc'd'} F^{(\text{S},k)}(Z) \eqend{,}
\end{equation}
with
\begin{equations}[F_S]
\begin{split}
F^{(\text{S},1)} &= \frac{\lambda}{(n-1)^2 (n-2)} \left( Z^2 I^{(2)}_\text{S} + 
Z I^{(1)}_\text{S} \right) \\
&\quad+ \frac{2 n + \lambda \beta}{(n-1) (n-2)} \left( Z^2 
\tilde{I}^{(2)}_\text{S} + Z \tilde{I}^{(1)}_\text{S} \right) \\
&\quad- \frac{4}{(n-1) (n-2)} Z I^{(1)}_\text{S} \eqend{,}
\end{split} \raisetag{1.1\baselineskip} \\
\begin{split}
F^{(\text{S},2)} &= - \frac{\lambda}{(n-1)^2 (n-2)} \left( Z I^{(3)}_\text{S} + 
2 I^{(2)}_\text{S} \right) \\
&\quad- \frac{2 n + \lambda \beta}{(n-1) (n-2)} \left( Z 
\tilde{I}^{(3)}_\text{S} + 2 \tilde{I}^{(2)}_\text{S} \right) \\
&\quad+ \frac{2}{(n-1) (n-2)} I^{(2)}_\text{S} \eqend{,}
\end{split} \\
F^{(\text{S},3)} &= \frac{1}{(n-1) (n-2)} \left[ \frac{\lambda}{n-1} 
I^{(4)}_\text{S} + \left( 2 n + \lambda \beta \right)
\tilde{I}^{(4)}_\text{S} \right] \eqend{,} \\
F^{(\text{S},4)} &= \frac{1}{(n-1) (n-2)} \left[ \frac{\lambda}{n-1} 
I^{(3)}_\text{S} + \left( 2 n + \lambda \beta \right)
\tilde{I}^{(3)}_\text{S} \right] \eqend{,} \\
F^{(\text{S},5)} &= \frac{1}{(n-1) (n-2)} \left[ \frac{\lambda}{n-1} 
I^{(2)}_\text{S} + \left( 2 n + \lambda \beta \right)
\tilde{I}^{(2)}_\text{S} \right] \eqend{,}
\end{equations}
where we have written $I^{(k)}_\text{S} \equiv I^{(k)}_{\mu_\text{S}}$ and 
$\tilde{I}^{(k)}_\text{S} \equiv \tilde{I}^{(k)}_{\mu_\text{S}}$.

\subsection{Vector sector}
\label{subsec:vector_sector}

We now treat the vector sector, for which the classical field is given by 
$h_{ab}^\text{(V)} = \nabla_a v_b + \nabla_b v_a$ in terms of a divergence-free 
vector field $v^a$ [see Eq.~\eqref{vector_sector}]. We assume $\alpha > 0$. 
Since the 
vector modes are pure gauge, we have $L_{ab}^{(\text{inv})cd} 
h_{cd}^{(\text{V})} = 0$ identically, and hence Eq.~\eqref{field_eq} reduces to
\begin{equation}
\label{vector_eqn}
\nabla_{(a} \left[ \nabla^2 + (n-1) H^2 - \alpha m^2 \right] v_{b)} = 0 
\eqend{,}
\end{equation}
which is equivalent to
\begin{equation}
\label{killing_field}
\left[ \nabla^2 + (n-1) H^2 - \alpha m^2 \right] v^a = u^a \eqend{,}
\end{equation}
for an arbitrary Killing vector $u^a$ of the background de~Sitter metric. All 
solutions of this equation are given by the sum of a solution of the homogeneous 
equation and one particular solution. The particular solution can be obtained 
by 
first noticing that Killing vectors satisfy
\begin{splitequation}
0 &= \nabla^a \left( \nabla_a u_b + \nabla_b u_a \right) - \nabla_b \nabla^a u_a 
\\
&= \left[ \nabla^2 + (n-1) H^2 \right] u_b \eqend{.}
\end{splitequation}
Then, it is clear that
\begin{equation}
\label{particular_solution}
v_a = - \frac{1}{\alpha m^2} u^a
\end{equation}
solves the inhomogeneous equation~\eqref{killing_field}. The particular 
solution~\eqref{particular_solution}, however, is a Killing vector itself and, 
therefore, can be discarded as it does not contribute to the vector 
perturbation. Hence, 
the relevant modes satisfy the homogeneous part of Eq.~\eqref{killing_field}, 
which reads
\begin{equation}
\label{vector_sector_field_eq}
\left[ \nabla^2 + (n-1) H^2 - \alpha m^2 \right] v^a = 0
\end{equation}
together with the constraint $\nabla_a v^a = 0$. One can also show that given 
any solution $h_{ab}^{(V+T)}$ of Eq.~\eqref{field_eq} satisfying $\nabla^a 
\nabla^b h_{ab}^{(V+T)} = 0$ and $h^{(V+T)} = 0$, the divergence $\nabla^b 
h_{ab}^{(
V+T)}$ can be reproduced by $h^{(V)}_{ab} = \nabla_a v_b + \nabla_b v_a$, where 
$v_a$ satisfies the homogeneous equation~\eqref{vector_sector_field_eq}. This 
implies that we can write $h^{(V+T)}_{ab} = h^{(V)}_{ab} + h^{(T)}_{ab}$ where 
$h_
{ab}^{(T)}$ is transverse and traceless.

From Eq.~\eqref{vector_sector_field_eq} one sees that $v^a$ corresponds to a 
Stueckelberg vector field for the gauge parameter $\xi \to \infty$ and the mass 
$M^2 = \alpha m^2 - 2 (n-1) H^2$. The canonical quantization of the 
Stueckelberg 
field on a de~Sitter background and the mode-sum construction of its two-point 
function in the Bunch-Davies vacuum was recently carried out in 
Ref.~\cite{froeb_higuchi_jmp_2014}. Instead of using the vector two-point 
function in this 
reference, we use the symplectic method to quantize the vector sector of the 
graviton two-point function and also verify the results of 
Ref.~\cite{froeb_higuchi_jmp_2014} that are relevant here.

We start the construction of a complete set of modes for 
Eq.~\eqref{vector_sector_field_eq} by decomposing the components of the vector 
field $v^a$ in their irreducible parts with respect to the spatial 
$\mathrm{O}(n-1)$ symmetry. Note 
that we use $\mu$ and $\nu$ for spatial indices. The constraint $\nabla_a v^a = 
0$ can be solved as
\begin{equations}[v_parts]
v_0 &= T \eqend{,} \label{v_S_0} \\
v_\mu &= (-H\eta)^{-1} W_\mu + \frac{\partial_\mu}{\laplace} \left( 
\partial_\eta - \frac{n-2}{\eta} \right) T \label{v_SV_i} \eqend{,}
\end{equations}
where the spatial vector $W_\mu$ is spatially divergence free, i.e., 
$\eta^{\mu\nu} \partial_\mu W_\nu = 0$, and $\laplace^{-1}$ is the inverse of 
the Laplace operator with vanishing boundary conditions at (spatial) infinity. 
This inverse 
is well defined if the function it acts on in Eq.~\eqref{v_SV_i} vanishes 
sufficiently fast at spatial infinity. We will assume this for the moment, and 
comment later on the condition for it to be justified. By substituting this 
decomposition into the field equation~\eqref{vector_sector_field_eq}, one 
obtains the following equations of motion:
\begin{equations}[vector_field_eqns]
\left( \partial^2_\eta - \frac{n-2}{\eta} \partial_\eta - \laplace + 
\frac{\alpha m^2 - n H^2}{H^2 \eta^2} \right) T &= 0 \eqend{,} \label{T_eq} \\
\left( \partial^2_\eta - \frac{n-2}{\eta} \partial_\eta - \laplace + 
\frac{\alpha m^2 - n H^2}{H^2 \eta^2} \right) W_\mu &= 0 \eqend{.} 
\label{W_i_eq}
\end{equations}

The divergence-free spatial vector $W_\mu$ can readily be expressed in the 
momentum space in terms of the polarization vectors $e_\mu^{(k)}(\vec{p})$, with 
$k \in \{1,\ldots,n-2\}$, satisfying $p^\mu e_\mu^{(k)}(\vec{p}) = 0$, which 
form a 
basis of the subspace of the momentum space orthogonal to $\vec{p}$. We 
normalize them by imposing the condition
\begin{equation}
\label{vector_polarization_normalization}
\eta^{\mu\nu} e_\mu^{(k)}(\vec{p}) e_\nu^{(l)*}(\vec{p}) = \delta_{kl} \eqend{,}
\end{equation}
where $\delta_{kl}$ is the Kronecker delta. Since the $e_\mu^{(k)}(\vec{p})$ 
form a basis of the the space of momenta orthogonal to $\vec{p}$, they also 
satisfy
\begin{equation}
\label{vector_polarization_sum}
\sum_{k=1}^{n-2} e_\mu^{(k)}(\vec{p}) e_\nu^{(k)*}(\vec{p}) = \tilde{P}_{\mu\nu} 
\eqend{,}
\end{equation}
where
\begin{equation}
\label{P_operator_momentum_space}
\tilde{P}_{\mu\nu} \equiv \eta_{\mu\nu} - \frac{p_\mu p_\nu}{\vec{p}^2}
\end{equation}
is the projection operator onto the subspace orthogonal to $\vec{p}$, 
corresponding to the projector onto the space of spatially transverse tensors
\begin{equation}
\label{P_operator}
P_{\mu\nu} \equiv \delta_{\mu\nu} - \frac{\partial_\mu \partial_\nu}{\laplace}
\end{equation}
in coordinate space.

Equations~\eqref{vector_field_eqns} have exactly the form of the 
Klein-Gordon equation~\eqref{KG_eq} with squared mass $M^2 = \alpha m^2 - n 
H^2$. Consequently the modes defining the Bunch-Davies vacuum are just the 
modes~\eqref{scalar_mode}. Hence we find \begin{equations}[vector_modes]
T_\vec{p}(\eta, \vec{x}) &\equiv \phi^{\alpha m^2 - n H^2}_\vec{p}(\eta, 
\vec{x}) \eqend{,} \label{T_modes} \\
W_{\vec{p}\mu}^{(k)}(\eta, \vec{x}) &\equiv e_\mu^{(k)}(\vec{p}) \phi^{\alpha 
m^2 - n H^2}_\vec{p}(\eta, \vec{x}) \eqend{.} \label{W_i_modes}
\end{equations}
The modes for the vector field $v^a$ can then be determined from the 
decomposition~\eqref{v_parts}. The modes that are spatial scalars are obtained 
by substituting Eq.~\eqref{T_modes} into Eqs.~\eqref{v_S_0} and~\eqref{v_SV_i} 
with $W_{\mu}
 = 0$, and read
\begin{equations}[v_S_modes]
v_{\vec{p}0}^{(\text{S})}(\eta, \vec{x}) &= \phi^{\alpha m^2 - n 
H^2}_\vec{p}(\eta, \vec{x}) \eqend{,} \label{v_S_0_mode} \\
\begin{split}
v_{\vec{p}\mu}^{(\text{S})}(\eta, \vec{x}) &= \frac{\partial_\mu}{\laplace} 
\left( \partial_\eta - \frac{n-2}{\eta} \right) \phi^{\alpha m^2 - n 
H^2}_\vec{p}(\eta, \vec{x}) \\
&= - \frac{\mathi p_\mu}{\vec{p}^2} \left( \partial_\eta - \frac{n-2}{\eta} 
\right) \phi^{\alpha m^2 - n H^2}_\vec{p}(\eta, \vec{x}) \eqend{.} 
\label{v_S_i_mode}
\end{split}
\end{equations}
The modes that are spatial vectors are obtained by substituting 
Eq.~\eqref{W_i_modes} into Eq.~\eqref{v_SV_i} with $T = 0$, and read
\begin{equations}[v_V_modes]
v_{\vec{p}0}^{(\text{V},k)}(\eta, \vec{x}) &= 0 \eqend{,} \label{v_V_0_mode} \\
v_{\vec{p}\mu}^{(\text{V},k)}(\eta, \vec{x}) &= - (H\eta)^{-1} 
e_\mu^{(k)}(\vec{p}) \phi^{\alpha m^2 - n H^2}_\vec{p}(\eta, \vec{x}) \eqend{.} 
\label{v_V_i_mode}
\end{equations}
Thus, the modes for the field $h_{ab}^{(V)}$ given by Eq.~\eqref{vector_sector} 
are
\begin{equation}
\label{h_VS_mode}
h_{\vec{p}ab}^{(\text{VS})} \equiv \nabla_a v_{\vec{p}b}^{(\text{S})} + \nabla_b 
v_{\vec{p}a}^{(\text{S})}
\end{equation}
and
\begin{equation}
\label{h_VV_mode}
h_{\vec{p}ab}^{(\text{VV},k)} \equiv \nabla_a v_{\vec{p}b}^{(\text{V},k)} + 
\nabla_b v_{\vec{p}a}^{(\text{V},k)} \eqend{.}
\end{equation}
Hence, the vector sector of the quantum operator $\hat{h}_{ab}$ can be expanded 
as follows:
\begin{splitequation}
\label{field_expansion_vector_sector}
\hat{h}_{ab}^{(\text{V})} &= \int \left[ \hat{a}^{(\text{VS})}(\vec{p}) 
h_{\vec{p}ab}^{(\text{VS})} + \textrm{H.c.} \right] \frac{\total^{n-1} 
p}{(2\pi)^{n-1}} \\
&ß\ + \sum_{k=1}^{n-2} \int \left[ \hat{a}^{(\text{VV})}_k(\vec{p}) 
h_{\vec{p}ab}^{(\text{VV},k)} + \textrm{H.c.} \right] \frac{\total^{n-1} 
p}{(2\pi)^{n-1}} \eqend{.}
\end{splitequation}

The commutation relations for the creation and annihilation operators in this 
expansion are again obtained from the symplectic product~\eqref{symplectic_prod} 
of the modes~\eqref{h_VS_mode} and~\eqref{h_VV_mode}. The symplectic 
product~\eqref{symplectic_prod} for the vector sector reduces to
\begin{equation}
\label{graviton_vector_symplectic_relation}
\symp{h^{(\text{V})(1)}_{ab}}{h^{(\text{V})(2)}_{cd}} = - m^2 
\symp{v_a^{(1)}}{v_b^{(2)}}_\text{V} \eqend{,}
\end{equation}
with the ``vector'' symplectic product $\symp{\,\cdot\,}{\,\cdot\,}_\text{V}$ 
given by
\begin{equation}
\symp{v_a^{(1)}}{v_b^{(2)}}_\text{V} \equiv \mathi \int_\Sigma \left( v^{(1)b*} 
\nabla^a v_b^{(2)} - v^{(2)b} \nabla^a v_b^{(1)*} \right) n_a \total \Sigma
\end{equation}
and conserved for solutions of Eq.~\eqref{vector_sector_field_eq}. We then 
calculate (e.g., by taking $\Sigma$ to be $\eta = \text{const}$) the symplectic 
product~\eqref{symplectic_prod} for the modes~\eqref{h_VS_mode} 
and~\eqref{h_VV_mode}
 and obtain
\begin{equations}
\begin{split}
\symp{h_{\vec{p}ab}^{(\text{VS})}}{h_{\vec{q}cd}^{(\text{VS})}} &= - m^2 \left[ 
\alpha m^2 - 2 (n-1) H^2 \right] \abs{\vec{p}}^{-2} \\
&\qquad\times (2\pi)^{n-1} \delta^{n-1}(\vec{p}-\vec{q}) \eqend{,}
\end{split} \label{h_VS_mode_norm} \raisetag{1.8\baselineskip} \\
\symp{h_{\vec{p}ab}^{(\text{VV},k)}}{h_{\vec{q}cd}^{(\text{VV},l)}} &= - m^2 
\delta_{kl} (2\pi)^{n-1} \delta^{n-1}(\vec{p}-\vec{q}) \label{h_VV_mode_norm} 
\eqend{,} \\
\symp{h_{\vec{p}ab}^{(\text{VS})}}{h_{\vec{q}cd}^{(\text{VV},k)}} &= 0 \eqend{.} 
\label{h_VS_h_VV}
\end{equations}
Thus, the commutation relations for the creation and annihilation operators of 
the vector sector are given by
\begin{equations}
\begin{split}
\left[ \hat{a}^{(\text{VS})}(\vec{p}), \hat{a}^{(\text{VS})\dagger}(\vec{q}) 
\right] &= - \frac{\vec{p}^2}{m^2 \left[ \alpha m^2 - 2 (n-1) H^2 \right]} \\
&\qquad\times (2\pi)^{n-1} \delta^{n-1}(\vec{p}-\vec{q}) \eqend{,}
\end{split} \label{commutator_a_a_dagger_VS_part} \\
\left[ \hat{a}^{(\text{VV})}_k(\vec{p}), \hat{a}^{(\text{VV})\dagger}_l(\vec{q}) 
\right] &= - \frac{1}{m^2} \delta_{kl} (2\pi)^{n-1} 
\delta^{n-1}(\vec{p}-\vec{q}) \eqend{,} \label{commutator_a_a_dagger_VV_part}
\end{equations}
with all other commutators vanishing.

The vector sector of the graviton two-point function can then be constructed 
from these modes according to Eq.~\eqref{wightman_mode_sum} and reads
\begin{splitequation}
\label{vector_sector_two_point_function_mode_sum}
\Delta^{(\text{V},m^2)}_{abc'd'}(x,x') &= N^{(\text{S})}_\alpha 
\Delta^{(\text{VS},m^2)}_{abc'd'}(x,x') \\
&\quad+ N^{(\text{V})} \Delta^{(\text{VV},m^2)}_{abc'd'}(x,x') \eqend{,}
\end{splitequation}
with the spatial scalar part
\begin{splitequation}
\label{Delta_VS}
&\Delta^{(\text{VS},m^2)}_{abc'd'}(x,x') \\
&\quad \equiv - \frac{(4\pi)^\frac{n}{2}}{4 H^{n+2}} \int \vec{p}^2 
h_{\vec{p}ab}^{(\text{VS})}(x)h_{\vec{p}c'd'}^{(\text{VS})*}(x') 
\frac{\total^{n-1} p}{(2\pi)^{n-1}} \eqend{,}
\end{splitequation}
the spatial vector part
\begin{splitequation}
\label{Delta_VV}
&\Delta^{(\text{VV},m^2)}_{abc'd'}(x,x') \\
&\equiv - \frac{(4\pi)^\frac{n}{2}}{H^n} \sum_{k=1}^{n-2} \int 
h_{\vec{p}ab}^{(\text{VV},k)}(x)h_{\vec{p}c'd'}^{(\text{VV},k)}(x') 
\frac{\total^{n-1} p}{(2\pi)^{n-1}} \eqend{,}
\end{splitequation}
and the constants
\begin{equation}
\label{N_S_alpha}
N^{(\text{S})}_\alpha \equiv \frac{4 H^4}{m^2 \left[ \alpha m^2 - 2 (n-1) H^2 
\right]} \frac{H^{n-2}}{(4\pi)^\frac{n}{2}}
\end{equation}
and
\begin{equation}
\label{N_V}
N^{(\text{V})} \equiv \frac{H^2}{m^2} \frac{H^{n-2}}{(4\pi)^\frac{n}{2}} 
\eqend{.}
\end{equation}

The explicit form of the mode functions~\eqref{v_S_modes} and~\eqref{v_V_modes} 
shows that the momentum integrals~\eqref{Delta_VS} and~\eqref{Delta_VV} converge 
in the IR if the squared mass $M^2 = \alpha m^2 - n H^2$ of the scalar mode 
functions is positive, since the factor of $p_\mu p_\nu/\vec{p}^4\sim 
1/\vec{p}^2$ gets canceled by the explicit factor of $\vec{p}^2$ in the momentum 
integral~\eqref{Delta_VS}. We thus assume $M^2 > 0$, i.e., $\alpha m^2 > n H^2$, 
and discuss this condition in more detail in Sec.~\ref{sec:ir_issues}. Using 
the explicit mode functions~\eqref{v_S_modes}, we obtain for the 
spatially scalar part
\begin{equation}
\Delta^{(\text{VS},m^2)}_{abc'd'}(x,x') = H^{-4} \nabla^{\phantom{()}}_{(c'} 
\nabla^{\phantom{()}}_{|(a} K^{(\text{S})}_{b)|d')}(x,x'),
\end{equation}
where
\begin{equations}
K^{(\text{S})}_{00'} &= \laplace I_\text{V}(Z) \eqend{,} \label{KS00} \\
K^{(\text{S})}_{0\nu'} &= \partial_{\nu'} \left( \partial_{\eta'} - 
\frac{n-2}{\eta'} \right) I_\text{V}(Z) \eqend{,} \label{KS0nu} \\
K^{(\text{S})}_{\mu0'} &= \partial_\mu \left( \partial_\eta - \frac{n-2}{\eta} 
\right) I_\text{V}(Z) \eqend{,} \label{KSmu0} \\
K^{(\text{S})}_{\mu\nu'} &= \frac{\partial_\mu \partial_{\nu'}}{\laplace} \left( 
\partial_\eta - \frac{n-2}{\eta} \right) \left( \partial_{\eta'} - 
\frac{n-2}{\eta'} \right) I_\text{V}(Z) \eqend{,} \label{KSmunu}
\end{equations}
with the function $I_\text{V}(Z) \equiv I_{\mu_\text{V}}(Z)$ given by 
Eq.~\eqref{I}. The parameter $\mu_\text{V}$ is defined by substituting $M^2 = 
\alpha m^2 - n H^2$ in Eq.~\eqref{mu} that gives the parameter $\mu$ in terms of 
$M^2$. 
Thus,
\begin{equation}
\label{mu_V}
\mu_\text{V} \equiv \sqrt{ \frac{(n+1)^2}{4} - \frac{\alpha m^2}{H^2} } 
\eqend{.}
\end{equation}
As for the spatially vector part, it follows from the mode 
functions~\eqref{v_V_modes} and the polarization 
sum~\eqref{vector_polarization_sum} that
\begin{equation}
\Delta^{(\text{VV},m^2)}_{abc'd'}(x,x') = H^{-4} \nabla^{\phantom{()}}_{(c'} 
\nabla^{\phantom{()}}_{|(a} K^{(\text{V})}_{b)|d')}(x,x')
\end{equation}
with
\begin{equations}
K^{(\text{V})}_{00'} &= K^{(\text{V})}_{0\nu'} = K^{(\text{V})}_{\mu0'} = 0 
\eqend{,} \\
K^{(\text{V})}_{\mu\nu'} &= - \frac{4}{\eta \eta'} P_{\mu\nu} I_\text{V}(Z) 
\eqend{.}
\end{equations}

In order to obtain a covariant expression for 
$\Delta_{abc'd'}^{(\text{V},m^2)}(x,x')$, we need to further simplify the 
spatially-scalar contribution to the vector two-point function, 
$K^{(\text{S})}_{ab'}$. For the temporal and mixed 
spatial-temporal components of $K^{(\text{S})}_{ab'}$, the derivatives acting on 
$I_\text{V}(Z)$ in Eqs.~\eqref{KS00}, \eqref{KS0nu}, and~\eqref{KSmu0} can 
readily be found using the derivatives of $Z$, which is defined by 
Eq.~\eqref{Z}, as
\begin{equations}[del_eta_Z]
\partial_\eta Z &= \frac{1}{\eta'} - \frac{Z}{\eta} \eqend{,} \\
\partial_{\eta'} Z &= \frac{1}{\eta} - \frac{Z}{\eta'} \eqend{,} \\
\partial_\eta \partial_{\eta'} Z &= - \frac{1}{\eta^2} - \frac{1}{(\eta')^2} + 
\frac{Z}{\eta \eta'} \eqend{,}
\end{equations}
and
\begin{equations}[del_i_Z]
\partial_\mu Z &= - \frac{r_\mu}{\eta \eta'} \eqend{,} \\
\partial_{\nu'} Z &= \frac{r_{\nu'}}{\eta \eta'} \eqend{,} \\
\partial_\mu \partial_{\nu'} Z &= - \frac{\eta_{\mu\nu'}}{\eta \eta'}
\end{equations}
with $r_\mu \equiv (\vec{x} - \vec{x}')_\mu$. The calculation for the purely 
spatial components, $K^{(\text{S})}_{\mu\nu'}$, given by Eq.~\eqref{KSmunu} is 
more complicated. First, note that the action of the Laplacian on the functions 
$I^{
(k)}_\mu$ defined by Eq.~\eqref{I_k} can be found using the 
relations~\eqref{del_i_Z} as
\begin{equation}
\label{zero_del_eta}
\laplace I^{(k)}_\mu = - \frac{(n-1)}{\eta \eta'} I^{(k+1)}_\mu + 
\frac{\vec{r}^2}{\eta^2 (\eta')^2} I^{(k+2)}_\mu \eqend{.}
\end{equation}
Similarly, the time derivatives of $I_{\mu}^{(k)}$ can be found by using the 
relations~\eqref{del_eta_Z}. Thus, we obtain
\begin{equation}
\label{one_del_eta}
\left( \eta \partial_\eta + \eta' \partial_{\eta'} \right) I^{(k)}_\mu = \eta 
\eta' \laplace I^{(k-1)}_\mu + (n-1) I^{(k)}_\mu
\end{equation}
and
\begin{splitequation}
\label{two_del_eta}
\eta \eta' \partial_\eta \partial_{\eta'} I^{(k)}_\mu &= \left[ 
\frac{(n-1)^2}{4} - \mu^2 \right] I^{(k)}_\mu - \frac{\eta^2 + (\eta')^2}{2} 
\laplace I^{(k)}_\mu \\
&\quad+ \frac{n+1}{2} \eta \eta' \laplace I^{(k-1)}_\mu + \frac{(\eta 
\eta')^2}{2} \laplace^2 I^{(k-2)}_\mu \eqend{,}
\end{splitequation}
where the equality~\eqref{hypergeometric_diff_eq_I} satisfied by the 
$I^{(k)}_\mu$ has also been used. These equations allow us to trade time 
derivatives for Laplacians. Thus, defining
\begin{equation}
\label{K}
K_{bd'} \equiv K^{(\text{S})}_{bd'} + 
\frac{N^{(\text{V})}}{N^{(\text{S})}_\alpha} K^{(\text{V})}_{bd'} \eqend{,}
\end{equation}
we find [using $\partial_{\nu'} I_\mu(Z) = - \partial_\nu I_\mu(Z)$]
\begin{splitequation}
\label{k_vector_munu}
K_{\mu\nu'} &= - \frac{\partial_\mu \partial_\nu}{\laplace} \partial_\eta 
\partial_{\eta'} I_\text{V}(Z) - \frac{(n-2)^2}{\eta \eta'} \frac{\partial_\mu 
\partial_\nu}{\laplace} I_\text{V}(Z) \\
&\quad+ \frac{n-2}{\eta \eta'} \frac{\partial_\mu \partial_\nu}{\laplace} \left( 
\eta \partial_\eta + \eta' \partial_{\eta'} \right) I_\text{V}(Z) \\
&\quad- \frac{\alpha m^2 - 2 (n-1) H^2}{H^2 \eta \eta'} \left( \eta_{\mu\nu} - 
\frac{\partial_\mu \partial_\nu}{\laplace} \right) I_\text{V}(Z) \eqend{.}
\end{splitequation}
Recalling that $I_\text{V}(Z) = I^{(0)}_\text{V}(Z)$, where $I_{\text{V}} = 
I_{\mu_{\text{V}}}$ with $\mu_\text{V}$ defined by Eq.~\eqref{mu_V}, we can 
exchange time derivatives for Laplacians using Eqs.~\eqref{one_del_eta} 
and~\eqref{two_del_eta}. All inverse Laplacians in Eq.~\eqref{k_vector_munu} 
cancel out in the process. Next, by using Eq.~\eqref{zero_del_eta} to eliminate 
the remaining Laplacians, we are left with
\begin{splitequation}
K_{\mu\nu'} &= - \frac{\alpha m^2 - 2 (n-1) H^2}{H^2 \eta \eta'} \eta_{\mu\nu} 
I^{(0)}_\text{V}(Z) \\
&\quad+ \partial_\mu \partial_\nu \bigg[ Z I^{(0)}_\text{V}(Z) + (n-3) 
I^{(-1)}_\text{V}(Z) \bigg] \eqend{.}
\end{splitequation}
We evaluate the remaining spatial derivatives using Eqs.~\eqref{del_i_Z}. The 
final result turns out to be de~Sitter--invariant, as expected, and has the form
\begin{splitequation}
K_{bd'} &= \left[ - (1-Z^2) I^{(2)}_\text{V}(Z) + (n-1) Z I^{(1)}_\text{V}(Z) 
\right] Z_{;bd'} \\
&\quad- \left[ Z I^{(2)}_\text{V}(Z) + (n-1) I^{(1)}_\text{V}(Z) \right] Z_{;b} 
Z_{;d'} \eqend{,}
\end{splitequation}
where we have made use of Eq.~\eqref{hypergeometric_diff_eq_I} to simplify the 
final expression. This result agrees with that derived in 
Ref.~\cite{froeb_higuchi_jmp_2014}, but the derivation presented here is 
shorter. 

Note that we were able to perform the exchange of derivatives only because the 
momentum integral was convergent thanks to the condition $\alpha m^2 > n H^2$ 
and because, thus, the inverse of the Laplacian in Eqs.~\eqref{KSmunu} and 
\eqref{k_vector_munu} was well defined. If this condition were not satisfied, 
the final result would have depended on how the momentum integral was 
regularized (e.g., by a cutoff at low momenta, or by an additional factor of 
$\abs{\vec{p}}^\omega$ in the integrand).

The final result for the vector sector two-point function then reads
\begin{equation}
\label{vector_sector_tp_AH}
\Delta^{(\text{V},m^2)}_{abc'd'}(x,x') = N^{(\text{S})}_\alpha \nabla_{(c'} 
\nabla_{|(a} K_{b)|d')} \eqend{.}
\end{equation}
Again, it is convenient for later use to have an explicit expression in terms of 
de~Sitter--invariant bitensors and scalar functions. We evaluate the derivatives 
using the relations~\eqref{Z_relations}. The result is
\begin{equation}
\label{vector_sector_two_point_function_m}
\Delta^{(\text{V},m^2)}_{abc'd'}(x,x') = N^{(\text{S})}_\alpha \sum_{k=1}^5 
T^{(k)}_{abc'd'} F^{(\text{V},k)}(Z) \eqend{,}
\end{equation}
where the bitensors $T^{(k)}_{abc'd'}$ are defined by Eq.~\eqref{T_set} and the 
coefficients are given by 
\begin{subequations}
\label{F_V}
\begin{align}
F^{(\text{V},1)} &= - Z I^{(2)}_\text{V} \eqend{,} \label{F_V_1} \\
F^{(\text{V},2)} &= I^{(3)}_\text{V} \eqend{,} \label{F_V_2} \\
F^{(\text{V},3)} &= - Z I^{(4)}_\text{V} - (n+1) I^{(3)}_\text{V} \eqend{,} 
\label{F_V_3} \\
F^{(\text{V},4)} &= - \frac{1}{4} (1-Z^2) I^{(4)}_\text{V} + \frac{n}{4} \left( 
Z I^{(3)}_\text{V} - I^{(2)}_\text{V} \right) \eqend{,} \label{F_V_4} \\
F^{(\text{V},5)} &= - \frac{1}{2} (1-Z^2) I^{(3)}_\text{V} + \frac{n}{2} Z 
I^{(2)}_\text{V} \eqend{.} \label{F_V_5}
\end{align}
\end{subequations}
As a check, we have verified that the two-point function 
$\Delta_{abc'd'}^{(V,m^2)}(x,x')$ given by 
Eq.~\eqref{vector_sector_two_point_function_m} is traceless in each index pair 
using the relations~\eqref{Z_relations}. [This must be the 
case because the vector $v^a$, in terms of which the vector sector of the 
gravitational perturbation is defined by Eq.~\eqref{vector_sector}, is 
divergence free.]

\subsection{Tensor sector}
\label{subsec:tensor_sector}

The tensor field $h^{(\text{T})}_{ab}$ is transverse and traceless, from which 
it follows that $G_b = 0$ [see Eq.~\eqref{gb_def}]. Consequently, only the 
invariant part of the field equation~\eqref{field_eq} contributes. Thus, we have
\begin{equation}
\label{tensor_sector_field_eq}
\left( \nabla^2 - 2 H^2 - m^2 \right) h^{(\text{T})}_{ab} = 0 \eqend{.}
\end{equation}
In order to find a complete set of modes, we proceed similarly to the case of the 
vector sector, decomposing the components of $h^{(\text{T})}_{ab}$ in their 
irreducible parts with respect to the spatial $\mathrm{O}(n-1)$ symmetry. 
Enforcing (covariant) transversality and tracelessness, we obtain
\begin{equations}[t_parts]
h^{(\text{T})}_{00} &= S \eqend{,} \label{h_T_00} \\
h^{(\text{T})}_{0\mu} &= \frac{\partial_\mu}{\laplace} \left( \partial_\eta - 
\frac{n-2}{\eta} \right) S - (H\eta)^{-1} V_\mu \eqend{,} \label{h_T_0i} \\
\begin{split}
h^{(\text{T})}_{\mu\nu} &= \frac{\partial_\mu \partial_\nu}{\laplace} S + 
\frac{Q_{\mu\nu}}{n-2} \left[ 1 - \frac{1}{\laplace} \left( \partial_\eta - 
\frac{n-2}{\eta} \right)^2 \right] S \\
&\quad- 2 (H\eta)^{-1} \left( \partial_\eta - \frac{n-1}{\eta} \right) 
\frac{\partial_{(\mu} V_{\nu)}}{\laplace} + (H\eta)^{-2} H_{\mu\nu} \eqend{,} 
\label{h_T_ij}
\end{split}
\end{equations}
with the spatial vector $V_\mu$ spatially divergence free, $\eta^{\mu\nu} 
\partial_\mu V_\nu = 0$, and the spatial tensor $H_{\mu\nu}$ spatially 
divergence free and traceless, $\eta^{\mu\nu} \partial_\mu H_{\nu\rho} = 0 = 
\eta^{\mu\nu} H_{\mu\nu}$. In order to shorten the formulas, we have defined the 
spatially traceless operator
\begin{equation}
\label{Q_operator}
Q_{\mu\nu} \equiv \eta_{\mu\nu} - (n-1) \frac{\partial_\mu 
\partial_\nu}{\laplace} \eqend{.}
\end{equation}
As in the vector case, the inverse Laplacian in the 
decomposition~\eqref{t_parts} is well defined only if the momentum integrals 
converge. This condition is satisfied if $m^2 > 0$ as we shall see below. 
Substituting the decomposition~\eqref{t_parts} back into the field 
equation~\eqref{tensor_sector_field_eq}, we obtain the following equations of 
motion:
\begin{equations}[tensor_field_eqns]
\left( \partial^2_\eta - \frac{n-2}{\eta} \partial_\eta - \laplace + 
\frac{m^2}{H^2 \eta^2} \right) S &= 0 \eqend{,} \label{S_eq} \\
\left( \partial^2_\eta - \frac{n-2}{\eta} \partial_\eta - \laplace + 
\frac{m^2}{H^2 \eta^2} \right) V_\mu &= 0 \eqend{,} \label{V_i_eq} \\
\left( \partial^2_\eta - \frac{n-2}{\eta} \partial_\eta - \laplace + 
\frac{m^2}{H^2 \eta^2} \right) H_{\mu\nu} &= 0 \eqend{.} \label{H_ij_eq}
\end{equations}

To express the solutions of these equations, we need to define a set of spatial 
polarization tensors $e_{\mu\nu}^{(k)}(\vec{p})$ with $k \in \{ 1, \ldots, 
n(n-3)/2 \}$ which are symmetric, transverse, $p^\mu e_{\mu\nu}^{(k)}(\vec{p}) = 
0$, 
traceless, $\eta^{\mu\nu} e_{\mu\nu}^{(k)}(\vec{p}) = 0$, and form a basis of 
the subspace of tensors satisfying these conditions. We normalize them by 
imposing the conditions
\begin{equation}
\eta^{\mu\nu} \eta^{\rho\sigma} e_{\mu\rho}^{(k)}(\vec{p}) 
e_{\nu\sigma}^{(l)*}(\vec{p}) = \delta_{kl} \eqend{.}
\end{equation}
They satisfy the following completeness relation since they form a basis:
\begin{equation}
\sum_{k=1}^{n(n-3)/2} e_{\mu\nu}^{(k)}(\vec{p}) e_{\rho\sigma}^{(k)*}(\vec{p}) = 
\tilde{P}_{\mu(\rho} \tilde{P}_{\sigma)\nu} - \frac{1}{n-2} \tilde{P}_{\mu\nu} 
\tilde{P}_{\rho\sigma} \eqend{,}
\end{equation}
where $\tilde{P}_{\mu\nu}$ is defined by Eq.~\eqref{P_operator_momentum_space}. 
Again, Eqs.~\eqref{tensor_field_eqns} have exactly the form of the Klein-Gordon 
equation~\eqref{KG_eq} with mass $M = m$, and, thus, the mode solutions 
corresponding to the Bunch-Davies vacuum are given by
\begin{equations}[tensor_modes]
S_\vec{p}(\eta, \vec{x}) &= \phi_\vec{p}^{m^2}(\eta, \vec{x}) \eqend{,} 
\label{S_mode} \\
V_{\vec{p}\mu}^{(k)}(\eta, \vec{x}) &= e_\mu^{(k)}(\vec{p}) 
\phi_\vec{p}^{m^2}(\eta, \vec{x}) \eqend{,} \label{V_mode} \\
H_{\vec{p}\mu\nu}^{(k)}(\eta, \vec{x}) &= e_{\mu\nu}^{(k)}(\vec{p}) 
\phi_\vec{p}^{m^2}(\eta, \vec{x}) \eqend{.} \label{H_mode}
\end{equations}
The modes for the field $h^{(\text{T})}_{ab}$ can then be determined from the 
decomposition~\eqref{t_parts}. We obtain
\begin{equations}[h_TT_modes]
h^{(\text{TT},k)}_{\vec{p}00}(\eta, \vec{x}) &= h^{(\text{TT},k)}_{\vec{p}0\mu}(\eta, \vec{x}) 
= 0 \eqend{,} \label{h_TT_00_0i_mode} \\
h^{(\text{TT},k)}_{\vec{p}\mu\nu}(\eta, \vec{x}) &= (H\eta)^{-2} 
e_{\mu\nu}^{(k)}(\vec{p}) \phi_\vec{p}^{m^2}(\eta, \vec{x}) \label{h_TT_ij_mode}
\end{equations}
for the spatial tensor modes by substituting Eq.~\eqref{H_mode} into 
Eq.~\eqref{t_parts} with $S = 0 = V_\mu$, and
\begin{equations}[h_TV_modes]
h^{(\text{TV},k)}_{\vec{p}00}(\eta, \vec{x}) &= 0 \eqend{,} \label{h_TV_00_mode} \\
h^{(\text{TV},k)}_{\vec{p}0\mu}(\eta, \vec{x}) &= - (H\eta)^{-1} e_\mu^{(k)}(\vec{p}) 
\phi_\vec{p}^{m^2}(\eta, \vec{x}) \eqend{,} \label{h_TV_0i_mode} \\
h^{(\text{TV},k)}_{\vec{p}\mu\nu}(\eta, \vec{x}) &= \frac{2 \mathi}{H \eta} 
\frac{p_{(\mu} e_{\nu)}^{(k)}(\vec{p})}{\vec{p}^2} \left( \partial_\eta - 
\frac{n-1}{\eta} \right) \phi_\vec{p}^{m^2}(\eta, \vec{x}) \label{h_TV_ij_mode}
\end{equations}
for the spatial vector modes by substituting Eq.~\eqref{V_mode} into 
Eq.~\eqref{t_parts} with $S = 0 = H_{\mu\nu}$. The spatial scalar modes are 
found by substituting Eq.~\eqref{S_mode} into Eq.~\eqref{t_parts} with $V_\mu = 
0 = H_{\mu\nu}
$, and using Eq.~\eqref{S_eq} to simplify the results, as follows:
\begin{equations}[h_TS_modes]
h^{(\text{TS})}_{\vec{p}00}(\eta, \vec{x}) &= \phi_\vec{p}^{m^2}(\eta, \vec{x}) 
\eqend{,} \label{h_TS_00_mode} \\
h^{(\text{TS})}_{\vec{p}0\mu}(\eta, \vec{x}) &= - \mathi \frac{p_\mu}{\vec{p}^2} \left( 
\partial_\eta - \frac{n-2}{\eta} \right) \phi_\vec{p}^{m^2}(\eta, \vec{x}) 
\eqend{,} \label{h_TS_0i_mode} \\
\begin{split}
h^{(\text{TS})}_{\vec{p}\mu\nu}(\eta, \vec{x}) &= \frac{p_\mu p_\nu}{\vec{p}^2} 
\phi_\vec{p}^{m^2}(\eta, \vec{x})- \frac{\tilde{Q}_{\mu\nu}}{\eta^2 \vec{p}^2} 
\\
&\times \left[ \eta \partial_\eta - (n-1) + \frac{m^2}{(n-2) H^2} \right] 
\phi_\vec{p}^{m^2}(\eta,\vec{x}) \eqend{,} \label{h_TS_ij_mode}
\end{split}
\end{equations}
where
\begin{equation}
\tilde{Q}_{\mu\nu} = \eta_{\mu\nu} - (n-1) \frac{p_\mu p_\nu}{\vec{p}^2}
\end{equation}
is the Fourier transform of $Q_{\mu\nu}$ defined by Eq.~\eqref{Q_operator}.

We expand the tensor sector of the graviton field operator, 
$\hat{h}_{ab}^{(\text{T})}$, as
\begin{splitequation}
\label{field_expansion_tensor_sector}
\hat{h}_{ab}^{(\text{T})} &= \int \left[ \hat{a}^{(\text{TS})}(\vec{p}) 
h_{\vec{p}ab}^{(\text{TS})} + \textrm{H.c.} \right] \frac{\total^{n-1} 
p}{(2\pi)^{n-1}} \\
&\ + \sum_{k=1}^{n-2} \int \left[ \hat{a}^{(\text{TV})}_k(\vec{p}) 
h_{\vec{p}ab}^{(\text{TV},k)} + \textrm{H.c.} \right] \frac{\total^{n-1} 
p}{(2\pi)^{n-1}} \\
&\ + \sum_{k=1}^{\mathclap{\textstyle\frac{n(n-3)}{2}}} \int \left[ 
\hat{a}^{(\text{TT})}_k(\vec{p}) h_{\vec{p}ab}^{(\text{TT},k)} + \textrm{H.c.} 
\right] \frac{\total^{n-1} p}{(2\pi)^{n-1}} \eqend{.}
\end{splitequation}
The commutation relations for the creation and annihilation operators in this 
expansion are obtained from the symplectic product~\eqref{symplectic_prod} of 
the modes~\eqref{h_TT_modes},~\eqref{h_TV_modes}, and~\eqref{h_TS_modes}. The 
calculation is again facilitated by taking the hypersurfaces $\Sigma$ to be 
$\eta = \text{const}$. Thus, we obtain
\begin{equations}
\symp{h_{\vec{p}ab}^{(\text{TT},k)}}{h_{\vec{q}cd}^{(\text{TT},l)}} &= 
\frac{1}{2} \delta_{kl} (2\pi)^{n-1} \delta^{n-1}(\vec{p}-\vec{q}) \eqend{,} 
\label{h_TT_norm} \\
\symp{h_{\vec{p}ab}^{(\text{TV},k)}}{h_{\vec{q}cd}^{(\text{TV},l)}} &= 
\frac{m^2}{\vec{p}^2} \delta_{kl} (2\pi)^{n-1} \delta^{n-1}(\vec{p}-\vec{q}) 
\eqend{,} \label{h_TV_norm} \\
\begin{split}
\symp{h_{\vec{p}ab}^{(\text{TS})}}{h_{\vec{q}cd}^{(\text{TS})}} &= \frac{(n-1) 
m^2 [ m^2 - (n-2) H^2 ]}{2 (n-2) (\vec{p}^2)^2} \\
&\quad\times (2\pi)^{n-1} \delta^{n-1}(\vec{p}-\vec{q}) \eqend{,} 
\label{h_TS_norm}
\end{split} \raisetag{1.1\baselineskip}
\end{equations}
and all other components vanish. Note that the symplectic norm of 
$h^{(\text{TS})}_{\vec{q}ab}$ vanishes for $m^2 = (n-2) H^2$. This is a 
consequence of a gauge symmetry for this value of 
mass~\cite{deser_nepomechie_ap_1984}. The 
commutators for the creation and annihilation operators can be found by using 
Eq.~\eqref{a_a_dagger_commutator} as
\begin{equations}
\left[ \hat{a}^{(\text{TT})}_k(\vec{p}), \hat{a}^{(\text{TT})\dagger}_l(\vec{q}) 
\right] &= 2 \delta_{kl} (2\pi)^{n-1} \delta^{n-1}(\vec{p}-\vec{q}) \eqend{,} 
\label{commutator_a_a_dagger_TT_part} \raisetag{1.1\baselineskip} \\
\left[ \hat{a}^{(\text{TV})}_k(\vec{p}), \hat{a}^{(\text{TV})\dagger}_l(\vec{q}) 
\right] &= \frac{\vec{p}^2}{m^2} \delta_{kl} (2\pi)^{n-1} 
\delta^{n-1}(\vec{p}-\vec{q}) \eqend{,} \label{commutator_a_a_dagger_TV_part} \\
\begin{split}
\left[ \hat{a}^{(\text{TS})}(\vec{p}), \hat{a}^{(\text{TS})\dagger}(\vec{q}) 
\right] &= \frac{2 (n-2) (\vec{p}^2)^2}{(n-1) m^2 \left[ m^2 - (n-2) H^2 
\right]} \\
&\quad\times (2\pi)^{n-1} \delta^{n-1}(\vec{p}-\vec{q}) \eqend{,} 
\label{commutator_a_a_dagger_TS_part}
\end{split}
\end{equations}
with all others vanishing. Notice that, if $0 < m^2 < (n-2) H^2$, then the 
spatial scalar sector has negative norm~\cite{higuchi_npb_1987}.

The tensor sector of the Wightman two-point function in the Bunch-Davies vacuum 
can thus be expressed as
\begin{splitequation}
\label{tensor_sector_two_point_function_mode_sum}
\Delta^{(\text{T},m^2)}_{abc'd'}(x,x') &= N^{(\text{V})} 
\Delta^{(\text{TT},m^2)}_{abc'd'}(x,x') \\
&\quad+ N^{(\text{V})} \Delta^{(\text{TV},m^2)}_{abc'd'}(x,x') \\
&\quad+ N^{(\text{S})} \Delta^{(\text{TS},m^2)}_{abc'd'}(x,x') \eqend{,}
\end{splitequation}
with the mode sums
\begin{splitequation}
\label{Delta_TT}
&\Delta^{(\text{TT},m^2)}_{abc'd'}(x,x') \\
&\quad\equiv 2 m^2 \frac{(4\pi)^\frac{n}{2}}{H^n} 
\sum_{k=1}^{\mathclap{\textstyle\frac{n(n-3)}{2}}} \int 
h_{\vec{p}ab}^{(\text{TT},k)}(x) h_{\vec{p}c'd'}^{(\text{TT},k)*}(x') 
\frac{\total^{n-1} p}{(2\pi)^{n-1}} \eqend{,}
\end{splitequation}
\begin{splitequation}
\label{Delta_TV}
&\Delta^{(\text{TV},m^2)}_{abc'd'}(x,x') \\
&\quad\equiv \frac{(4\pi)^\frac{n}{2}}{H^n} \sum_{k=1}^{n-2} \int \vec{p}^2 
h_{\vec{p}ab}^{(\text{TV},k)}(x) h_{\vec{p}c'd'}^{(\text{TV},k)*}(x') 
\frac{\total^{n-1} p}{(2\pi)^{n-1}} \eqend{,}
\end{splitequation}
and
\begin{splitequation}
\label{Delta_TS}
&\Delta^{(\text{TS},m^2)}_{abc'd'}(x,x')\\
&\quad\equiv \frac{(4\pi)^\frac{n}{2}}{H^{n+2}} \int (\vec{p}^2)^2 
h_{\vec{p}ab}^{(\text{TS})}(x) h_{\vec{p}c'd'}^{(\text{TS})*}(x') 
\frac{\total^{n-1} p}{(2\pi)^{n-1}} \eqend{,}
\end{splitequation}
and the constant
\begin{splitequation}
\label{N_S}
N^{(\text{S})} &\equiv \lim_{\alpha \to 2 (n-1)/(n-2)} N^{(\text{S})}_\alpha \\
&= \frac{2 (n-2) H^4}{(n-1)m^2 \left[ m^2 - (n-2) H^2 \right]} 
\frac{H^{n-2}}{(4\pi)^\frac{n}{2}} \eqend{,}
\end{splitequation}
with $N^{(\text{S})}_\alpha$ defined by Eq.~\eqref{N_S_alpha} and 
$N^{(\text{V})}$ given by Eq.~\eqref{N_V}. The explicit form of the mode 
functions~\eqref{h_TT_modes},~\eqref{h_TV_modes}, and~\eqref{h_TS_modes} shows 
that the momentum integrals~\eqref{Delta_TT},~\eqref{Delta_TV}, 
and~\eqref{Delta_TS} converge in the IR if the mass $m^2$ of the scalar mode 
functions is positive. This is because the factor of $p_\mu p_\nu/\vec{p}^4 \sim 
1/\vec{p}^2$ in the spatial vector part is canceled by the explicit factor of 
$\vec{p}^2$ in the momentum integral~\eqref{Delta_TV}, and the explicit factor 
of $1/(\vec{p}^2)^2$ in the spatial scalar part is also canceled in the momentum 
integral~\eqref{Delta_TS}. Hence, we assume $m^2 > 0$ for the time being and 
treat the limit $m^2 \to 0$ in Sec.~\ref{sec:ir_issues}.

Thus, by substituting the mode functions~\eqref{h_TT_modes},~\eqref{h_TV_modes}, 
and~\eqref{h_TS_modes} into Eqs.~\eqref{Delta_TT}, \eqref{Delta_TV}, and 
\eqref{Delta_TS}, respectively, we obtain the spatial tensor, spatial vector 
and spatial scalar part of the tensor sector of the graviton two-point function. 
The spatial scalar part is given by
\begin{widetext}
\begin{equations}
\Delta^{(\text{TS},m^2)}_{0000}(x,x') &= \laplace^2 I_\text{T}(Z) \eqend{,} \\
\Delta^{(\text{TS},m^2)}_{000\sigma'}(x,x') &= \left( \partial_{\eta'} - 
\frac{n-2}{\eta'} \right) \partial_{\sigma'} \laplace I_\text{T}(Z) \eqend{,} \\
\Delta^{(\text{TS},m^2)}_{00\rho'\sigma'}(x,x') &= (\eta')^{-2} \left[ \eta' 
\partial_{\eta'} - (n-1) + \frac{m^2}{(n-2) H^2} \right] Q_{\rho'\sigma'} 
\laplace I_\text{T}(Z) + \partial_{\rho'} \partial_{\sigma'} \laplace 
I_\text{T}(Z) \eqend{,} \\
\Delta^{(\text{TS},m^2)}_{0\nu0\sigma'}(x,x') &= \left( \partial_\eta - 
\frac{n-2}{\eta} \right) \left( \partial_{\eta'} - \frac{n-2}{\eta'} \right) 
\partial_\nu \partial_{\sigma'} I_\text{T}(Z) \eqend{,} \\
\Delta^{(\text{TS},m^2)}_{0\nu\rho'\sigma'}(x,x') &= \left( \partial_\eta - 
\frac{n-2}{\eta} \right) \left\{ (\eta')^{-2} \left[ \eta' \partial_{\eta'} - 
(n-1) + \frac{m^2}{(n-2) H^2} \right] Q_{\rho'\sigma'} + \partial_{\rho'} 
\partial_{\sigma'} \right\} \partial_\nu I_\text{T}(Z) \eqend{,} \\
\begin{split}
\Delta^{(\text{TS},m^2)}_{\mu\nu\rho'\sigma'}(x,x') &= \left\{ 
\frac{Q_{\mu\nu}}{\eta^2} \left[ \eta \partial_\eta - (n-1) + \frac{m^2}{(n-2) 
H^2} \right] + \partial_\mu \partial_\nu \right\} \\
&\quad\times \left\{ \frac{Q_{\rho'\sigma'}}{(\eta')^2} \left[ \eta' 
\partial_{\eta'} - (n-1) + \frac{m^2}{(n-2) H^2} \right] + \partial_{\rho'} 
\partial_{\sigma'} \right\} I_\text{T}(Z) \eqend{,}
\end{split}
\end{equations}
The nonzero components of the spatial vector part are given by
\begin{equations}
\Delta^{(\text{TV},m^2)}_{00c'd'}(x,x') &= 0 \eqend{,} \\
\Delta^{(\text{TV},m^2)}_{0\nu0\sigma'}(x,x') &= - \frac{1}{H^2 \eta \eta'} 
P_{\nu\sigma} \laplace I_\text{T}(Z) \eqend{,} \\
\Delta^{(\text{TV},m^2)}_{0\nu\rho'\sigma'}(x,x') &= \frac{2}{H^2 \eta \eta'} 
\left( \partial_{\eta'} - \frac{n-1}{\eta'} \right) P_{\nu(\rho} 
\partial_{\sigma)} I_\text{T}(Z) \eqend{,} \\
\Delta^{(\text{TV},m^2)}_{\mu\nu\rho'\sigma'}(x,x') &= \frac{4}{H^2 \eta \eta'} 
\left( \partial_\eta - \frac{n-1}{\eta} \right) \left( \partial_{\eta'} - 
\frac{n-1}{\eta'} \right) \frac{\partial_{(\mu} P_{\nu)(\rho} 
\partial_{\sigma)}}{\laplace} I_\text{T}(Z) \eqend{,}
\end{equations}
and the spatial tensor part is given by
\begin{equation}
\Delta^{(\text{TT},m^2)}_{\mu\nu\rho'\sigma'}(x,x') = \frac{2 m^2}{H^2 \eta^2 
(\eta')^2} \left( P_{\mu(\rho} P_{\sigma)\nu} - \frac{1}{n-2} P_{\mu\nu} 
P_{\rho\sigma} \right) I_\text{T}(Z) \eqend{,}
\end{equation}
\end{widetext}
with all components with at least one temporal index vanishing. We recall that 
the operators $P_{\mu\nu}$ and $Q_{\mu\nu}$ are defined by 
Eqs.~\eqref{P_operator} and~\eqref{Q_operator}, respectively. The function 
$I_\text{T}(Z) \equiv I_{\mu_\text{T}}(Z)$ appearing above is defined by 
Eq.~\eqref{I} with the parameter $\mu_\text{T}$ given by
\begin{equation}
\label{mu_T}
\mu_\text{T} \equiv \sqrt{\frac{(n-1)^2}{4} - \frac{m^2}{H^2}} \eqend{.}
\end{equation}
For the $00c'd'$ and the $0\nu0\sigma'$ components, the time derivatives acting 
on $I_\text{T}(Z)$ can be evaluated directly with the help of 
Eqs.~\eqref{del_eta_Z} and~\eqref{del_i_Z}. For the remaining components, we 
proceed as in the 
case of the vector sector, using Eqs.~\eqref{zero_del_eta}--\eqref{two_del_eta} 
to trade time derivatives for Laplacians. This procedure results in
\begin{widetext}
\begin{splitequation}
\Delta^{(\text{TS},m^2)}_{0\nu\rho'\sigma'} + 
\frac{N^{(\text{V})}}{N^{(\text{S})}} 
\Delta^{(\text{TV},m^2)}_{0\nu\rho'\sigma'} &= (\eta')^{-1} \partial_\nu 
Q_{\rho\sigma} \laplace \left\{ \frac{\vec{r}^2}{2 \eta \eta'} 
I^{(0)}_\text{T}(Z) 
+ \left[ \frac{m^2}{(n-2) H^2} - (n-2) \right] I^{(-1)}_\text{T}(Z) \right\} \\
&\quad- \frac{1}{\eta \eta'} \frac{m^2 - (n-2) H^2}{(n-2) H^2} \left( 
\partial_{\eta'} - \frac{n-1}{\eta'} \right) \left[ \eta_{\rho\sigma} 
\partial_\nu - (n-1) \eta_{\nu(\rho} \partial_{\sigma)} \right] 
I^{(0)}_\text{T}(Z) \\
&\quad+ \left( \partial_\eta - \frac{n-2}{\eta} \right) \partial_\nu 
\partial_\rho \partial_\sigma I^{(0)}_\text{T}(Z)
\end{splitequation}
and
\begin{splitequation}
\label{tensor_sector_2pf_spatial}
&\Delta^{(\text{TS},m^2)}_{\mu\nu\rho'\sigma'} + 
\frac{N^{(\text{V})}}{N^{(\text{S})}} \left[ 
\Delta^{(\text{TV},m^2)}_{\mu\nu\rho'\sigma'} + 
\Delta^{(\text{TT},m^2)}_{\mu\nu\rho'\sigma'} \right] = \frac{(n-2) \tilde{m}^2 
( 1 - \tilde{m}^2 )}{\eta^2 (\eta')^2} [ \eta_{\mu\nu} \eta_{\rho\sigma} - (n-1) 
\eta_{\mu(\rho} \eta_{\sigma)\nu} ] I^{(0)}_\text{T}(Z) \\
&\qquad+ 2 \frac{\eta_{\mu\nu} \eta_{\rho\sigma} \laplace}{(n-1) \eta \eta'} 
\left\{ I^{(1)}_\text{T}(Z) + ( 1 - \tilde{m}^2 ) \left[ Z I^{(0)}_\text{T}(Z) + 
(n-2) I^{(-1)}_\text{T}(Z) \right] \right\} + \partial_\mu \partial_\nu 
\partial_\rho \partial_\sigma I^{(0)}_\text{T}(Z) \\
&\qquad+ \frac{(\eta')^2 + \eta^2}{\eta^2 (\eta')^2} \frac{\eta_{\mu\nu} 
\eta_{\rho\sigma} \laplace}{(n-1)} \left\{ Z I^{(1)}_\text{T}(Z) + \left[ 
\tilde{m}^2 - (n-1) \right] I^{(0)}_\text{T}(Z) \right\} - \left( \eta_{\mu\nu} 
\partial_\rho \partial_\sigma + \eta_{\rho\sigma} \partial_\mu \partial_\nu 
\right) \frac{1}{\eta \eta'} I^{(1)}_\text{T}(Z) \\
&\qquad- 2 \frac{( 1 - \tilde{m}^2 )}{\eta \eta'} \left[ \eta_{\mu\nu} 
\partial_\rho \partial_\sigma + \eta_{\rho\sigma} \partial_\mu \partial_\nu - 
(n-1) \partial_{(\mu} \eta_{\nu)(\rho} \partial_{\sigma)} \right] \left[ Z 
I^{(0)}_\text{T}(Z) + (n-2) I^{(-1)}_\text{T}(Z) \right] \\
&\qquad+ \left( \frac{\eta_{\rho\sigma} \partial_\mu \partial_\nu}{\eta^2} + 
\frac{\eta_{\mu\nu} \partial_\rho \partial_\sigma}{(\eta')^2} \right) \left\{ Z 
I^{(1)}_\text{T}(Z) + \left[ (n-1) - \tilde{m}^2 \right] I^{(0)}_\text{T}(Z) 
\right\} \\
&\qquad+ \frac{\eta^2 + (\eta')^2}{(n-1) \eta^2 (\eta')^2} Q_{\mu\nu} 
Q_{\rho\sigma} \laplace \left\{ Z I^{(1)}_\text{T}(Z) + \left[ (n-1) - 
\tilde{m}^2 \right] I^{(0)}_\text{T}(Z) \right\} \\
&\qquad- \frac{1}{(n-1) \eta \eta'} Q_{\mu\nu} Q_{\rho\sigma} \laplace \left\{ 2 
I^{(1)}_\text{T}(Z) + \left[ (n+1) - 2 \tilde{m}^2 \right] Z I^{(0)}_\text{T}(Z) 
+ \left[ (n-2) (n+1) - (3n-5) \tilde{m}^2 \right] I^{(-1)}_\text{T}(Z) \right\} 
\eqend{,}
\end{splitequation}
\end{widetext}
with the abbreviation $\tilde{m}^2 \equiv m^2/[(n-2) H^2]$. Except for the 
components~\eqref{tensor_sector_2pf_spatial} with all indices spatial, the 
remaining derivatives can now be evaluated, and the result can be cast into a 
de~Sitter-
invariant form. It is most useful to present this result in terms of 
de~Sitter--invariant bitensors and scalar coefficient functions. Thus, we find
\begin{equation}
\label{tensor_sector_two_point_function_m}
\Delta^{(\text{T},m^2)}_{abc'd'}(x,x') = - \frac{N^{(\text{S})}}{n-2} 
\sum_{k=1}^5 T^{(k)}_{abc'd'} F^{(\text{T},k)}(Z)
\end{equation}
with the bitensor set~\eqref{T_set} and the coefficients 
\begin{equations}[F_T]
\begin{split}
F^{(\text{T},1)} &= (1-Z^2)^2 I^{(4)}_\text{T} - 2 (n+1) (1-Z^2) Z 
I^{(3)}_\text{T} \\
&\quad- (n+1) (2 - n Z^2) I^{(2)}_\text{T} \eqend{,}
\end{split} \label{F_T_1} \\
\begin{split}
F^{(\text{T},2)} &= - (1-Z^2) I^{(4)}_\text{T} + 2 (n+1) Z I^{(3)}_\text{T} \\
&\quad+ n (n+1) I^{(2)}_\text{T} \eqend{,}
\end{split} \label{F_T_2} \raisetag{1.1\baselineskip} \\
\begin{split}
F^{(\text{T},3)} &= [ 1 - (n-1) Z^2 ] I^{(4)}_\text{T} - 2 (n^2-1) Z 
I^{(3)}_\text{T} \\
&\quad- n (n^2-1) I^{(2)}_\text{T} \eqend{,}
\end{split}
\label{F_T_3} \\
\begin{split}
F^{(\text{T},4)} &= - \frac{n-1}{2} (1-Z^2) Z I^{(4)}_\text{T} + \frac{n 
(n^2-1)}{2} Z I^{(2)}_\text{T} \\
&\quad- \frac{n+1}{2} [ n-2 - 2 (n-1) Z^2 ] I^{(3)}_\text{T} \eqend{,}
\end{split} \label{F_T_4} \raisetag{1.1\baselineskip} \\
\begin{split}
F^{(\text{T},5)} &= - \frac{n-1}{2} (1-Z^2)^2 I^{(4)}_\text{T} + (n^2-1) (1-Z^2) 
Z I^{(3)}_\text{T} \\
&\quad+ \frac{n (n+1)}{2} [ 1 - (n-1) Z^2 ] I^{(2)}_\text{T} \eqend{,}
\end{split} \label{F_T_5} \raisetag{1.1\baselineskip}
\end{equations}
where we have used Eq.~\eqref{hypergeometric_diff_eq_I} to simplify the 
expressions. It can readily be verified that the right-hand side of 
Eq.~\eqref{tensor_sector_two_point_function_m} is transverse (in spacetime) and 
traceless with the 
aid of Eqs.~\eqref{Z_relations}.

One can show that the purely spatial 
components~\eqref{tensor_sector_2pf_spatial} agree with 
Eq.~\eqref{tensor_sector_two_point_function_m} by proving that
\begin{equation}
\label{eq_to_show_AH}
\Delta^{(\text{T},m^2)}_{\mu\nu\rho'\sigma'}(x,x') + \frac{N^{(\text{S})}}{n-2} 
\sum_{k=1}^5 T^{(k)}_{\mu\nu\rho'\sigma'} F^{(\text{T},k)}(Z) = 0 \eqend{.}
\end{equation}
Since the mode sums~\eqref{Delta_TT},~\eqref{Delta_TV}, and~\eqref{Delta_TS} are 
convergent for small $p$ as long as $m^2 > 0$ (which we have assumed above), 
Eq.~\eqref{eq_to_show_AH} holds if
\begin{equation}
\laplace \left[ \Delta^{(\text{T},m^2)}_{\mu\nu\rho'\sigma'}(x,x') + 
\frac{N^{(\text{S})}}{n-2} \sum_{k=1}^5 T^{(k)}_{\mu\nu\rho'\sigma'}
F^{(\text{T},k)}(Z) \right] = 0 \eqend{.}
\end{equation}
This equation can readily be checked by explicit calculation.

\subsection{Overall result and the massless limit}
\label{subsec:massless_limit}

In the preceding sections, we have derived the expressions for the scalar, 
vector, and tensor sectors of the graviton two-point function in a two-parameter 
family of covariant gauges in terms of mode sums. The vector and tensor sectors 
have been regularized by a Fierz-Pauli mass term. As already mentioned above, 
while the expression for the scalar sector, 
Eq.~\eqref{scalar_sector_two_point_function_mode_sum}, converges for $m = 0$ and 
$\beta > 0$, the ones obtained for the vector and tensor sectors, 
Eqs.~\eqref{vector_sector_two_point_function_mode_sum} 
and~\eqref{tensor_sector_two_point_function_mode_sum}, are only convergent for 
$\alpha m^2 > n H^2$ and $m^2 > 0$, respectively. Our aim now is to show that 
despite this fact, the sum of the vector and tensor sectors (and thus the whole 
graviton two-point function) is finite in the limit $m \to 0$. We note that, 
although the two-point function for the divergence-free vector field $v_a$ 
satisfying Eq.~\eqref{vector_sector_field_eq} is defined only for $\alpha m^2 > 
n H^2$, the expression for the two-point function for $\nabla_a v_b + \nabla_b 
v_a$ given by Eqs.~\eqref{vector_sector_tp_AH} and~\eqref{F_V} is well defined 
for $0 < \alpha m^2 < 2 (n-1) H^2$ as well. Since $n H^2 < 2 (n-1) H^2$ for $n > 
2$, we can vary $\alpha m^2$ from above $n H^2$ to $0$ continuously. (This does 
not mean, however, that the momentum integrals for the vector sector is 
convergent with $0 < \alpha m^2 < n H^2$. We will find in the next section that 
they are IR-divergent. Thus, we are analytically continuing the vector sector 
obtained for $\alpha m^2 > n H^2$ to all positive values of $m^2$.)

Let us first analyze the behavior of $\Delta^{(\text{V},m^2)}_{abc'd'}$ and 
$\Delta^{(\text{T},m^2)}_{abc'd'}$ for small $m^2$. We begin by noting that the 
final result for the vector sector, 
Eq.~\eqref{vector_sector_two_point_function_m}, diverges like $1/m^2$ for $m \to 
0$ because of the constant factor $N^{(\text{S})}_\alpha$ defined by 
Eq.~\eqref{N_S_alpha}. For this reason, we expand the coefficients 
$F^{(\text{V},k)}$ to first order in $m^2$. Similarly, the final result 
for the tensor sector, Eq.~\eqref{tensor_sector_two_point_function_m}, diverges 
like $1/m^2$ for $m \to 0$ because of the constant $N^{(\text{S})}$ defined by 
Eq.~\eqref{N_S}, so we also expand the coefficients $F^{(\text{T},k)}$ to first 
order in $m^2$. The coefficients $F^{(\text{T},k)}$ and $F^{(\text{V},k)}$ are 
given in terms of the derivatives of $I_\text{T}(Z)$ and $I^{(1)}_\text{V}(Z)$ 
with respect to $Z$ through Eqs.~\eqref{F_T} and~\eqref{F_V}, respectively. 
Using hypergeometric identities~\cite{dlmf}, we see that, as $m \to 0$, the 
integral $I_\text{T}(Z)$ [with $\mu_\text{T}$ given by Eq.~\eqref{mu_T}] 
diverges as 
\begin{equation}
\label{I_T_small_m}
I_\text{T}(Z) = \frac{\Gamma(n)}{\Gamma\left( \frac{n}{2} \right)} 
\frac{H^2}{m^2} + \bigo{m^0} \eqend{.}
\end{equation}
The divergent term in Eq.~\eqref{I_T_small_m}, however, is constant such that 
all derivatives $I^{(k)}_\text{T}(Z)$ with $k \geq 1$ are finite in the massless 
limit. Similarly, as $m \to 0$, the integral $I^{(1)}_\text{V}(Z)$ [with 
$\mu_\text{V}$ given in Eq.~\eqref{mu_V}] diverges as
\begin{equation}
\label{I_V_small_m}
I^{(1)}_\text{V}(Z) = \frac{\Gamma(n+2)}{2 \Gamma\left( \frac{n}{2} + 1 \right)} 
\frac{H^2}{\alpha m^2} + \bigo{m^0} \eqend{,}
\end{equation}
and all higher derivatives $I^{(k)}_\text{V}(Z)$ with $k \geq 2$ are finite. The 
coefficients for the vector and tensor sectors depend only on 
$I^{(k)}_\text{V}(Z)$ and $I^{(k)}_\text{T}(Z)$ with $k \geq 2$. Therefore it is 
enough to 
expand the integrals $I^{(k)}_\mu$ with $k \geq 2$ up to linear order. By noting 
that 
\begin{equations}
\lim_{m \to 0} \mu_\text{V} &= \frac{n+1}{2} \eqend{,} \\
\lim_{m \to 0} \mu_\text{T} &= \frac{n-1}{2} \eqend{,}
\end{equations}
and using the definition~\eqref{I_tilde_k} for the $\mu$-derivative of 
$I_\mu(Z)$, we find the small-$m$ limit of the integrals $I^{(k)}_\text{V}$ and 
$I^{(k)}_\text{T}$ with $k \geq 2$ as
\begin{equations}[I_k_expansion]
I^{(k)}_\text{V} &= I^{(k)}_{(n+1)/2} + \frac{\alpha m^2}{H^2} 
\tilde{I}^{(k)}_{(n+1)/2} + \bigo{m^4} \eqend{,} \\
I^{(k)}_\text{T} &= I^{(k)}_{(n-1)/2} + \frac{m^2}{H^2} 
\tilde{I}^{(k)}_{(n-1)/2} + \bigo{m^4} \eqend{.}
\end{equations}

Now we write the vector sector and tensor sector of the two-point function, 
Eqs.~\eqref{vector_sector_two_point_function_m} 
and~\eqref{tensor_sector_two_point_function_m}, respectively, for small $m^2$ as
\begin{splitequation}
\label{vector_sector_two_point_function_small_m}
\Delta^{(\text{V},m^2)}_{abc'd'}(x,x') &= \left[ \frac{H^2}{m^2} + 
\frac{\alpha}{2 (n-1)} \right] \Delta^{(\text{V},\text{div})}_{abc'd'}(x,x') \\
&\quad+ \alpha \Delta^{(\text{V},\text{fin})}_{abc'd'}(x,x') + \bigo{m^2} 
\raisetag{\baselineskip}
\end{splitequation}
and
\begin{splitequation}
\label{tensor_sector_two_point_function_small_m}
\Delta^{(\text{T},m^2)}_{abc'd'}(x,x') &= \left( \frac{H^2}{m^2} + \frac{1}{n-2} 
\right) \Delta^{(\text{T},\text{div})}_{abc'd'}(x,x') \\
&\quad+ \Delta^{(\text{T},\text{fin})}_{abc'd'}(x,x') + \bigo{m^2} \eqend{,} 
\raisetag{\baselineskip}
\end{splitequation}
where we have introduced the following $m$-independent and de~Sitter--invariant 
bitensors:
\begin{splitequation}
\label{vector_tensor_sector_small_m_bitensors}
&\Delta^{(\text{T/V},\text{div/fin})}_{abc'd'}(x,x') \\
&= - \frac{2}{n-1} \frac{H^{n-2}}{(4\pi)^\frac{n}{2}} \sum_{k=1}^5 
T^{(k)}_{abc'd'} F^{(\text{T/V},\text{div/fin},k)}(Z) \eqend{.}
\end{splitequation}
Here the coefficients $F^{(\text{T/V},\text{div/fin},k)}(Z)$ defined by
\begin{splitequation}
F^{(\text{T/V},k)}(Z) &= F^{(\text{T/V},\text{div},k)}(Z) \\
&\ + \frac{m^2}{H^2} F^{(\text{T/V},\text{fin},k)}(Z) + \bigo{m^4} \eqend{.}
\end{splitequation}
They are obtained by substituting Eqs.~\eqref{I_k_expansion} into 
Eqs.~\eqref{vector_sector_two_point_function_m} 
and~\eqref{tensor_sector_two_point_function_m}. We first simplify the ``div'' 
coefficients, $F^{(\text{T/V},\text{div},k)}(Z)$, further with the aid of the 
recursion relation~\eqref{hypergeometric_diff_eq_I}. Since the limit $m \to 0$ 
exists for the $I^{(k)}_\text{V}$ with $k \geq 2$ and the $I^{(k)}_\text{T}$ 
with $k \geq 1$, we can write down these recursion relations for these cases 
simply by letting $m^2 = 0$. However, we also need these relations for $k = 1$ 
in the vector sector and $k = 0$ in the tensor sector. For these cases of the 
recursion relation~\eqref{hypergeometric_diff_eq_I}, we first substitute the 
small-$m$ expansions~\eqref{I_T_small_m} and~\eqref{I_V_small_m} into 
Eq.~\eqref{hypergeometric_diff_eq_I} and then take the limit $m \to 0$. This 
procedure results in
\begin{equations}[hypergeometric_diff_eq_massless]
(1-Z^2) I^{(2)}_{(n-1)/2} - n Z I^{(1)}_{(n-1)/2} &= 
\frac{\Gamma(n)}{\Gamma\left( \frac{n}{2} \right)} \eqend{,} \\
(1-Z^2) I^{(3)}_{(n+1)/2} - (n+2) Z I^{(2)}_{(n+1)/2} &= \frac{\Gamma(n+2)}{2 
\Gamma\left( \frac{n}{2} + 1 \right)} \eqend{,}
\end{equations}
which have an additional term on the right-hand side arising as the product of a 
divergent term proportional to $H^2/m^2$ and the first-order term in the 
expansion of $I_\mu(Z)$ in $m^2$. Furthermore, by using the series expansion of 
the hypergeometric functions~\cite{dlmf} we obtain the relation
\begin{equation}
\label{n_+_1_n_-_1_relation}
n I^{(1)}_{(n-1)/2} = (1-Z^2) I^{(2)}_{(n+1)/2} - \frac{\Gamma(n)}{\Gamma\left( 
\frac{n}{2} \right)} Z \eqend{,}
\end{equation}
which we can use to replace all occurrences of the $I^{(k)}_{(n-1)/2}$ with $k 
\geq 1$ by the $I^{(k)}_{(n+1)/2}$. Thus, we find the ``div'' coefficients in 
Eq.~\eqref{vector_tensor_sector_small_m_bitensors} as
\begin{equations}[F_T_div]
F^{(\text{T},\text{div},1)} &= Z I^{(2)}_{(n+1)/2} \eqend{,} \label{F_T_1_div} 
\\
F^{(\text{T},\text{div},2)} &= - I^{(3)}_{(n+1)/2} \eqend{,} \label{F_T_2_div} 
\\
F^{(\text{T},\text{div},3)} &= Z I^{(4)}_{(n+1)/2} + (n+1) I^{(3)}_{(n+1)/2} 
\eqend{,} \label{F_T_3_div} \\
F^{(\text{T},\text{div},4)} &= Z I^{(3)}_{(n+1)/2} + \frac{n+1}{2} 
I^{(2)}_{(n+1)/2} \eqend{,} \label{F_T_4_div} \\
F^{(\text{T},\text{div},5)} &= Z I^{(2)}_{(n+1)/2} + \frac{\Gamma(n+2)}{4 
\Gamma\left( \frac{n}{2} + 1 \right)} \eqend{,} \label{F_T_5_div}
\end{equations}
and $F^{(\text{V},\text{div},k)} = - F^{(\text{T},\text{div},k)}$, which is 
obtained using Eq.~\eqref{n_+_1_n_-_1_relation}. This last relation implies that 
$\Delta^{(\text{V},\text{div})}_{abc'd'}(x,x') + 
\Delta^{(\text{T},\text{div})}_{abc'd'}(x,x') = 0$ and, consequently, that the 
limit
\begin{splitequation}
\label{vectortensor_two_point_function_small_m}
\Delta^{(\text{VT})}_{abc'd'}(x,x') &\equiv \lim_{m \to 0} \left[ 
\Delta^{(\text{V},m^2)}_{abc'd'}(x,x') + \Delta^{(\text{T},m^2)}_{abc'd'}(x,x') 
\right] \\
&= \left[ \frac{1}{n-2} - \frac{\alpha}{2 (n-1)} \right] 
\Delta^{(\text{T},\text{div})}_{abc'd'}(x,x') \\
&\quad+ \Delta^{(\text{T},\text{fin})}_{abc'd'}(x,x') + \alpha 
\Delta^{(\text{V},\text{fin})}_{abc'd'}(x,x')
\end{splitequation}
\vspace{0.4cm}
is finite.

The ``fin'' coefficients, $F^{(\text{T/V},\text{fin},k)}(Z)$, can be computed 
using the recursion relation~\eqref{hypergeometric_diff_eq_tilde_I}. For the 
$\tilde{I}^{(k)}_\text{V}$ with $k \geq 2$ and the $\tilde{I}^{(k)}_\text{T}$ 
with $k \geq 1$, we may simply substitute $m^2 = 0$ to find these recursion 
relations. For the case $k = 1$ in the vector sector and $k = 0$ in the tensor 
sector, which we also need, we first show that for small $m$
\begin{equations}[I_0_T_and_I_1_V_massless_lim]
\left[ \mu_\text{T}^2 - \frac{(n-1)^2}{4} \right] \tilde{I}^{(0)}_\text{T} &= 
\frac{H^2}{m^2} \frac{\Gamma(n)}{\Gamma\left( \frac{n}{2} \right)} + \bigo{m^2} 
\eqend{,} \raisetag{1.5\baselineskip} \\
\left[ \mu_\text{V}^2 - \frac{(n-1)^2}{4} - n \right] \tilde{I}^{(1)}_\text{V} 
&= \frac{H^2}{\alpha m^2} \frac{\Gamma(n+2)}{2 \Gamma\left( \frac{n}{2}+1 
\right)} + \bigo{m^2} \eqend{.}
\end{equations}
The contribution~\eqref{I_0_T_and_I_1_V_massless_lim} present on the left-hand 
side of Eq.~\eqref{hypergeometric_diff_eq_tilde_I} exactly cancels the divergent 
term on the right-hand side in the $m \to 0$ limit [given by 
Eqs.~\eqref{I_T_small_m} and~\eqref{I_V_small_m}]. Let us define
\begin{widetext}
\begin{splitequation}
\label{ii_tensor_def}
\varPi^{(0)}_{(n-1)/2}(Z) &\equiv \lim_{m \to 0} \left( I^{(0)}_\text{T}(Z) - 
\frac{\Gamma(n)}{\Gamma\left( \frac{n}{2} \right)} \frac{H^2}{m^2} \right) = 
\lim_{m \to 0} \left( I^{(0)}_\text{T}(Z) - I^{(0)}_\text{T}(-1) \right) - 
\left( \psi(n-1) + \gamma + \frac{1}{n-1} \right) 
\frac{\Gamma(n-1)}{\Gamma\left( \frac{n}{2} \right)} \\
&= \int_{-1}^Z I^{(1)}_{(n-1)/2}(Z') \total Z' - \left( \psi(n-1) + \gamma + 
\frac{1}{n-1} \right) \frac{\Gamma(n-1)}{\Gamma\left( \frac{n}{2} \right)}
\end{splitequation}
and
\begin{splitequation}
\label{ii_vector_def}
\varPi^{(1)}_{(n+1)/2}(Z) &\equiv \lim_{m \to 0} \left( I^{(1)}_\text{V}(Z) - 
\frac{\Gamma(n+2)}{2 \Gamma\left( \frac{n}{2} + 1 \right)} \frac{H^2}{\alpha 
m^2} \right) = \lim_{m \to 0} \left( I^{(1)}_\text{V}(Z) - I^{(1)}_\text{V}(-1) 
\right) - \left( \psi(n+1) + \gamma + \frac{1}{n+1} \right) 
\frac{\Gamma(n+1)}{2 \Gamma\left( \frac{n}{2} + 1 \right)} \\
&= \int_{-1}^Z I^{(2)}_{(n+1)/2}(Z') \total Z' - \left( \psi(n+1) + \gamma + 
\frac{1}{n+1} \right) \frac{\Gamma(n+1)}{2 \Gamma\left( \frac{n}{2} + 1 \right)} 
\eqend{.} \raisetag{1.3\baselineskip}
\end{splitequation}
\end{widetext}
Then the limit $m \to 0$ of the relation~\eqref{hypergeometric_diff_eq_tilde_I} 
for $k = 0$ in the tensor case and $k = 1$ in the vector case can be written as
\begin{equations}[ii_identities]
(1-Z^2) \tilde{I}^{(2)}_{(n-1)/2} - n Z \tilde{I}^{(1)}_{(n-1)/2} &= 
\varPi^{(0)}_{(n-1)/2} \eqend{,} \raisetag{1.3\baselineskip} \\
(1-Z^2) \tilde{I}^{(3)}_{(n+1)/2} - (n+2) Z \tilde{I}^{(2)}_{(n+1)/2} &= 
\varPi^{(1)}_{(n+1)/2} \eqend{.}
\end{equations}
Finally, using these relations we obtain for the ``fin'' coefficients as 
follows:
\begin{equations}[F_V_fin]
F^{(\text{V},\text{fin},1)} &= - Z \tilde{I}^{(2)}_{(n+1)/2} \eqend{,} 
\label{F_V_1_fin} \\
F^{(\text{V},\text{fin},2)} &= \tilde{I}^{(3)}_{(n+1)/2} \eqend{,} 
\label{F_V_2_fin} \\
F^{(\text{V},\text{fin},3)} &= - Z \tilde{I}^{(4)}_{(n+1)/2} - (n+1) 
\tilde{I}^{(3)}_{(n+1)/2} \eqend{,} \label{F_V_3_fin} \\
F^{(\text{V},\text{fin},4)} &= - Z \tilde{I}^{(3)}_{(n+1)/2} - \frac{n+1}{2} 
\tilde{I}^{(2)}_{(n+1)/2} - \frac{1}{4} I^{(2)}_{(n+1)/2}
\eqend{,} \label{F_V_4_fin} \\
F^{(\text{V},\text{fin},5)} &= - Z \tilde{I}^{(2)}_{(n+1)/2} - \frac{1}{2} 
\varPi^{(1)}_{(n+1)/2} \eqend{,} \label{F_V_5_fin}
\end{equations}
and
\begin{equations}[F_T_fin]
\begin{split}
F^{(\text{T},\text{fin},1)} &= \tilde{I}^{(2)}_{(n-1)/2} - 
\varPi^{(0)}_{(n-1)/2} \\
&\quad- \frac{2}{n-2} Z I^{(1)}_{(n-1)/2} - \frac{(n-1) 
\Gamma(n-2)}{\Gamma\left( \frac{n}{2} \right)} \eqend{,}
\end{split} \label{F_T_1_fin} \\
\begin{split}
F^{(\text{T},\text{fin},2)} &= - Z \tilde{I}^{(3)}_{(n-1)/2} - (n+1) 
\tilde{I}^{(2)}_{(n-1)/2} \\
&\quad+ \frac{1}{n-2} I^{(2)}_{(n-1)/2} \eqend{,}
\end{split} \label{F_T_2_fin} \\
\begin{split}
F^{(\text{T},\text{fin},3)} &= \tilde{I}^{(4)}_{(n-1)/2} + (n-1) Z 
\tilde{I}^{(3)}_{(n-1)/2} \\
&\quad+ (n^2-1) \tilde{I}^{(2)}_{(n-1)/2} - \frac{n-1}{n-2} I^{(2)}_{(n-1)/2} 
\eqend{,}
\end{split} \label{F_T_3_fin} \\
\begin{split}
F^{(\text{T},\text{fin},4)} &= \tilde{I}^{(3)}_{(n-1)/2} + \frac{n-1}{2} \left( 
Z \tilde{I}^{(2)}_{(n-1)/2} + n \tilde{I}^{(1)}_{(n-1)/2}
\right) \\
&\quad+ \frac{n-1}{2 (n-2)} \left[ Z I^{(2)}_{(n-1)/2} + (n-2) I^{(1)}_{(n-1)/2} 
\right] \eqend{,}
\end{split} \label{F_T_4_fin} \\
\begin{split}
F^{(\text{T},\text{fin},5)} &= \tilde{I}^{(2)}_{(n-1)/2} + \frac{n-1}{2} 
\varPi^{(0)}_{(n-1)/2} \\
&\quad+ \frac{n-1}{n-2} \left( Z I^{(1)}_{(n-1)/2} + \frac{\Gamma(n)}{2 
\Gamma\left( \frac{n}{2} \right)} \right) \eqend{.}
\end{split} \label{F_T_5_fin}
\end{equations}

The full two-point function for the graviton in the massless limit is given by
\begin{equation}
\label{graviton_two_point_function_m_0}
\Delta_{abc'd'}(x,x') = \Delta_{abc'd'}^{(\text{S})}(x,x') + 
\Delta_{abc'd'}^{(\text{VT})}(x,x') \eqend{,}
\end{equation}
with the scalar part $\Delta_{abc'd'}^{(\text{S})}(x,x')$ given by 
Eqs.~\eqref{scalar_sector_2pf_bitensors} and~\eqref{F_S}. The right-hand side of 
Eq.~\eqref{graviton_two_point_function_m_0} can naturally be expressed in terms 
of the 
bitensor set~\eqref{T_set}. Indeed, by combining 
Eqs.~\eqref{vectortensor_two_point_function_small_m} 
and~\eqref{vector_tensor_sector_small_m_bitensors}, the two-point 
function~\eqref{graviton_two_point_function_m_0} can be given as
\begin{equation}
\label{full_two_point_function}
\Delta_{abc'd'}(x,x') = \frac{H^{n-2}}{(4\pi)^\frac{n}{2}} \sum_{k=1}^5 
T^{(k)}_{abc'd'} F^{(k)}(Z)
\end{equation}
with the coefficients
\begin{splitequation}
\label{F_k}
F^{(k)}(Z) &\equiv F^{(\text{S},k)}(Z) \\
&- \frac{2}{n-1} \left[ \frac{1}{n-2} F^{(\text{T},\text{div},k)}(Z) + 
F^{(\text{T},\text{fin},k)}(Z) \right] \\
&+ \frac{\alpha}{n-1} \left[ \frac{1}{n-1} F^{(\text{T},\text{div},k)}(Z) - 2 
F^{(\text{V},\text{fin},k)}(Z) \right] \eqend{.}
\end{splitequation}
The functions $F^{(\text{S},k)}$, $F^{(\text{T},\text{div},k)}$, 
$F^{(\text{V},\text{fin},k)}$ and $F^{(\text{T},\text{fin},k)}$ are given by 
Eqs.~\eqref{F_S}, \eqref{F_T_div}, \eqref{F_V_fin}, and~\eqref{F_T_fin}, 
respectively.

Equation~\eqref{full_two_point_function} shows that, for the two-parameter 
family of covariant gauges considered here, there is a de~Sitter--invariant 
two-point function constructed by the mode-sum method which is free of IR 
divergences, 
even though it is necessary to perform analytic continuation of the vector 
sector, which is of pure-gauge form, in the mass parameter $m$ from $\alpha m^2 
> nH^2$ to all positive $m^2$. In the case of the scalar field theory with 
finite 
mass, it is known that, if the spacetime is global de~Sitter space, there is a 
de~Sitter--invariant 
state~\cite{chernikov_tagirov_ap_1968,schomblond_spindel_ahp_1976} which 
coincides with the Bunch-Davies vacuum in the Poincar{\'e} 
patch~\cite{bunch_davies_prsa_1978} and Euclidean vacuum obtain by analytic 
continuation from the sphere~\cite{gibbons_hawking_prd_1977}. For gravitons, it 
is known that on the sphere (in the Euclidean approach) the family of covariant 
gauges employed here leads to a propagator which is both IR finite and, after 
analytic continuation, de~Sitter--invariant in the full de~Sitter 
space~\cite{higuchi_kouris_cqg_2001b}. When restricted to the Poincar{\'e} 
patch, it is straightforward to check that it agrees with our 
result~\eqref{full_two_point_function}, where the explicit expressions in four 
dimensions are given in Appendix~\ref{appendix:n_4_expressions}. Thus, our 
result gives strong evidence for the existence of an 
IR finite, de~Sitter--invariant Hadamard state in the full de~Sitter space whose 
associated Wightman two-point function coincides with 
Eq.~\eqref{full_two_point_function} in the Poincar{\'e} patch, just as in the 
massive scalar field case, as claimed in Ref.~\cite{morrison_arxiv_2013}.

\subsection{Behavior for large separations with \texorpdfstring{$n = 4$}{n = 4}}
\label{subsec:large_z}

Large separations between the two points $x$ and $x'$ of the two-point function 
corresponds to large $Z$~\eqref{Z}, where large timelike separations entail $Z 
\to +\infty$ while large spacelike separations result in $Z \to -\infty$. 
(Note, however, that there is no geodesic connecting the two points if $Z < 
-1$.) It is known that for $n = 4$ the graviton two-point function grows in 
these limits, with the exact nature of the growth depending on the gauge 
parameters~\cite{higuchi_kouris_cqg_2001b} and taking the form of a linearized 
gauge transformation, such that two-point functions of gauge-invariant 
quantities do not 
grow~\cite{higuchi_kouris_cqg_2000,higuchi_kouris_cqg_2001b,higuchi_arxiv_2012,
morrison_arxiv_2013}. The aim of this section is to to clarify the nature of 
that growth for our result~\eqref{full_two_point_function}.

Since the four-dimensional case is the most interesting physically, we will 
specialize to $n = 4$ in this section. Some results here can also be found in 
Ref.~\cite{higuchi_kouris_cqg_2001b}. We first note that, since the function 
$I_\text{S}(Z)$ is essentially the two-point function for the scalar field of 
mass $\sqrt{(n-1)\beta} H$ if $\beta > 0$ [see Eqs.~\eqref{B_eq} 
and~\eqref{Psi_eq}], the scalar-sector two-point function 
$\Delta^{(\text{S})}_{abc'd'}(x,x')$ tends to zero as $\abs{Z} \to \infty$ as 
long as $\beta > 0$. We work under this assumption for the rest of this section. 
Thus, we only need to examine the vector and tensor sectors, which are 
independent of $\beta$. We could then expand the coefficients $F^{(k)}$ for 
large $Z$, whose explicit expressions are given in 
Appendix~\ref{appendix:n_4_expressions}. However, since the bitensor 
set~\eqref{T_set} is not normalized and thus itself changes with distance, we 
need to convert it to a normalized set. Such a set can be constructed from the 
tangent vectors $n_a$ and $n_{a'}$ to the geodesic connecting $x$ and $x'$, 
which can be expressed via
\begin{equation}
n_a = - \frac{Z_{;a}}{H \sqrt{1-Z^2}}
\end{equation}
as a function of $Z$, and the parallel propagator $g_{ab'}$ which reads
\begin{equation}
g_{ab'} = H^{-2} \left( Z_{;ab'} - \frac{Z_{;a} Z_{;b'}}{1+Z} \right) \eqend{.}
\end{equation}
These bivectors were first used by Allen and Jacobson in 
Ref.~\cite{allen_jacobson_cmp_1986}. From the relations
\begin{equation}
n_a n^a = 1 \eqend{,} \quad n^a g_{ab'} = - n_{b'} \eqend{,} \quad g_{ab'} 
g^{b'c} = \delta_a^c
\end{equation}
[which can be derived, e.g., from Eqs.~\eqref{Z_relations}], one finds that they 
are, indeed, normalized. We now introduce the set of bitensors
\begin{equations}[S_set]
S^{(1)}_{abc'd'} &\equiv g_{ab} g_{c'd'} \label{S_1} \eqend{,} \\
S^{(2)}_{abc'd'} &\equiv g_{ab} n_{c'} n_{d'} + g_{c'd'} n_a n_b \label{S_2} 
\eqend{,} \\
S^{(3)}_{abc'd'} &\equiv n_a n_b n_{c'} n_{d'} \label{S_3} \eqend{,} \\
S^{(4)}_{abc'd'} &\equiv 4 n_{(a} g_{b)(c'} n_{d')} \label{S_4} \eqend{,} \\
S^{(5)}_{abc'd'} &\equiv 2 g_{a(c'} g_{d')b} \label{S_5} \eqend{,}
\end{equations}
and expand the two-point function~\eqref{full_two_point_function} in this basis, 
obtaining
\begin{equation}
\Delta_{abc'd'}(x,x') = \frac{H^{n-2}}{(4\pi)^\frac{n}{2}} \sum_{k=1}^5 
S^{(k)}_{abc'd'} G^{(k)}(Z)
\end{equation}
with the coefficients
\begin{equations}[S_coeffs]
G^{(1)}(Z) &= F^{(1)}(Z) \eqend{,} \\
G^{(2)}(Z) &= (1-Z^2) F^{(2)}(Z) \eqend{,} \\
\begin{split}
G^{(3)}(Z) &= (1-Z^2)^2 F^{(3)}(Z) + 2 (1-Z)^2 F^{(5)}(Z) \\
&\quad+ 4 (1-Z) (1-Z^2) F^{(4)}(Z) \eqend{,} \raisetag{0.9\baselineskip}
\end{split} \\
G^{(4)}(Z) &= (1-Z^2) F^{(4)}(Z) + (1-Z) F^{(5)}(Z) \eqend{,} \\
G^{(5)}(Z) &= F^{(5)}(Z) \eqend{.}
\end{equations}

The explicit expressions for the coefficients $G^{(\text{TV},k)}(Z)$ are given 
in Appendix~\ref{appendix:n_4_expressions}. By expanding them for large $Z$, we 
obtain
\begin{widetext}
\begin{equations}[G_k_largeZ]
\begin{split}
G^{(\text{TV},1)}(Z) &= 2 - \frac{4}{3} \left( 1 - \frac{3}{5} \alpha \right) 
\left[ \frac{11}{6} + \frac{1}{Z} + \left( 1 - \frac{1}{3 Z^2}
\right) \ln \left( \frac{1-Z}{2} \right) \right] + \bigo{Z^{-2}} + \bigo{Z^{-3} 
\ln Z} \eqend{,}
\end{split} \\
\begin{split}
G^{(\text{TV},2)}(Z) &= - \frac{2}{3} + \frac{4}{3} \left( 1 - \frac{3}{5} 
\alpha \right) \left[ \frac{5}{6} + \frac{1}{Z} + \left( 1 -
\frac{2}{Z^2} \right) \ln \left( \frac{1-Z}{2} \right) \right] + \bigo{Z^{-2}} + 
\bigo{Z^{-3} \ln Z} \eqend{,}
\end{split} \\
G^{(\text{TV},3)}(Z) &= 4 G^{(\text{TV},4)}(Z) - 4 G^{(\text{TV},2)}(Z) 
\eqend{,} \\
\begin{split}
G^{(\text{TV},4)}(Z) &= \frac{3}{5} \alpha Z - \frac{11}{3} + \frac{100}{27 Z} - 
\left( 1 - \frac{3}{5} \alpha \right) \left[ - \frac{13}{3}
+ \frac{46}{27 Z} + \left( - 2 + \frac{20}{9 Z} - \frac{4}{9 Z^2} \right) \ln 
\left( \frac{1-Z}{2} \right) \right] \\
&\quad+ \bigo{Z^{-2}} + \bigo{Z^{-3} \ln Z} \eqend{,}
\end{split} \\
G^{(\text{TV},5)}(Z) &= - 2 G^{(\text{TV},1)}(Z) - \frac{1}{2} 
G^{(\text{TV},2)}(Z) \eqend{.}
\end{equations}
\end{widetext}
Note that while we can remove the terms which grow logarithmically for large $Z$ 
by taking $\alpha = 5/3$, there will be a divergence linear in $Z$ unless 
$\alpha = 0$. As shown in Refs.~\cite{higuchi_kouris_cqg_2000, 
higuchi_kouris_cqg_2001b}, however, this growth does not contribute to 
gauge-invariant observables in linearized gravity, such as the linearized Weyl
tensor~\cite{kouris_cqg_2001, mora_tsamis_woodard_prd_2012, 
froeb_roura_verdaguer_jcap_2014}.

\section{IR issues}
\label{sec:ir_issues}

Let us summarize our mode-sum construction of the Wightman two-point function of 
gravitons in the family of covariant gauges parametrized by $\alpha$ and $\beta$ 
[see Eq.~\eqref{gb_def}] in the Poincar{\'e} patch of de Sitter space. We 
introduced the Fierz-Pauli mass term [see Eq.~\eqref{L_mass}] as an infrared 
regularization and observed that the two-point function can be expressed as the 
sum of three sectors: the scalar, vector and tensor sectors. We first observed 
that the scalar sector can directly be obtained in the massless limit with the 
assumption that $\beta > 0$. We then found that there is no IR problem in the 
mode-sum construction of the vector and tensor sectors if $\alpha m^2 > n H^2$ 
and $m^2 > 0$, respectively, with the assumption $\alpha > 0$. Then, the vector 
sector was found to be well defined for $0 < \alpha m^2 \leq n H^2$ as well. We 
used this expression of the vector sector, added it to the tensor sector and 
took the limit $m \to 0$. We found that this limit was finite. In this manner, 
we obtained the graviton two-point function, 
Eq.~\eqref{full_two_point_function}, which is well defined in the IR and 
invariant under the symmetry of the background de~Sitter space. In this section, 
we discuss the IR issues arising in the momentum integral of the vector and 
tensor sectors. In particular, we find that the momentum integral for the mixed 
spatial-temporal components of the vector sector is IR-divergent for $0 < \alpha 
m^2 < n H^2$. We then discuss the finiteness of the limit $m \to 0$ for the sum 
of the vector and tensor sectors, which was shown in 
Sec.~\ref{subsec:massless_limit}, in the context of the momentum integration for 
the purely spatial components. We note that those components are IR finite for 
all positive $m^2$ for both vector and tensor sectors.

\subsection{The IR issues for \texorpdfstring{$0 < \alpha m^2 < n H^2$}{0 < 
\textalpha m\texttwosuperior{} < n H\texttwosuperior}}
\label{subsec:small_m}

If $m^2 > 0$, all the components of the tensor sector of the Wightman two-point 
function converge for small momenta and, thus, are IR finite. On the other hand, 
we assumed $\alpha m^2 > n H^2$ with $\alpha > 0$ to guarantee the convergence 
of the momentum integration in the IR for the vector sector. However, the 
derivatives in $h_{ab}^{(\text{V})} = \nabla_a v_b + \nabla_b v_a$ somewhat 
mitigate the IR behavior, and one can readily verify that the momentum integrals 
for most components of $\Delta_{abc'd'}^{(\text{V})}(x,x')$ are IR finite as 
long as $\alpha m^2 > 0$. However, the exceptions are the mixed 
spatial-temporal components with indices $0\mu0\nu'$, for which the momentum 
integral is IR divergent if $0 < \alpha m^2 < n H^2$. We discuss this problem 
here.

We first perform the angular part of the integrals in momentum space in 
Eqs.~\eqref{Delta_VS} and~\eqref{Delta_VV} through the formula
\begin{splitequation}
\label{angular_integrals}
\int f(\abs{\vec{p}}) \mathe^{\mathi \vec{p} \vec{x}} 
\frac{\total^{n-1}p}{(2\pi)^{n-1}} &= \frac{1}{(2\pi)^\frac{n-1}{2} 
r^\frac{n-3}{2}} \\
&\quad\times \int_0^{\infty} f(p) \bessel{\frac{n-3}{2}}(pr) p^\frac{n-1}{2} 
\total p \eqend{,}
\end{splitequation}
where $f(p)$ is any function depending only on the absolute value $p = 
\abs{\vec{p}}$, $\bessel\rho(z)$ is the Bessel function of order $\rho$, and $r 
\equiv \abs{\vec{x}}$. Then the spatial scalar and spatial vector contributions 
to $\Delta_{0\mu0\nu'}^{(\text{V})}(x.x')$ can be written as
\begin{equations}[Delta_V_0i0i_integrated]
\Delta^{(\text{VS},m^2)}_{0\mu0\nu'}(x,x') &= \int_0^\infty 
\tilde{\Delta}^{(\text{VS},m^2)}_{0\mu0\nu'}(x,x';p) \total p \eqend{,} 
\label{Delta_VS_0i0i_integrated} \\
\Delta^{(\text{VV},m^2)}_{0\mu0\nu'}(x,x') &= \int_0^\infty 
\tilde{\Delta}^{(\text{VV},m^2)}_{0\mu0\nu'}(x,x';p) \total p \eqend{,} 
\label{Delta_VV_0i0i_integrated}
\end{equations}
respectively, so that the vector sector of the two-point function is given as
\begin{equation}
\label{Delta_V_integrated_AH}
\Delta_{0\mu0\nu'}^{(\text{V},m^2)}(x,x') = \int_0^\infty 
\tilde{\Delta}^{(\text{V},m^2)}_{0\mu0\nu'}(x,x';p) \total p \eqend{,}
\end{equation}
where
\begin{splitequation}
\tilde{\Delta}^{(\text{V},m^2)}_{0\mu0\nu'}(x,x';p) &\equiv 
N^{(\text{S})}_\alpha \tilde{\Delta}^{(\text{VS},m^2)}_{0\mu0\nu'}(x,x';p) \\
&\quad+ N^{(\text{V})} \tilde{\Delta}^{(\text{VV},m^2)}_{0\mu0\nu'}(x,x';p) 
\eqend{.}
\end{splitequation}

We use the expansion for small arguments of the Hankel and Bessel 
functions~\cite{dlmf} to approximate the integrands of 
Eqs.~\eqref{Delta_V_0i0i_integrated} for small $p$ as
\begin{splitequation}
\label{Delta_tilde_VS_0i0i_expanded}
&\tilde{\Delta}^{(\text{VS},m^2)}_{0\mu0\nu'}(x,x';p) = - \frac{1}{2} \left( 
\frac{n+1}{2} - \mu_\text{V} - \frac{\alpha m^2}{2 H^2} \right) \\
&\quad\times \bigg\{ \left( \frac{n+1}{2} - \mu_\text{V} - \frac{\alpha m^2}{2 
H^2} \right) \left[ 1 + \frac{(p\eta)^2 + (p\eta')^2}{4 (\mu_\text{V} - 1)} 
\right] \\
&\quad\qquad- (2 \mu_\text{V} - 3) \frac{(p\eta)^2 + (p\eta')^2}{2 (\mu_\text{V} 
- 1)} + \bigo{p^3} \bigg\} \\
&\quad\times K_{\mu\nu'}^{(\text{S})}(\eta,\eta',pr) p^{n-2-2\mu_\text{V}}
\end{splitequation}
and
\begin{splitequation}
\label{Delta_tilde_VV_0i0i_expanded}
&\tilde{\Delta}^{(\text{VV},m^2)}_{0\mu0\nu'}(x,x';p) = - \frac{1}{2} \left( 
\frac{n+1}{2} - \mu_\text{V} \right) \bigg\{ \left( \frac{n+1}{2} - \mu_\text{V} 
\right) \\
&\quad\times \left[ 1 + \frac{(p\eta)^2 + (p\eta')^2}{4 (\mu_\text{V} - 1)} 
\right] + \frac{(p\eta)^2 + (p\eta')^2}{2 (\mu_\text{V} - 1)} + \bigo{p^3} 
\bigg\} \\
&\quad\times K_{\mu\nu'}^{(\text{V})}(\eta,\eta',pr) p^{n-2-2\mu_\text{V}} 
\eqend{,}
\end{splitequation}
where the parameter $\mu_\text{V}$ is given by Eq.~\eqref{mu_V}, and we have 
defined the spatial bivectors
\begin{splitequation}
&K_{\mu\nu'}^{(\textrm{S})}(\eta,\eta',pr) \equiv 
\frac{H^{1-n}}{(4\pi)^\frac{1}{2}} \frac{\Gamma^2( \mu_\text{V} )}{\Gamma\left( 
\frac{n+3}{2} \right)} (H^2\eta\eta')^\frac{n-5}{2} \left( \frac{4}{\eta\eta'} 
\right)^{\mu_\text{V}} \\
&\quad\times \left\{ \left[ (n+1) - \frac{1}{2}(pr)^2 \right] \eta_{\mu\nu'} + 
(pr)^2 \frac{r_\mu r_{\nu'}}{\vec{r}^2} \right\} \raisetag{1.3\baselineskip}
\end{splitequation}
and
\begin{splitequation}
&K_{\mu\nu'}^{(\textrm{V})}(\eta,\eta',pr) \equiv 
\frac{H^{1-n}}{(4\pi)^\frac{1}{2}} \frac{\Gamma^2( \mu_\text{V} )}{\Gamma\left( 
\frac{n+3}{2} \right)} (H^2\eta\eta')^\frac{n-5}{2} \left( \frac{4}{\eta\eta'} 
\right)^{\mu_\text{V}} \\
&\quad\times \left\{ \left[ (n-2)(n+1) - \frac{n}{2} (pr)^2 \right] 
\eta_{\mu\nu'} + (pr)^2 \frac{r_\mu r_{\nu'}}{\vec{r}^2} \right\} \eqend{.}
\end{splitequation}

As $\alpha m^2 - nH^2 \to 0^+$, the power of $p$ in 
Eqs.~\eqref{Delta_tilde_VS_0i0i_expanded} 
and~\eqref{Delta_tilde_VV_0i0i_expanded} tends to $-1$. This means that the $p$ 
integral diverges like $(\alpha m^2 - n H^2)^{-1}$ in this limit. 
However, $\tilde{\Delta}_{0\mu 0\nu'}^{(\textrm{V},m^2)}(x,x';p)$ behaves like 
$\alpha m^2 - n H^2$ in this limit, thus canceling this pole. This agrees with 
the fact that the vector sector given by Eqs.~\eqref{vector_sector_tp_AH} 
and~\eqref{F_V} has no singularity at $\alpha m^2 = n H^2$. Nevertheless, the 
$p$ integral in Eq.~\eqref{Delta_V_integrated_AH} is divergent for $0 < \alpha 
m^2 < n H^2$ because the power of $p$ in 
Eqs.~\eqref{Delta_tilde_VS_0i0i_expanded} 
and~\eqref{Delta_tilde_VV_0i0i_expanded} is less than $-1$. In particular, for 
small $m^2$, to lowest order in $p$ we find from 
Eqs.~\eqref{Delta_tilde_VS_0i0i_expanded} 
and~\eqref{Delta_tilde_VV_0i0i_expanded} that both expressions behave as
\begin{equation}
p^{n-2-2\mu_\text{V}} = p^{-3 + 2 \frac{\alpha m^2}{(n+1) H^2} + \bigo{m^4}} 
\eqend{.}
\end{equation}
Thus, the $p$ integral in Eq.~\eqref{Delta_V_integrated_AH} has a power-law 
divergence in the IR for $0 < \alpha m^2 < n H^2$. It is interesting to note 
that the coefficient multiplying the leading divergence is proportional to 
$\alpha^2 m^4$. Thus, if one takes the $m \to 0$ limit before the $p$ 
integration, or equivalently, if one regularizes the $p$ integration, takes the 
$m \to 0$ limit and then removes the IR regularization for the $p$ integral, the 
leading power-law IR singularity we encounter here will vanish. However, as we 
shall see in the next subsection, this strategy will not yield a finite result 
in the total two-point function. As for the next-to-leading order terms in 
Eqs.~\eqref{Delta_tilde_VV_0i0i_expanded} 
and~\eqref{Delta_tilde_VS_0i0i_expanded}, they both converge 
for $m^2 > 0$ but diverge as $m \to 0$. Nevertheless, these terms cancel each 
other, rendering the final expression for 
$\Delta^{(\text{V},m^2)}_{0\mu0\nu'}(x,x')$ finite for $m = 0$ except for the 
leading divergence mentioned above. It is unclear how to justify discarding 
this divergence and using the expression of the vector sector, 
Eq.~\eqref{vector_sector_two_point_function_m}, analytically continued in $m^2$ 
from $\alpha m^2 > n H^2$ to study the $m \to 0$ limit, but we conjecture that 
the subtraction of this divergence will not affect physical observables since 
the vector sector of the free graviton field comprises part of the gauge sector 
of the theory. We also conjecture that one will obtain the result of 
Sec.~\ref{subsec:vector_sector} for all $\alpha m^2 > 0$ without encountering IR 
divergences if the vector sector is constructed by the mode-sum method in global 
de~Sitter space, because the mode sum in global de~Sitter space is discrete.

\subsection{The \texorpdfstring{$m \to 0$}{m \unichar{"2192} 0} limit}
\label{subsec:m_=_0}

In the previous subsection, we found that the $p$ integrals in our graviton 
two-point function are IR finite for $m^2 > 0$ if the IR divergence of the 
$0\mu0\nu'$ component of the vector sector for $0 < \alpha m^2 < n H^2$ is dealt 
with by analytic continuation in $m^2$. Then, the $m \to 0$ limit of our 
two-point function is finite because of cancellation of terms behaving like 
$1/m^2$ between the vector and tensor sectors, and is given by 
Eqs.~\eqref{full_two_point_function} and~\eqref{F_k}. In this subsection we 
investigate this cancellation by analyzing the $p$ integration closely for the 
purely spatial components. We note that these components of the vector sector as 
well as the tensor sector are IR finite for $m^2 > 0$ (with $\alpha > 0$). We 
will show how the well-known IR divergences of the two-point function in the 
transverse-traceless-synchronous gauge~\cite{ford_parker_prd_1977b}, which 
contains only physical degrees of freedom, are canceled by gauge contributions. 
We will also find that the $m \to 0$ limit and the $p$ integral do not commute 
and that, if the $m \to 0$ limit is taken before the $p$ integration, our 
two-point function will be IR-divergent unless $\alpha$ takes a particular 
value.

We first perform the angular integrals in the expressions for their spatial 
tensor, spatial vector, and spatial scalar parts, 
Eqs.~\eqref{Delta_TT}--\eqref{Delta_TS}, respectively, with the aid of 
Eq.~\eqref{angular_integrals} and express 
them as follows:
\begin{equations}[Delta_T_ijij_integrated]
\Delta^{(\text{TT},m^2)}_{\mu\nu\rho'\sigma'}(x,x') &= \int_0^\infty 
\tilde{\Delta}^{(\text{TT},m^2)}_{\mu\nu\rho'\sigma'}(x,x';p) \total p \eqend{,} 
\label{Delta_TT_ijij_integrated} \\
\Delta^{(\text{TV},m^2)}_{\mu\nu\rho'\sigma'}(x,x') &= \int_0^\infty 
\tilde{\Delta}^{(\text{TV},m^2)}_{\mu\nu\rho'\sigma'}(x,x';p) \total p \eqend{,} 
\label{Delta_TV_ijij_integrated} \\
\Delta^{(\text{TS},m^2)}_{\mu\nu\rho'\sigma'}(x,x') &= \int_0^\infty 
\tilde{\Delta}^{(\text{TS},m^2)}_{\mu\nu\rho'\sigma'}(x,x';p) \total p \eqend{.} 
\label{Delta_TS_ijij_integrated}
\end{equations}
We note that the $m \to 0$ limit of 
$\Delta^{(\text{TT},m^2)}_{\mu\nu\rho'\sigma'}(x,x')$ is the two-point function 
in the transverse-traceless-synchronous gauge whereas the contributions 
$\Delta^{(\text{TV},m^2)}_{\mu\nu\rho'\sigma'}(x,x')
$ and $\Delta^{(\text{TS},m^2)}_{\mu\nu\rho'\sigma'}(x,x')$ are of pure-gauge 
form in this limit. The behavior for small $m^2$ and $p$ of the integrands in 
Eqs.~\eqref{Delta_T_ijij_integrated} is given by
\begin{splitequation}
\label{Delta_tilde_TT_ijij_small_m}
&\tilde{\Delta}^{(\text{TT},m^2)}_{\mu\nu\rho'\sigma'}(x,x';p) = - 
\frac{n(n-3)}{n-2} \frac{m^2}{H^2} \\
&\quad\times K_{\mu\nu\rho'\sigma'}^{(\mu_\text{T})}(\eta,\eta') p^{-1+\frac{2 
m^2}{(n-1) H^2}} \left[ 1 + \bigo{p^2} \right] \eqend{,}
\end{splitequation}
\begin{splitequation}
\label{Delta_tilde_TV_ijij_small_m}
&\tilde{\Delta}^{(\text{TV},m^2)}_{\mu\nu\rho'\sigma'}(x,x';p) = - 2 \left( 
\frac{n-1}{2} + \mu_\text{T} \right)^2 \\
&\quad\times K_{\mu\nu\rho'\sigma'}^{(\mu_\text{T})}(\eta,\eta') p^{-1+\frac{2 
m^2}{(n-1) H^2}} \left[ 1 + \bigo{p^2} \right] \eqend{,}
\end{splitequation}
and
\begin{splitequation}
\label{Delta_tilde_TS_ijij_small_m}
&\tilde{\Delta}^{(\text{TS},m^2)}_{\mu\nu\rho'\sigma'}(x,x';p) = - \left[ 
\frac{n-1}{2} + \mu_\text{T} - \frac{m^2}{(n-2) H^2} \right]^2 \\
&\quad\times K_{\mu\nu\rho'\sigma'}^{(\mu_\text{T})}(\eta,\eta') p^{-1+\frac{2 
m^2}{(n-1)H^2}} \left[ 1 + \bigo{p^2} \right] \eqend{,}
\end{splitequation}
where $\mu_\text{T}$ is defined by Eq.~\eqref{mu_T}, and we have defined the 
following traceless spatial bitensor:
\begin{splitequation}
\label{K_ijij}
K_{\mu\nu\rho'\sigma}^{(\alpha)}(\eta,\eta') &\equiv 
\frac{H^{1-n}}{(4\pi)^\frac{1}{2}} \frac{\Gamma^2(\alpha)}{\Gamma\left( 
\frac{n+3}{2} \right)} (H^2\eta\eta')^\frac{n-5}{2} \left( \frac{4}{\eta\eta'} 
\right)^\alpha \\
&\quad \times \left[ \eta_{\mu\nu} \eta_{\rho'\sigma'} - (n-1) \eta_{\mu(\rho'} 
\eta_{\sigma')\nu} \right] \eqend{.}
\end{splitequation}
The purely spatial components of the tensor sector are thus obtained by 
integrating the following tensor over $p$:
\begin{splitequation}
\tilde{\Delta}^{(\text{T},m^2)}_{\mu\nu\rho'\sigma'}(x,x';p) &\equiv 
N^{(\text{V})} \tilde{\Delta}^{(\text{TT},m^2)}_{\mu\nu\rho'\sigma'}(x,x';p) \\
&\quad+ N^{(\text{V})} 
\tilde{\Delta}^{(\text{TV},m^2)}_{\mu\nu\rho'\sigma'}(x,x';p) \\
&\quad+ N^{(\text{S})} 
\tilde{\Delta}^{(\text{TS},m^2)}_{\mu\nu\rho'\sigma'}(x,x';p) \eqend{,}
\end{splitequation}
where $N^{(\textrm{V})}$ and $N^{(\textrm{S})}$, which are proportional to 
$H^2/m^2$, are given by Eqs.~\eqref{N_V} and~\eqref{N_S}, respectively. The 
contribution from the spatial tensor part, which corresponds to the two-point 
function 
in the transverse-traceless-synchronous gauge, is
\begin{splitequation}
\label{Delta_tilde_TT_ijij_small_m_AH}
N^{(V)} \tilde{\Delta}^{(\text{TT},m^2)}_{\mu\nu\rho'\sigma'}(x,x';p) &= - 
\frac{n(n-3)}{n-2} \frac{H^{n-2}}{(4\pi)^{\frac{n}{2}}} 
K_{\mu\nu\rho'\sigma'}^{(\frac{n-1}{2})}(\eta,\eta') \\
&\quad\times p^{-1+\frac{2 m^2}{(n-1) H^2}} \left[ 1 + \bigo{p^2} \right] 
\eqend{,}
\end{splitequation}
where the terms of order $m^2$ have been neglected except in $p^{-1+\frac{2 
m^2}{(n-1) H^2}}$. By adding the spatial vector and spatial scalar 
contributions, both of which become pure gauge in the limit $m \to 0$, we find
\begin{splitequation}
\label{Delta_tilde_T_ijij_small_m}
\tilde{\Delta}^{(\text{T},m^2)}_{\mu\nu\rho'\sigma'}(x,x';p) &= (n-3) 
\frac{H^{n-2}}{(4\pi)^{\frac{n}{2}}} 
K_{\mu\nu\rho'\sigma'}^{(\frac{n-1}{2})}(\eta,\eta') \\
&\quad\times p^{-1+\frac{2 m^2}{(n-1) H^2}} \left[ 1 + \bigo{p^2} \right] 
\eqend{.}
\end{splitequation}
Thus, in the tensor sector the gauge contribution overcompensates the one from 
the transverse-traceless-synchronous modes, and changes the sign of the IR 
divergence.

The purely spatial components of the vector sector of the graviton two-point 
function can be expressed as a $p$ integral in the same fashion as for the 
tensor sector. By combining Eqs.~\eqref{Delta_VS} and~\eqref{Delta_VV} with 
Eq.~\eqref{angular_integrals}, one can express 
$\Delta_{\mu\nu\rho'\sigma'}^{(\text{VS},m^2)}$ and 
$\Delta_{\mu\nu\rho'\sigma'}^{(\text{VV},m^2)}$ in the following form:
\begin{equations}[Delta_V_ijij_integrated]
\Delta^{(\text{VS},m^2)}_{\mu\nu\rho'\sigma'}(x,x') &= \int_0^\infty 
\tilde{\Delta}^{(\text{VS},m^2)}_{\mu\nu\rho'\sigma'}(x,x';p) \total p \eqend{,} 
\label{Delta_VS_ijij_integrated} \\
\Delta^{(\text{VV},m^2)}_{\mu\nu\rho'\sigma'}(x,x') &= \int_0^\infty 
\tilde{\Delta}^{(\text{VV},m^2)}_{\mu\nu\rho'\sigma'}(x,x';p) \total p \eqend{.}
\label{Delta_VV_ijij_integrated}
\end{equations}
If $m^2$ and $p$ are small, the integrands in 
Eqs.~\eqref{Delta_V_ijij_integrated} can be approximated as follows:
\begin{splitequation}
\label{Delta_tilde_VS_ijij_small_m}
&\tilde{\Delta}^{(\text{VS},m^2)}_{\mu\nu\rho'\sigma'}(x,x';p) = (n-1) \left[ 1 
- \frac{1}{(n-1)^2} \frac{\alpha m^2}{H^2} \right] \eta \eta' \\
&\quad\times K_{\mu\nu\rho'\sigma'}^{(\mu_\text{V})}(\eta,\eta') p^{-1+\frac{2 
\alpha m^2}{(n+1) H^2}} \left[ 1 + \bigo{p^2} \right]
\end{splitequation}
and
\begin{splitequation}
\label{Delta_tilde_VV_ijij_small_m}
&\tilde{\Delta}^{(\text{VV},m^2)}_{\mu\nu\rho'\sigma'}(x,x';p) = 2 \eta \eta' \\
&\quad\times K_{\mu\nu\rho'\sigma'}^{(\mu_\text{V})}(\eta,\eta') p^{-1+\frac{2 
\alpha m^2}{(n+1) H^2}} \left[ 1 + \bigo{p^2} \right] \eqend{.}
\end{splitequation}
The purely spatial components of the vector sector are obtained by integrating 
over $p$ the tensor
\begin{splitequation}
\tilde{\Delta}^{(\text{V},m^2)}_{\mu\nu\rho'\sigma'}(x,x';p) &\equiv 
N^{(\text{S})}_\alpha 
\tilde{\Delta}^{(\text{VS},m^2)}_{\mu\nu\rho'\sigma'}(x,x';p) \\
&\quad+ N^{(\text{V})} 
\tilde{\Delta}^{(\text{VV},m^2)}_{\mu\nu\rho'\sigma'}(x,x';p) \eqend{,}
\end{splitequation}
which for small $m^2$ and $p$ reads
\begin{splitequation}
\label{Delta_tilde_V_ijij_small_m}
&\tilde{\Delta}^{(\text{V},m^2)}_{\mu\nu\rho'\sigma'}(x,x';p) = - \alpha 
\frac{n-1}{n+1} (n-3) \frac{H^{n-2}}{(4\pi)^\frac{n}{2}} \\
&\quad\times K_{\mu\nu\rho'\sigma'}^{(\frac{n-1}{2})}(\eta,\eta') p^{-1+\frac{2 
\alpha m^2}{(n+1) H^2}} \left[ 1 + \bigo{p^2} \right] \eqend{.}
\end{splitequation}

The $p$ dependence of 
$\tilde{\Delta}_{\mu\nu\rho'\sigma'}^{(\text{T},m^2)}(x,x';p)$ given by 
Eq.~\eqref{Delta_tilde_T_ijij_small_m} and the one of 
$\tilde{\Delta}_{\mu\nu\rho'\sigma'}^{(\text{V},m^2)}(x,x';p)$ given by 
Eq.~\eqref{Delta_tilde_V_ijij_small_m} are of the form $p^{-1 + c m^2}$, where 
$c$ is a positive constant. Each of these give an IR-divergent contribution 
proportional to $1/(c m^2)$ when integrated over $p$. These IR contributions, 
however, cancel exactly for any $\alpha > 0$ when they are added together, in 
agreement with our conclusions in Sec.~\ref{subsec:massless_limit}: the total 
contribution diverging like $1/m^2$ is proportional to
\begin{equation}
\label{cancellation_vector_tensor}
(n-3) \cdot \frac{(n-1) H^2}{2 m^2} - \alpha \frac{n-1}{n+1} (n-3) \cdot 
\frac{(n+1) H^2}{2 \alpha m^2} = 0 \eqend{.}
\end{equation}
It is interesting to note that the $m \to 0$ limit and the $p$ integration do 
not commute. If one takes the $m \to 0$ limit first, then the cancellation shown 
in Eq.~\eqref{cancellation_vector_tensor} does not work for all $\alpha$ 
because 
there will be no distinction between the small-$p$ behavior of the form 
$p^{-1+\frac{2 m^2}{(n-1) H^2}}$ (for the tensor sector) and that of the form 
$p^{-1+\frac{2 \alpha m^2}{(n+1) H^2}}$ (for the vector sector) if the limit $m 
\to 0$ is 
taken before the $p$ integration. In this case the IR-divergent contribution to 
the two-point function with an IR momentum cutoff at $p = \lambda$ will be given 
by
\begin{splitequation}
\label{log-IR-div}
\Delta_{\mu\nu\rho'\sigma'}(x,x') &\approx \frac{H^{n-2}}{(4\pi)^\frac{n}{2}} 
K_{\mu\nu\rho'\sigma'}^{(\frac{n-1}{2})}(\eta,\eta')\\
&\times (n-3) \left( 1 - \alpha \frac{n-1}{n+1} \right) \ln(1/\lambda) \eqend{.}
\end{splitequation}
Thus, if the $m \to 0$ limit is taken before the $p$ integration, our two-point 
function is IR-divergent unless $\alpha = (n+1)/(n-1)$. (Recall also that the IR 
divergence in the vector sector for $0 < \alpha m^2 < n H^2$ will be absent if 
the $m \to 0$ limit is taken before the $p$ integration.) Especially, it will be 
IR-divergent for $\alpha = 0$, which corresponds to the Landau/exact gauge used 
in Refs.~\cite{miao_tsamis_woodard_jmp_2011, khaya_miao_woodard_jmp_2012, 
mora_tsamis_woodard_jmp_2012}, where the two-point function constructed using a 
different method was found to be IR-divergent in the cutoff regularization. 
However, the coefficient for the IR divergence we find is different from the one 
found 
in these works. In four dimensions, for example, the method of 
Ref.~\cite{mora_tsamis_woodard_jmp_2012} gives (see 
Ref.~\cite{miao_mora_tsamis_woodard_2014})
\begin{equation}
\Delta^{\text{MTW}}_{\mu\nu\rho'\sigma'}(x,x') \approx - 5 \frac{H^2}{(4\pi)^2} 
K_{\mu\nu\rho'\sigma'}^{(\frac{3}{2})}(\eta,\eta') \ln(1/\lambda) \eqend{,}
\end{equation}
whereas the coefficient for the IR divergence for our two-point function will be 
$1$ instead of $-5$ if the $m \to 0$ limit is taken before the $p$ integration.


\section{Summary and Discussion}
\label{sec:discussion}

We have, for the first time, derived the graviton two-point function by an 
explicit mode-sum construction in the Poincar{\'e} patch of de~Sitter space, for 
general spacetime dimension $n$, in a general linear covariant gauge with 
parameters $\alpha$ and $\beta$ [see Eqs.~\eqref{L_gf} and~\eqref{gb_def}]. We 
added a Fierz-Pauli mass term proportional to $m^2$ to control IR divergences 
and found that the two-point function can be expressed as the sum of the 
scalar, vector and tensor sectors. Since there is no IR problem in the scalar 
sector if $\beta > 0$~\cite{higuchi_kouris_cqg_2001b, 
mora_tsamis_woodard_jmp_2012}, we made this assumption and found the scalar 
sector directly with $m = 0$. As for the vector sector, we found that all 
relevant momentum integrals are convergent only for $\alpha m^2 > nH^2$. We 
performed the mode-sum construction with this assumption and found that the 
resulting vector sector of the two-point function is well defined for any value 
of $m^2 > 0$ (with $\alpha > 0$) if this sector is analytically continued in 
$m^2$. The tensor sector, which contains the contribution from the 
transverse-traceless-synchronous modes, was found to be IR finite 
as long as $m^2 > 0$. We then added together the tensor sector and the 
(analytically-continued) vector sector and found that the $m \to 0$ limit, i.e., 
the limit in which the IR regulator is removed, is finite. In this manner, we 
obtained a de~Sitter invariant two-parameter family of graviton two-point 
function in the Poincar{\'e} patch of de~Sitter space by the mode-sum method. We 
also verified that our result in four dimensions agrees with the two-point 
function obtained by analytic continuation from the 
sphere~\cite{higuchi_kouris_cqg_2001a, higuchi_kouris_cqg_2001b} (see also 
Ref.~\cite{morrison_arxiv_2013}). We verified that the resulting two-point 
function grows for large (time- or spacelike) separation but noted that this 
growth does not affect local gauge-invariant observables in linearized gravity 
as was shown in Ref.~\cite{higuchi_kouris_cqg_2001b}.

We also considered the procedure of setting $m = 0$ from the start and 
regularizing the IR divergences by an IR momentum cutoff as is usually done in 
the literature. We found that the graviton two-point function will be 
IR-divergent for a general value of $\alpha$ because the IR divergences do not 
cancel between the vector and tensor sector in this case. This conclusion is in 
agreement with the results of Refs.~\cite{miao_tsamis_woodard_jmp_2011, 
khaya_miao_woodard_jmp_2012, mora_tsamis_woodard_jmp_2012} obtained for the 
gauge parameter $\alpha = 0$ (the Landau or exact gauge) although there is 
disagreement in the exact value of the IR-divergent term. However, since the IR 
divergence of the vector sector is proportional to $\alpha$, there is one 
special value of $\alpha$ where these divergences cancel exactly, namely $\alpha 
= (n+1)/(n-1)$ in $n$ dimensions. For this value of $\alpha$, one encounters no 
IR problems at all. (We note, however, that some authors object to the gauge 
with nonzero $\alpha$~\cite{miao_tsamis_woodard_jmp_2009}.) If $\alpha$ is 
greater than this value, the IR divergences coming from the vector sector are 
not large enough to cancel the ones coming from the tensor sector, while for 
smaller $\alpha$ they are too large. In particular, for the Landau gauge $\alpha 
= 0$ the IR-divergent contribution coming from the vector sector and the spatial 
vector and spatial scalar parts of the tensor sector were found to 
overcompensate the ``physical'' IR divergence coming from the spatial 
transverse-traceless tensor part of the tensor sector and changes the overall 
sign of the IR divergence. 

In Ref.~\cite{higuchi_marolf_morrison_cqg_2011}, it was explicitly shown that 
these divergences are of gauge form, and can be transformed away by a ``large'' 
gauge transformation, at least in noninteracting linearized gravity (see also 
Refs.~\cite{allen_npb_1987,higuchi_kouris_cqg_2000}). If one considers these IR 
divergences to be unphysical and expects them not to contribute to correlation 
functions of gauge-invariant observables, the change of sign in the IR 
divergence does not matter (and in fact, one can then use the 
de~Sitter--invariant two-point function from the outset). On the other hand, if 
one is skeptical of large gauge transformations and attributes physical reality 
to the IR divergences, this change of sign appears to cause difficulties 
because, then, one would obtain different answers depending on whether one uses 
the covariant two-point function with a gauge-fixing term or that corresponding 
to just the physical, spatial transverse-traceless metric perturbations.

Although it was shown in Ref.~\cite{higuchi_arxiv_2012} that, as far as the 
computation of the correlation functions corresponding to compactly supported 
gauge-invariant quantities goes, the covariant graviton two-point function we 
have derived is physically equivalent at tree level to the de~Sitter-breaking 
ones, the same result may not be true for noncompactly supported observables. 
Especially, we find it plausible that a naive calculation of observables whose 
support extends to spatial infinity can give different results using our 
two-point function versus the de~Sitter--breaking two-point functions of, e.g., 
Refs.~\cite{miao_tsamis_woodard_jmp_2011, khaya_miao_woodard_jmp_2012, 
mora_tsamis_woodard_jmp_2012}. Such observables must be defined carefully 
(e.g., by taking their support to be contained in a sphere of radius 
$\mathsf{R}$ and showing that the limit $\mathsf{R} \to \infty$ is 
well defined), and while studies of this sort have been undertaken in both 
anti--de~Sitter and Minkowski spacetimes (see, e.g., 
Refs.~\cite{hollands_ishibashi_marolf_cqg_2005a, 
hollands_ishibashi_marolf_prd_2005b,barnich_trossaert_jhep_2011, 
strominger_jhep_2014, he_lysov_mitra_strominger_jhep_2015}), leading to the 
discovery of asymptotic symmetries and conserved charges at infinity, we are not 
aware of a similar analysis for the de Sitter case.

\acknowledgments

M.~F.\ acknowledges financial support through ERC Starting Grant QC\&C 259562. 
W.~L.\ acknowledges financial support from Conselho Nacional de Desenvolvimento 
Cient{\'\i}fico e Tecnol{\'o}gico --- Brazil (CNPq) under Grant 
No.~201270/2014-5.

\appendix
\section{Formulas for \texorpdfstring{$n = 4$}{n = 4}}
\label{appendix:n_4_expressions}

Here we present the explicit form of the coefficient functions in the massless 
two-point function~\eqref{full_two_point_function} in dimension $n = 4$.

The functions $I^{(k)}_{(n+1)/2}$ with $k \geq 2$ present in the tensor-sector 
coefficients $F^{(\text{T},\text{div},k)}(Z)$, Eq.~\eqref{F_T_div}, can be 
directly evaluated for $n = 4$ from the definition~\eqref{I_k}. For the 
coefficients 
$F^{(\text{T},\text{fin},k)}(Z)$, given by Eq.~\eqref{F_T_fin}, we also need the 
functions $I^{(k)}_{(n-1)/2}$ with $k \geq 1$, which can also be evaluated 
directly from the definition~\eqref{I_k}, the function 
$\varPi^{(0)}_{(n-1)/2}(Z)$, 
which can be calculated from its definition~\eqref{ii_tensor_def}, and 
the functions $\tilde{I}^{(k)}_{(n-1)/2}$ with $k \geq 1$, which were defined by 
Eq.~\eqref{I_tilde_k}. For the evaluation of $\tilde{I}^{(k)}_{(n-1)/2}$ for $n 
= 4$, we first write the relevant hypergeometric function as a series in four 
dimensions and then differentiate term by term since the series is absolutely 
convergent. This results in
\begin{widetext}
\begin{splitequation}
\label{I_tilde_k_4D_tensor}
\tilde{I}^{(1)}_{3/2}(Z) &= - \frac{1}{6} \lim_{\mu \to 3/2} 
\frac{\partial}{\partial \mu} \sum_{\ell = 0}^\infty \frac{\Gamma\left( 
\frac{5}{2} + \mu + \ell \right) \Gamma\left( \frac{5}{2} - \mu + \ell 
\right)}{\Gamma\left( 3 + \ell \right) \ell!} \left( \frac{1+Z}{2} \right)^\ell 
\\
&= - \frac{1}{6} \sum_{\ell = 0}^\infty \frac{3 \ell^2 + 12 \ell + 11}{(\ell+1) 
(\ell+2)} \left( \frac{1+Z}{2} \right)^\ell = - \frac{2 (1+2Z)}{3 (1-Z) (1+Z)} + 
\frac{2 (2+Z)}{3 (1+Z)^2} \ln \left( \frac{1-Z}{2} \right) \eqend{,}
\end{splitequation}
\end{widetext}
and the $\tilde{I}^{(k)}_{3/2}(Z)$ with $k > 1$ can be obtained by simple 
differentiation of this expression. For the vector-sector coefficients 
$F^{(\text{V},\text{fin},k)}(Z)$, Eq.~\eqref{F_V_fin}, we need the functions 
$I^{(k)}_{(n+1)/2}$ with $k \geq 2$, which can be evaluated directly from the 
definition~\eqref{I_k}, the function $\varPi^{(1)}_{(n+1)/2}(Z)$, which can also 
be calculated from its definition~\eqref{ii_vector_def}, and the functions 
$\tilde{I}^{(k)}_{(n+1)/
2}$ with $k \geq 2$, for which a similar calculation as in 
Eq.~\eqref{I_tilde_k_4D_tensor} shows
\begin{splitequation}
\label{I_tilde_k_4D_vector}
\tilde{I}^{(2)}_{5/2}(Z) &= \frac{2 (8Z^3+3Z^2-15Z-6)}{5 (1-Z)^2 (1+Z)^2} \\
&\quad+ \frac{2 (3Z^2+9Z+8)}{5 (1+Z)^3} \ln \left( \frac{1-Z}{2} \right) 
\eqend{.}
\end{splitequation}
Finally, for the scalar-sector coefficients $F^{(\text{S},k)}(Z)$ given by 
Eq.~\eqref{F_S} the four-dimensional limit is straightforward for general $\beta 
> 0$. However, the limit $\beta \to 0$ exists for the coefficients as we have
\begin{equations}
I^{(k)}_\text{S}(Z) &\to I^{(k)}_{3/2}(Z) \eqend{,} \\
\tilde{I}^{(k)}_\text{S}(Z) &\to \tilde{I}^{(k)}_{3/2}(Z)
\end{equations}
for $k \geq 1$. Simple expressions are also found for the case $\beta = 2/3$, 
where we obtain by a similar reasoning as above
\begin{equations}
I^{(1)}_\text{S}(Z) &= \frac{2}{(1-Z)^2} \eqend{,} \\
\tilde{I}^{(1)}_\text{S}(Z) &= - \frac{2}{(1-Z)^2} - \frac{2}{(1+Z)^2} \ln 
\left( \frac{1-Z}{2} \right) \eqend{,}
\end{equations}
and for $k > 1$ we again simply take derivatives. The greatest simplification 
occurs in the limit $\beta \to \infty$, for which it was shown in 
Ref.~\cite{froeb_higuchi_jmp_2014} that $I^{(k)}_\text{S}(Z)$ tends to zero 
exponentially fast, 
and the same is seen to be true for $\tilde{I}^{(k)}_\text{S}(Z)$.

\begin{widetext}
By combining the calculations described above, in four dimensions the 
scalar-sector coefficients $F^{(\text{S},k)}$ are given for $\beta = 0$ by
\begin{equations}[F_S_4D_beta_0]
F^{(\text{S},1)} &= - \frac{4 Z (5Z^4-5Z^3-24Z^2+7Z+5)}{9 (1-Z)^3 (1+Z)^2} - 
\alpha \frac{4 Z}{9 (1-Z)^3} + \frac{16 Z}{9 (1+Z)^3} \ln \left( \frac{1-Z}{2} 
\right) \eqend{,} \\
F^{(\text{S},2)} &= \frac{2 (7Z^5-7Z^4-104Z^3+8Z^2-71Z+23)}{9 (1-Z)^4 (1+Z)^3} + 
\alpha \frac{4}{3 (1-Z)^4} + \frac{16}{3 (1+Z)^4} \ln \left( \frac{1-Z}{2} 
\right) \eqend{,} \\
F^{(\text{S},3)} &= \frac{4 (7Z^5-19Z^4-386Z^3+98Z^2-293Z+17)}{9 (1-Z)^5 
(1+Z)^4} - \alpha \frac{4 (5-Z)}{3 (1-Z)^5} - \frac{16 (5+Z)}{3 (1+Z)^5} \ln 
\left( \frac{1-Z}{2} \right) \eqend{,} \\
F^{(\text{S},4)} &= - \frac{4 (Z^4-5Z^3-69Z^2+17Z-16)}{9 (1-Z)^4 (1+Z)^3} - 
\alpha \frac{4 (4-Z)}{9 (1-Z)^4} + \frac{16 (4+Z)}{9 (1+Z)^4} \ln \left( 
\frac{1-Z}{2} \right) \eqend{,} \\
F^{(\text{S},5)} &= \frac{2 (Z^3+3Z^2+27Z-7)}{9 (1-Z)^3 (1+Z)^2} - \alpha 
\frac{2 (3-Z)}{9 (1-Z)^3} - \frac{8 (3+Z)}{9 (1+Z)^3} \ln \left( \frac{1-Z}{2} 
\right) \eqend{,}
\end{equations}
and for $\beta = 2/3$ by
\begin{equations}[F_S_4D_beta_23]
F^{(\text{S},1)} &= - \frac{8 Z}{3 (1-Z)^3} + (9-\alpha) \left[ \frac{2 Z 
(5Z^3-3Z^2+7Z-1)}{9 (1-Z)^3 (1+Z)^2} - \frac{4 (1-Z) Z}{9 (1+Z)^3} \ln \left( 
\frac{1-Z}{2} \right) \right] \eqend{,} \\
F^{(\text{S},2)} &= \frac{4 (5+Z)}{3 (1-Z)^4} - (9-\alpha) \left[ \frac{8 
(3Z^4-3Z^3+13Z^2-3Z+2)}{9 (1-Z)^4 (1+Z)^3} + \frac{8 (2-Z)}{9 (1+Z)^4} \ln 
\left( \frac{1-Z}{2} \right) \right] \eqend{,} \\
F^{(\text{S},3)} &= - \frac{32}{(1-Z)^5} + (9-\alpha) \left[ \frac{8 
(7Z^4-66Z^3+16Z^2-54Z+1)}{9 (1-Z)^5 (1+Z)^4} + \frac{32}{3 (1+Z)^5} \ln \left( 
\frac{1-Z}{2} \right) \right] \eqend{,} \\
F^{(\text{S},4)} &= - \frac{8}{(1-Z)^4} + (9-\alpha) \left[ \frac{8 (-3 + Z - 11 
Z^2 + Z^3)}{9 (1-Z)^4 (1+Z)^3} - \frac{8}{3 (1+Z)^4} \ln \left( \frac{1-Z}{2} 
\right) \right] \eqend{,} \\
F^{(\text{S},5)} &= - \frac{8}{3 (1-Z)^3} - (9-\alpha) \left[ \frac{16 Z}{9 
(1-Z)^3 (1+Z)^2} - \frac{8}{9 (1+Z)^3} \ln \left( \frac{1-Z}{2} \right) \right] 
\eqend{,}
\end{equations}
while they vanish in the limit $\beta \to \infty$. For the vector- and 
tensor-sector coefficients $F^{(\text{TV},k)} \equiv F^{(k)}(Z) - 
F^{(\text{S},k)}(Z)$, we obtain
\begin{equations}[F_VT_4D]
\begin{split}
F^{(\text{TV},1)}(Z) &= - \frac{2 (-27 + 89 Z - 81 Z^2 + 27 Z^3)}{27 (1-Z)^3} + 
(5-3\alpha) \frac{2 Z (-4 + 19 Z - 73 Z^2 - 15 Z^3 + 33 Z^4)}{135 (1-Z)^3 
(1+Z)^2} \\
&\quad- (5-3\alpha) \frac{4 Z (8 + 9 Z + 3 Z^2)}{45 (1+Z)^3} \ln \left( 
\frac{1-Z}{2} \right) \eqend{,}
\end{split} \\
\begin{split}
F^{(\text{TV},2)}(Z) &= \frac{2 (17 - 12 Z + 3 Z^2)}{9 (1-Z)^4} - (5-3\alpha) 
\frac{2 (21 - 79 Z - 14 Z^2 - 14 Z^3 + Z^4 + 5 Z^5)}{45 (1-Z)^4 (1+Z)^3} \\
&\quad- (5-3\alpha) \frac{4 (5 + 4 Z + Z^2)}{15 (1+Z)^4} \ln \left( 
\frac{1-Z}{2} \right) \eqend{,}
\end{split} \\
\begin{split}
F^{(\text{TV},3)}(Z) &= \frac{2 (-121 + 149 Z - 75 Z^2 + 15 Z^3)}{9 (1-Z)^5} + 
(5-3\alpha) \frac{4 (25 + 29 Z + 15 Z^2 + 3 Z^3)}{15 (1+Z)^5} \ln \left( 
\frac{1-Z}{2} \right) \\
&\quad- (5-3\alpha) \frac{2 (-105 + 483 Z + 39 Z^2 + 295 Z^3 + 5 Z^4 - 95 Z^5 - 
3 Z^6 + 21 Z^7)}{45 (1-Z)^5 (1+Z)^4} \eqend{,}
\end{split} \\
\begin{split}
F^{(\text{TV},4)}(Z) &= \frac{8 (-5 + 3 Z) (7 - 7 Z + 3 Z^2)}{27 (1-Z)^4} - 
(5-3\alpha) \frac{2 (70 - 58 Z + 297 Z^2 + 52 Z^3 - 148 Z^4 - 18 Z^5 + 45 
Z^6)}{135 (1-Z)^4 (1+Z)^3} \\
&\quad- (5-3\alpha) \frac{2 (40 + 55 Z + 36 Z^2 + 9 Z^3)}{45 (1+Z)^4} \ln \left( 
\frac{1-Z}{2} \right) \eqend{,}
\end{split} \\
\begin{split}
F^{(\text{TV},5)}(Z) &= \frac{(-159 + 341 Z - 297 Z^2 + 99 Z^3)}{27 (1-Z)^3} - 
(5-3\alpha) \frac{(-63 + 221 Z + 118 Z^2 - 250 Z^3 - 63 Z^4 + 117 Z^5)}{135 
(1-Z)^3 (1+Z)^2} \\
&\quad+ (5-3\alpha) \frac{2 (5 + 3 Z) (3 + 4 Z + 3 Z^2)}{45 (1+Z)^3} \ln \left( 
\frac{1-Z}{2} \right) \eqend{,}
\end{split}
\end{equations}
and we see that the logarithmic terms are absent for $\alpha = 5/3$, which is 
exactly the value of $\alpha$ (in four dimensions) for which the potential IR 
divergences in the momentum integrals cancel if one sets $m = 0$ before the 
momentum 
integration is performed, as explained in Sec.~\ref{subsec:m_=_0}.

In order to compare the results above with those of 
Ref.~\cite{higuchi_kouris_cqg_2001b}, we need to switch to the Allen-Jacobson 
basis~\cite{allen_jacobson_cmp_1986} and use the bitensor set defined by 
Eqs.~\eqref{S_set}. Using the 
relations~\eqref{S_coeffs} to convert the coefficients, we obtain for the 
scalar-sector coefficients $G^{(\text{S},k)}$ for $\beta = 0$
\begin{equations}[G_S_4D_beta_0]
G^{(\text{S},1)}(Z) &= - \frac{4 Z (5Z^4-5Z^3-24Z^2+7Z+5)}{9 (1-Z)^3 (1+Z)^2} - 
\alpha \frac{4 Z}{9 (1-Z)^3} + \frac{16 Z}{9 (1+Z)^3} \ln \left( \frac{1-Z}{2} 
\right) \eqend{,} \\
G^{(\text{S},2)}(Z) &= \frac{2 (7Z^5-7Z^4-104Z^3+8Z^2-71Z+23)}{9 (1-Z)^3 
(1+Z)^2} + \alpha \frac{4 (1+Z)}{3 (1-Z)^3} + \frac{16 (1-Z)}{3 (1+Z)^3} \ln 
\left( \frac{1-Z}{2} \right) \eqend{,} \\
G^{(\text{S},3)}(Z) &= \frac{8 (37 - 192 Z + 192 Z^2 - 310 Z^3 - 21 Z^4 + 6 
Z^5)}{9 (1-Z)^3 (1+Z)^2} + \alpha \frac{8 (-17 - 8 Z + Z^2)}{9 (1-Z)^3} - 
\frac{32 (1-Z)^2}{9 (1+Z)^3} \ln \left( \frac{1-Z}{2} \right) \eqend{,} \\
G^{(\text{S},4)}(Z) &= - \frac{2 (5 + Z) (-5 + Z - 23 Z^2 + 3 Z^3)}{9 (1-Z)^3 
(1+Z)^2} + \alpha \frac{2 (-11 - 2 Z + Z^2)}{9 (1-Z)^3} + \frac{8 (5+Z) (1-Z)}{9 
(1+Z)^3} \ln \left( \frac{1-Z}{2} \right) \eqend{,} \\
G^{(\text{S},5)}(Z) &= \frac{2 (Z^3+3Z^2+27Z-7)}{9 (1-Z)^3 (1+Z)^2} - \alpha 
\frac{2 (3-Z)}{9 (1-Z)^3} - \frac{8 (3+Z)}{9 (1+Z)^3} \ln \left( \frac{1-Z}{2} 
\right) \eqend{,}
\end{equations}
and for $\beta = 2/3$
\begin{equations}[G_S_4D_beta_23]
G^{(\text{S},1)}(Z) &= - \frac{8 Z}{3 (1-Z)^3} + (9-\alpha) \left[ \frac{2 Z 
(5Z^3-3Z^2+7Z-1)}{9 (1-Z)^3 (1+Z)^2} - \frac{4 (1-Z) Z}{9 (1+Z)^3} \ln \left( 
\frac{1-Z}{2} \right) \right] \eqend{,} \\
G^{(\text{S},2)}(Z) &= \frac{4 (1+Z) (5+Z)}{3 (1-Z)^3} - (9-\alpha) \left[ 
\frac{8 (3Z^4-3Z^3+13Z^2-3Z+2)}{9 (1-Z)^3 (1+Z)^2} + \frac{8 (2-Z) (1-Z)}{9 
(1+Z)^3} \ln \left( \frac{1-Z}{2} \right) \right] \eqend{,} \\
G^{(\text{S},3)}(Z) &= - \frac{16 (13 + 10 Z + Z^2)}{3 (1-Z)^3} + (9-\alpha) 
\left[ \frac{8 (-11 - 42 Z - 24 Z^2 - 22 Z^3 + 3 Z^4)}{9 (1-Z)^3 (1+Z)^2} + 
\frac{16 (1-Z)^2}{9 (1+Z)^3} \ln \left( \frac{1-Z}{2} \right) \right] \eqend{,} 
\\
G^{(\text{S},4)}(Z) &= - \frac{16 (2+Z)}{3 (1-Z)^3} + (9-\alpha) \left[ \frac{8 
(-3 - Z - 9 Z^2 + Z^3)}{9 (1-Z)^3 (1+Z)^2} - \frac{16 (1-Z)}{9 (1+Z)^3} \ln 
\left( \frac{1-Z}{2} \right) \right] \eqend{,} \\
G^{(\text{S},5)}(Z) &= - \frac{8}{3 (1-Z)^3} - (9-\alpha) \left[ \frac{16 Z}{9 
(1-Z)^3 (1+Z)^2} - \frac{8}{9 (1+Z)^3} \ln \left( \frac{1-Z}{2} \right) \right] 
\eqend{.}
\end{equations}
As for the vector- and tensor-sector coefficients $G^{(\text{TV},k)}$, we find
\begin{equations}[G_VT_4D]
\begin{split}
G^{(\text{TV},1)}(Z) &= - \frac{2 (-27 + 89 Z - 81 Z^2 + 27 Z^3)}{27 (1-Z)^3} + 
(5-3\alpha) \frac{2 Z (-4 + 19 Z - 73 Z^2 - 15 Z^3 + 33 Z^4)}{135 (1-Z)^3 
(1+Z)^2} \\
&\quad- (5-3\alpha) \frac{4 Z (8 + 9 Z + 3 Z^2)}{45 (1+Z)^3} \ln \left( 
\frac{1-Z}{2} \right) \eqend{,}
\end{split} \\
\begin{split}
G^{(\text{TV},2)}(Z) &= \frac{2 (1+Z) (17 - 12 Z + 3 Z^2)}{9 (1-Z)^3} - 
(5-3\alpha) \frac{2 (21 - 79 Z - 14 Z^2 - 14 Z^3 + Z^4 + 5 Z^5)}{45 (1-Z)^3 
(1+Z)^2} \\
&\quad- (5-3\alpha) \frac{4 (1-Z) (5 + 4 Z + Z^2)}{15 (1+Z)^3} \ln \left( 
\frac{1-Z}{2} \right) \eqend{,}
\end{split} \\
G^{(\text{TV},3)}(Z) &= 4 G^{(\text{TV},4)}(Z) - 4 G^{(\text{TV},2)}(Z) 
\eqend{,} \\
\begin{split}
G^{(\text{TV},4)}(Z) &= \frac{-439 + 668 Z - 478 Z^2 + 180 Z^3 - 27 Z^4}{27 
(1-Z)^3} - (5-3\alpha) \frac{2 (1-Z) (25 + 26 Z + 9 Z^2)}{45 (1+Z)^3} \ln \left( 
\frac{1-Z}{2} \right) \\
&\quad- (5-3\alpha) \frac{77 + 168 Z + 491 Z^2 - 264 Z^3 - 109 Z^4 + 144 Z^5 - 
27 Z^6}{135 (1-Z)^3 (1+Z)^2} \eqend{,}
\end{split} \\
G^{(\text{TV},5)}(Z) &= - 2 G^{(\text{TV},1)}(Z) - \frac{1}{2} 
G^{(\text{TV},2)}(Z) \eqend{.}
\end{equations}
\end{widetext}
The results of Ref.~\cite{higuchi_kouris_cqg_2001b} are given by coefficients 
$f^{(\text{TV},k)}$ and $f^{(\text{S},k)}$ multiplying certain linear 
combinations of the bitensor set~\eqref{S_set}. Hence, the split between the 
scalar and the 
vector and tensor sectors in the Euclidean approach used in 
Ref.~\cite{higuchi_kouris_cqg_2001b} is different from ours. This is expected 
because each of our scalar, vector and tensor sectors satisfy the field 
equations separately whereas 
those in the Euclidean approach do not.

Converting their expression into the bitensor set~\eqref{S_set}, we obtain
\begin{equations}
\begin{split}
G^{(\text{TV},1)} + G^{(\text{S},1)} &= \frac{1}{16} f^{(\text{TV},1)} - 
\frac{1}{2} f^{(\text{TV},2)} + \frac{1}{16} f^{(\text{S},1)} \\
&\quad- \frac{1}{2} f^{(\text{S},2)} - \frac{1}{2} f^{(\text{S},4)} + 
f^{(\text{S},5)} \eqend{,} \raisetag{1.1\baselineskip}
\end{split} \\
G^{(\text{TV},2)} + G^{(\text{S},2)} &= - \frac{1}{4} f^{(\text{TV},1)} - 
\frac{1}{4} f^{(\text{S},1)} + f^{(\text{S},4)} \eqend{,} \\
G^{(\text{TV},3)} + G^{(\text{S},3)} &= f^{(\text{TV},1)} + 4 f^{(\text{TV},3)} 
+ f^{(\text{S},1)} + 4 f^{(\text{S},3)} \eqend{,} \\
G^{(\text{TV},4)} + G^{(\text{S},4)} &= f^{(\text{TV},3)} + f^{(\text{S},3)} 
\eqend{,} \\
G^{(\text{TV},5)} + G^{(\text{S},5)} &= f^{(\text{TV},2)} + f^{(\text{S},2)} 
\eqend{,}
\end{equations}
and using that their $z$ is related to $Z$ by $z = (1+Z)/2$, we find full 
agreement for the case $\beta = 2/3$ where explicit expressions were presented. 
Furthermore, since only the scalar-sector two-point function depends on $\beta$, 
one 
readily finds from Eq.~\eqref{scalar_sector_two_point_function} that there is 
also agreement for all other values of $\beta>0$.

\bibliography{fhlrev_v10}

\end{document}